\RequirePackage{rotating}
\documentclass[fleqn,usenatbib]{mnras2}
\usepackage{lastpage}
\usepackage{verbatim}

\usepackage{txfonts}

\usepackage{graphicx}	
\usepackage{amssymb}	

\usepackage{dcolumn}
\usepackage{bm}
\usepackage[utf8x]{inputenc}
\usepackage[T1]{fontenc}
\usepackage{soul}
\usepackage[normalem]{ulem}

\usepackage{fancyhdr}
\usepackage{longtable}
\usepackage{tabularx}
\usepackage{booktabs, threeparttable, stackengine}
\usepackage{gensymb}
\usepackage{caption}
\usepackage{xcolor}
\usepackage{gensymb}
\usepackage{rotating}
\usepackage{float}
\usepackage{natbib}
\usepackage{siunitx}
\usepackage{subcaption}
\usepackage{tablefootnote}

\usepackage{floatpag}
\usepackage{lipsum}  
\usepackage{afterpage}
\usepackage{enumitem}
\DeclareCaptionFormat{cont}{#1 (cont.)#2#3\par}

\newcolumntype{d}[1]{D{.}{\cdot}{#1}}

\newcolumntype{.}{D{.}{.}{-1}}

\newcommand{\lsun}{L$_\odot$}
\newcommand{\msun}{M$_\odot$}

\newcommand{\lbol}{\emph{L}$_{\rm{bol}}$}

\newcommand{\mum}{$\mu$m}

\newcommand{\kms}{\ensuremath{\textrm{km\,s}^{-1}}}

\newcommand{\hii}{H{\sc ii}}
\newcommand{\uchii}{UC\,H{\sc ii}}
\newcommand{\hchii}{HC\,H{\sc ii}}

\newcommand{\nelectron}{$\textit{n}_{\textup{e}}$}
\newcommand{\Telectron}{$\textit{T}_{\textup{e}}$}
\newcommand{\nly}{$\textit{N}_{\textup{Ly}}$}
\newcommand{\vt}{$\textit{$\nu$}_{\textup{t}}$}

\newcommand{\miriad}{\texttt{MIRIAD}}

\title[HC\,\hii\,regions]{SCOTCH $-$ III: Complete search for Hypercompact\,\hii\ regions in the fourth quadrant  }

\author[A.L Patel et al.]{A.\,L.\,Patel,$^{1}$\thanks{E-mail: alp48@kent.ac.uk}
J.\,S.\,Urquhart,$^{1}$\thanks{E-mail: j.s.urquhart@kent.ac.uk}
A.\,Y.\,Yang,$^{2,3}$\thanks{E-mail: yangay@nao.cas.cn}
L.\,K.\,Morgan,$^{4}$
K.\,M.\,Menten,$^{5}$
\newauthor M.\,A.\,Thompson,$^{6}$
T.\,Moore,$^{7}$
I.\,Grozdanova,$^{1}$
S.\,Khan,$^{5}$
T. Csengeri,$^{8}$
\\
$^{1}$ Centre for Astrophysics and Planetary Science, University of Kent, Canterbury, CT2\,7NH, UK \\
$^{2}$ National Astronomical Observatories, Chinese Academy of Sciences, A20 Datun Road, Chaoyang District, Beijing 100101, P. R. China\\
$^{3}$ Key Laboratory of Radio Astronomy and Technology, Chinese Academy of Sciences, A20 Datun Road, Chaoyang District, Beijing, 100101, P.R. China\\
$^{4}$ Green Bank Observatory, 155 Observatory Rd, Green Bank, WV 24944, USA\\
$^{5}$ Max Planck Institute for Radio Astronomy, Auf dem H\"ugel 69, 53121 Bonn, Germany\\
$^{6}$ School of Physics and Astronomy, University of Leeds, Leeds LS2 9JT, UK\\
$^{7}$ Astrophysics Research Institute, Liverpool John Moores University, Liverpool Science Park, 146 Brownlow Hill, Liverpool, L3\,5RF, UK\\
$^{8}$ Laboratoire d’astrophysique de Bordeaux, Univ. Bordeaux, CNRS, B18N, allée Geoffroy Saint-Hilaire, F-33615 Pessac, France\\
}

\date{Accepted XXX. Received YYY; in original form ZZZ}

\pubyear{2021}

\begin{document}
\label{firstpage}
\pagerange{\pageref{firstpage}--\pageref{lastpage}}
\maketitle

\begin{abstract}
We present high-frequency (18–24\,GHz) radio continuum observations towards 335 methanol masers, excellent signposts for young, embedded high-mass protostars. These complete the search for hypercompact \hii\ (\hchii) regions towards young high-mass star-forming clumps within the fourth quadrant of the Galactic plane. \hchii\ regions are the earliest observable signatures of radio continuum emission from high-mass stars ionizing their surroundings, though their rarity and short lifetimes make them challenging to study. We have observed methanol maser sites at 20-arcsec resolution and identified 121 discrete high-frequency radio sources. Of these, 42 compact sources are embedded in dense clumps and coincide with methanol masers, making them as excellent HC\,\hii\ region candidates. These sources were followed up at higher resolution (0.5-arcsec) for confirmation. We constructed spectral energy distributions across 5–24\,GHz to determine their physical properties, fitting either a simple \hii\ region model or a power-law as needed. This analysis identified 20 \hchii\ regions, 9 intermediate objects, 3 \uchii\ regions, and 3 radio jet candidates. Combining these results with previous findings, the SCOTCH survey has identified 33 \hchii\ regions, 15 intermediate objects, 9 \uchii\ regions, and 4 radio jet candidates, tripling the known number of \hchii\ regions. Eleven of these sources remain optically thick at 24\,GHz. This survey provides a valuable sample of the youngest \hii\ regions and insights into early massive star formation.

\end{abstract}




\section{Introduction}
The study of embedded \hii\ regions has been a topic of interest in astrophysics for several decades \citep{wood1989a,wood1989b,kurtz1994,churchwell2002,Kurtz2005,giveon2005b}. These compact regions of ionized gas surround newly formed massive stars and provide a unique window into the early stages of massive star formation and evolution. The youngest \hii\ region stage is commonly known as a hyper-compact (HC)\,\hii\ region and is characterised by its high electron density (\nelectron) of $\geq$ $10^5$ cm$^{-3}$, emission measure (EM) of $\geq$ $10^8$ pc cm$^{-6}$, and physical diameter $\leq$ 0.05\,pc \citep{kurtz1999,hoare2007}. HC\,\hii\ regions are also associated with strong radiation fields, complex kinematics and radio recombination lines (RRLs) with a line width of frequently $\Delta$V $\geq$ 40\,\kms\ \citep{sewilo2004,murphy2010}. The HC\,\hii\ region stage provides the earliest indication that a massive star is beginning to ionize its surroundings, and it represents a key stage in the development of high mass stars. However, the rarity of  HC\,\hii\ regions and their rapid evolution make them challenging to study \citep{Comeron1996,Gonzalez2005}.

Young \hii\ regions are deeply embedded in dense molecular clumps, making them opaque at visible and near-infrared wavelengths. However, their dusty envelopes are optically thin to thermal bremsstrahlung emission from the ionized gas, allowing it to penetrate through the dense molecular material. Therefore, most surveys of young \hii\ regions have been been conducted via observations of radio continuum emission, many at frequencies around and higher than 5\,GHz   \citep[e.g.][]{wood1989b,kurtz1994,purcell2013,Irabor2023}. 

To date, many hundreds of young \hii\ regions have been identified and characterised (e.g., \citealt{urquhart2013_cornish, Kalcheva2018, Irabor2023}). However, as a result of their small sizes and high densities, HC\,\hii\ regions are optically thick at 5\,GHz, making them difficult to detect and limiting the number of known HC\,\hii\ regions to just 23 (\citealt{Yang2019, Yang2021}). Furthermore, given that the majority of the known HC\,\hii\ regions were serendipitous discoveries in 5-GHz surveys, the sample is biased towards the brightest and most evolved regions, which may not be representative of this stage.

\begin{figure*}
    \centering
    \includegraphics[width=0.95\textwidth]{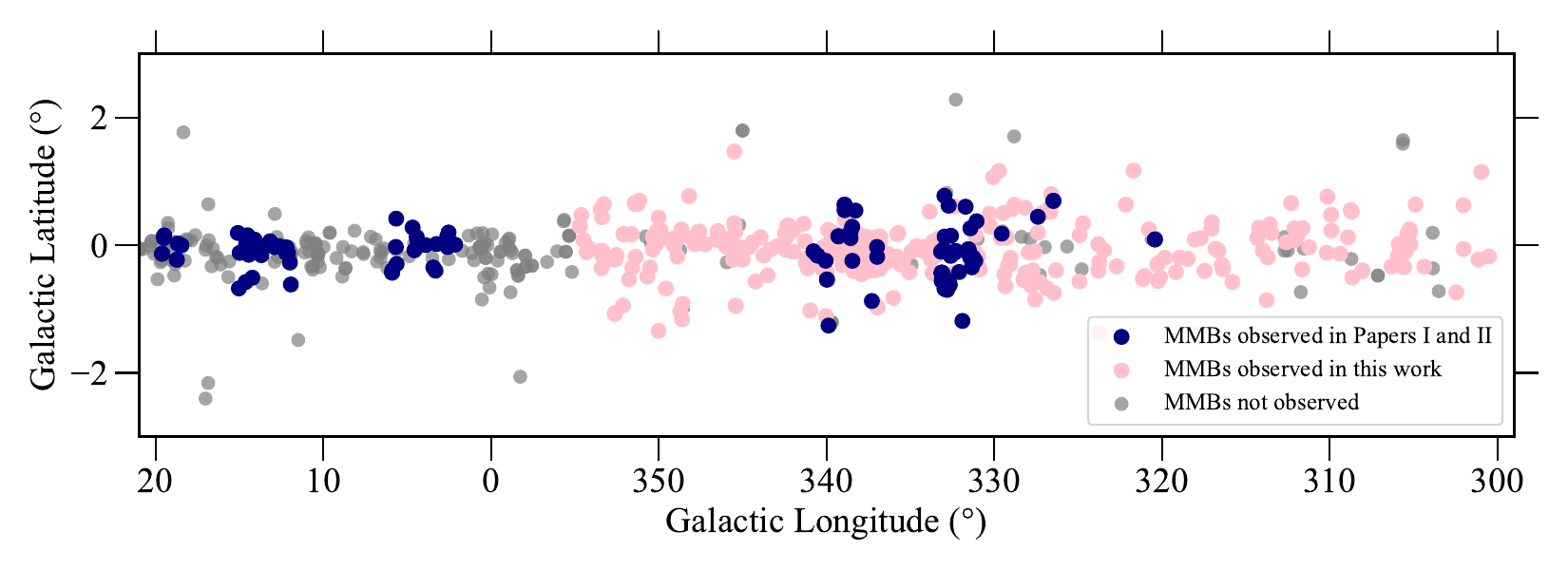}
    \caption{The distribution of MMB methanol masers (grey filled circles) observed as part of the SCOTCH project. The blue filled circles represent the MMB masers observed in Papers\,I and II while the pink filled circles are the methanol masers observed in this work. Sources between $20\degr - 2\degr$ are excluded in the analysis of this paper.}
    \label{fig:dist_of_sources}
\end{figure*}

The Search for Clandestine Optically Thick Compact \hii\ Regions 
(SCOTCH) is a project that has been developed with two key objectives: First, to increase the number of known HC\,\hii\ regions and characterize their physical properties. Second, to produce a more representative sample of HC\,\hii\ regions by targeting methanol masers, which are considered the earliest signposts of high-mass star formation. In \citealt{Patel2023, Patel2024}; (hereafter Papers I and II, respectively), we have identified and characterised a sample of 13 new HC\,\hii\ regions, based on the analysis of archival 23.7\,GHz observations toward a sample of 141 methanol masers from the Methanol Multi-Beam survey (MMB; \citealt{green2010_mmb}). 
Papers I and II observed 141 MMB sources, 57 of which are between Galactic longitude $320\degr$ to $340\degr$, and the remaining 84 are between Galactic longitude $2\degr$ and $20\degr$. Figure\,\ref{fig:dist_of_sources} shows the distribution of all MMB sources in the region covered by the SCOTCH survey; blue filled circles identify the MMB sources observed as part of Papers I and II, the pink filled circles indicate the methanol masers observed in this work, while those in grey have not been observed.

In this paper, we present both moderate and higher resolution radio continuum observations at 18 and 24\,GHz of a more complete sample of 6.7\,GHz MMB masers located within the range 300\degr < $l$ < 355\degr, where $l$ is Galactic longitude. The structure of the paper is as follows: the observations, data reduction, imaging and source extraction are described in Section\,\ref{sec:observationsanddatareduction}. Section\,\ref{sec:results} presents the detection statistics for both resolutions and frequencies. In Section\,\ref{sec:sedmodelandphysicalproperties}, we derive the physical properties of our \hii\ regions and compare their characteristics with known compact \hii\ regions. We compare our sample of \hii\ regions with the global population of compact \hii\ regions, summarizing the key similarities and differences across the different classes. In Section\,\ref{sec:conclusions} we summaries our results and highlight the main findings.

\section{Observations and Data Reduction}
\label{sec:observationsanddatareduction}
In this study, we complete the search for HC\,\hii\ regions towards methanol masers located in the fourth quadrant of the Galactic plane ($300\degr \leq \ell \leq 355\degr$), for which no high-frequency (i.e., 18-24\,GHz) continuum observations were previously available. Radio continuum observations were made during July 2022 and August 2023 using the Australia Telescope Compact Array (ATCA). The observations were conducted at 18\,GHz and 24\,GHz, which corresponds to wavelengths of 
$\sim$1.7 \& $\sim$1.4 cm, respectively.

\subsection{Observational setup}

We use the ATCA in two array configurations: the hybrid H214 for low-angular resolution observations of all target sources, and the linear 6A for high-angular resolution follow-up of candidate HC\,\hii\ regions identified in the low-resolution data. These configurations provide an angular resolution of $\sim$ 20 and 0.5\,arcsec respectively. The data were correlated with
the Compact Array Broadband Backend \citep[CABB,][]{wilson2011}. The CABB was configured with 2 $\times$ 2\,GHz continuum bands with 32 $\times$ 64\,MHz channels (\citealt{wilson2011}). The continuum bands were centred at 18 and 24\,GHz providing coverage from 17 to 19\,GHz and 23 to 25\,GHz. We simultaneously measured 5 zoom windows centred at 17.3, 18.0, 18.7, 23.4 and 24.5\,GHz targeting the H72$\alpha$, H71$\alpha$, H70$\alpha$, H65$\alpha$ and H64$\alpha$ RRLs, respectively \citep{Brown1978}. Here we present the results obtained from the two continuum bands, the results from the RRLs will be presented in a future paper. In Table\,\ref{tab:array_config}, we present a summary of the array configurations, observation dates and numbers of sources observed.

The low-resolutions observations were carried out over four days in June and July 2022. The high-resolution observations were collected over three days in August the following year. The observational parameters for both resolutions and frequencies are given in Table\,\ref{tab:obs_parameters}. The set up and procedure were the same for all observations with the sources being divided into three 20-degree blocks and observations of each field typically consisting of five or six $\sim2$-minute snapshot observations, which were spaced over a range of hour angles to maximize $uv$-coverage. To correct for fluctuations in the phase and amplitude caused by atmospheric and instrumental effects, each block was sandwiched between two short observations ($\sim$ 1\,min) of a nearby phase calibrator (typically within 10\degr\ of the targets). The primary flux calibrator (1934-648) and bandpass calibrator (1252$-$055) were observed once during each set of observation for approximately 10-mins each, to allow the absolute calibration of the flux density and bandpass. The only difference between the observations is that the low-resolution sources were observed over a single day with a total integration time of $\sim$10 minutes, while the high-resolution sources were observed over three days with a total integration time of $\sim$30 minutes. This sensitivity is sufficient to detect an \hii\ region powered by a zero-age main sequence (ZAMS) star of spectral type B0.5 or earlier at a distance of 20\,kpc \citep{Anderson2014}.

\begin{figure}
    \centering
    \includegraphics[width=0.455\textwidth]{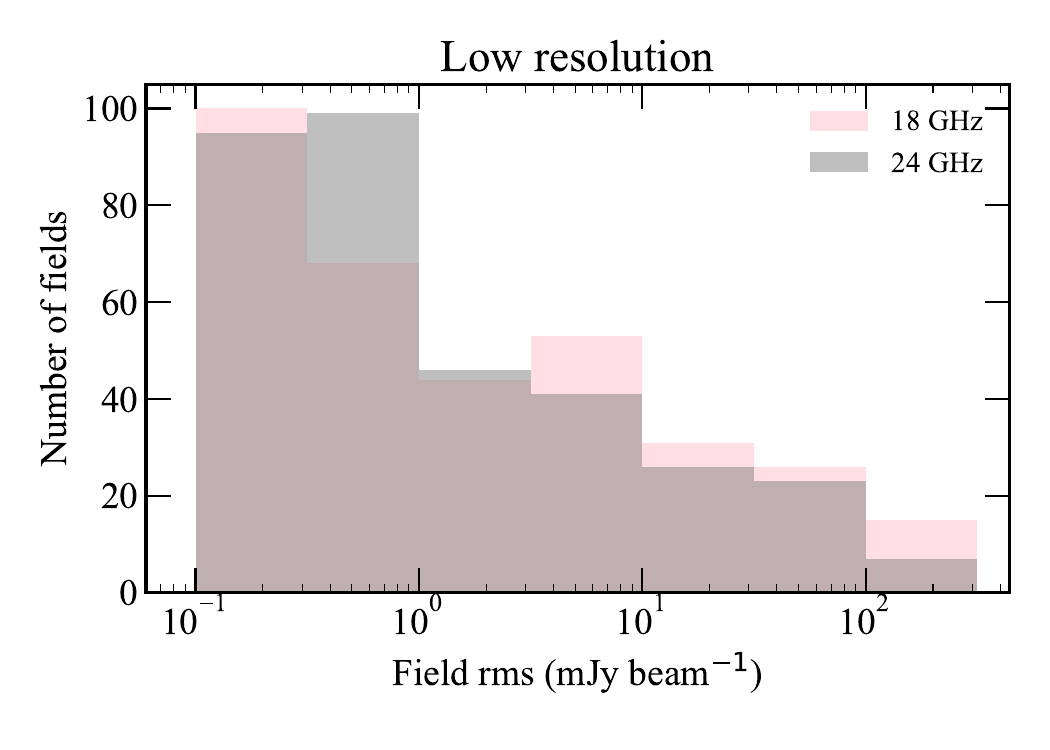}
    \includegraphics[width=0.455\textwidth]{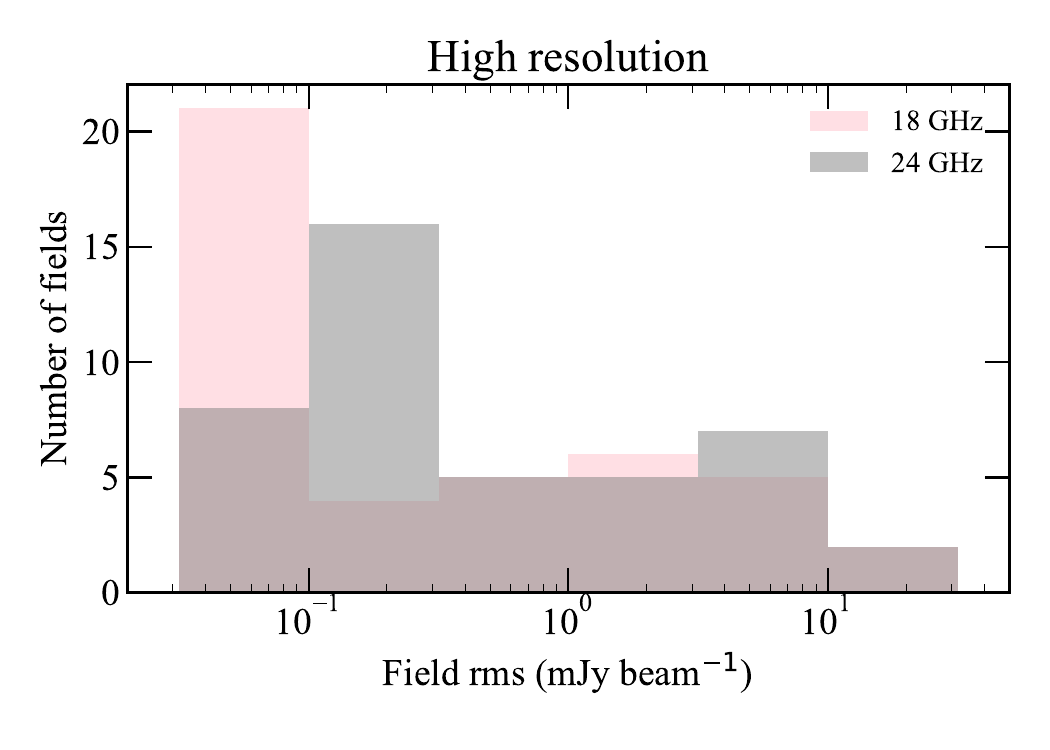}
    
     \caption{Histograms of the number of fields observed as a function of the map RMS noise at both frequencies and resolutions. The data have been binned using a value of 0.5\,dex.}
    \label{fig:rms_hist}
\end{figure}

\begin{figure*}
    \centering
    \includegraphics[width=0.45\textwidth, height=0.35\textwidth]{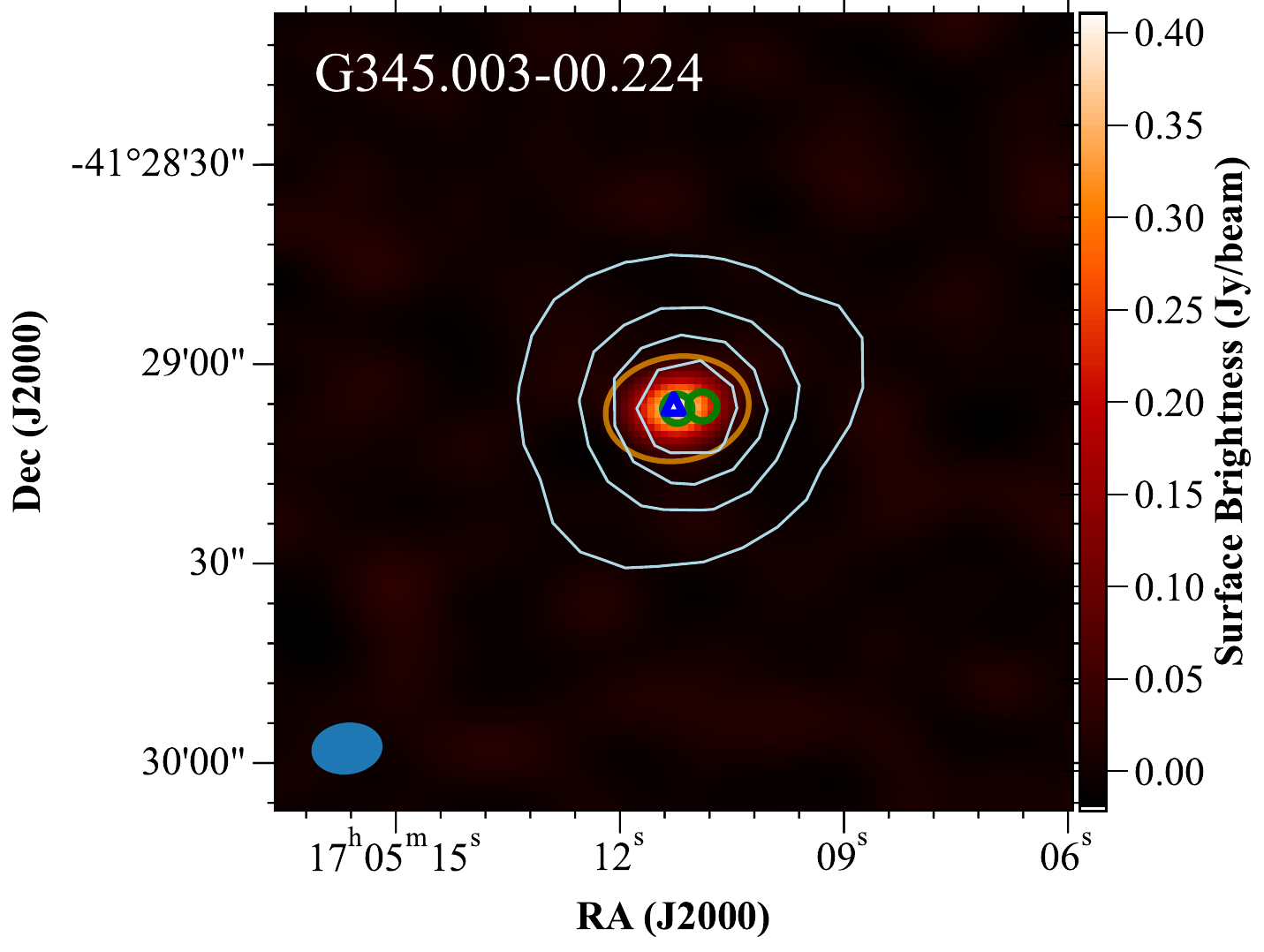}
    \includegraphics[width=0.45\textwidth, height=0.35\textwidth]{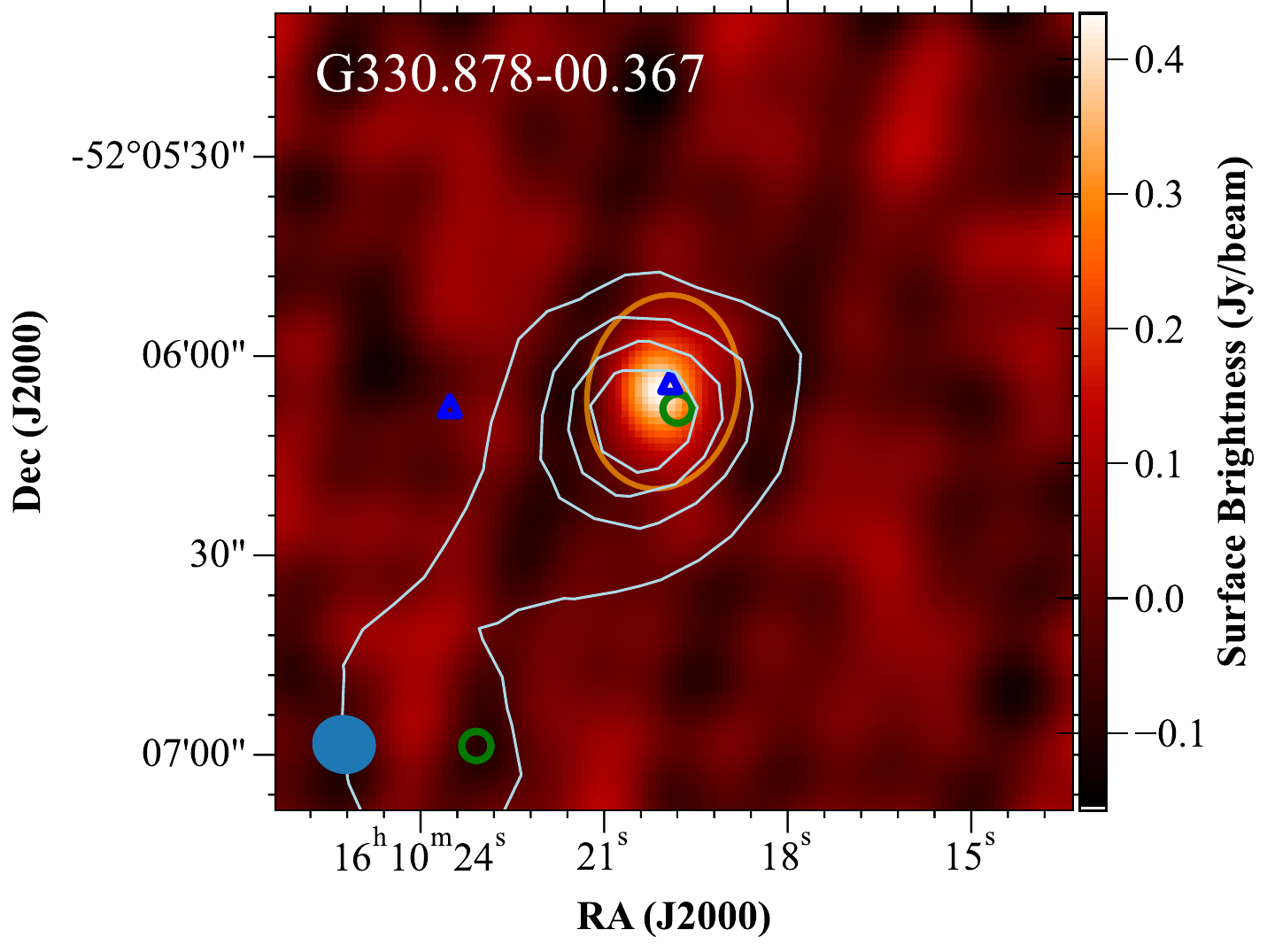}
    \includegraphics[width=0.45\textwidth, height=0.35\textwidth]{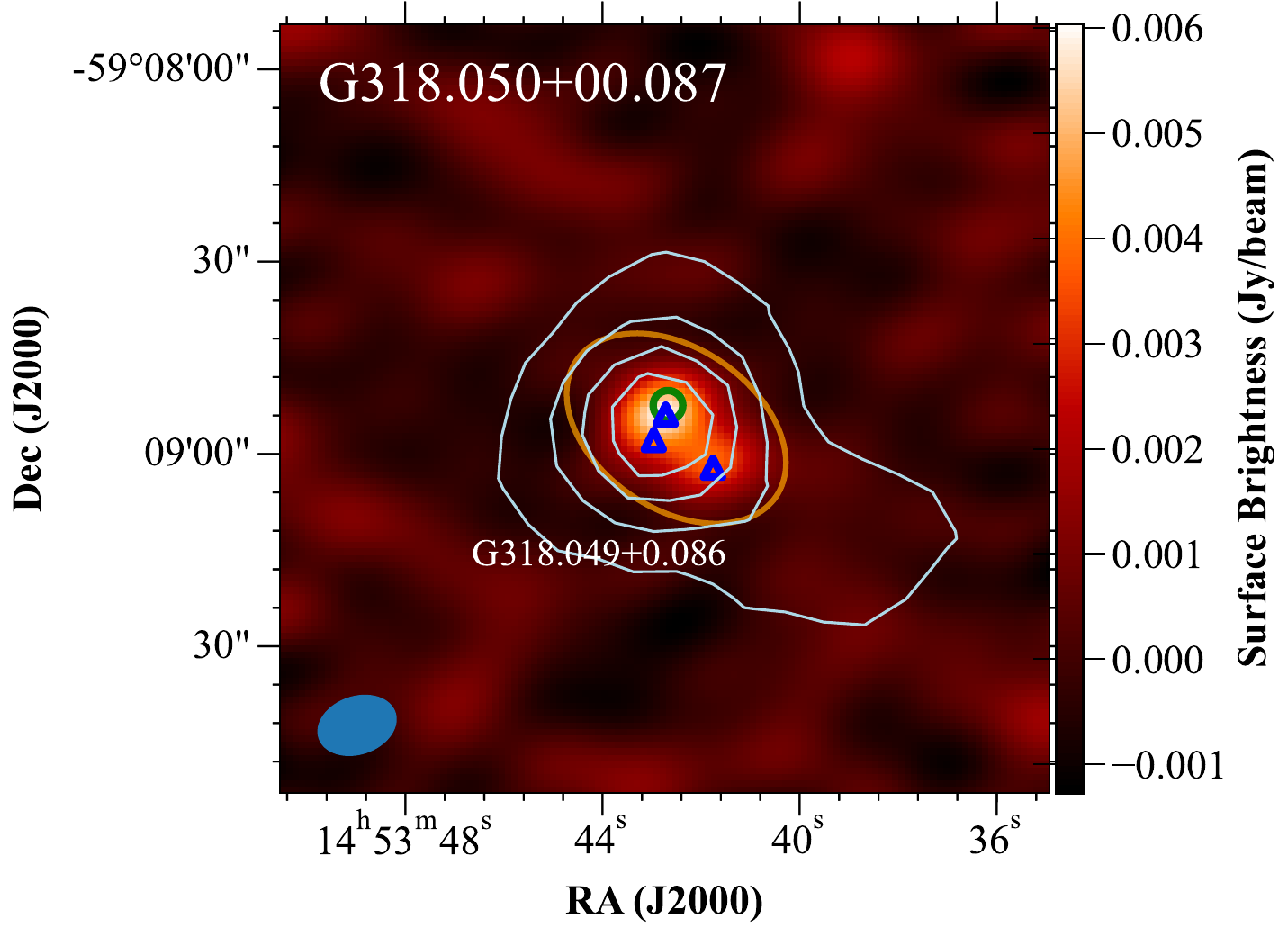}
    \includegraphics[width=0.45\textwidth, height=0.35\textwidth]{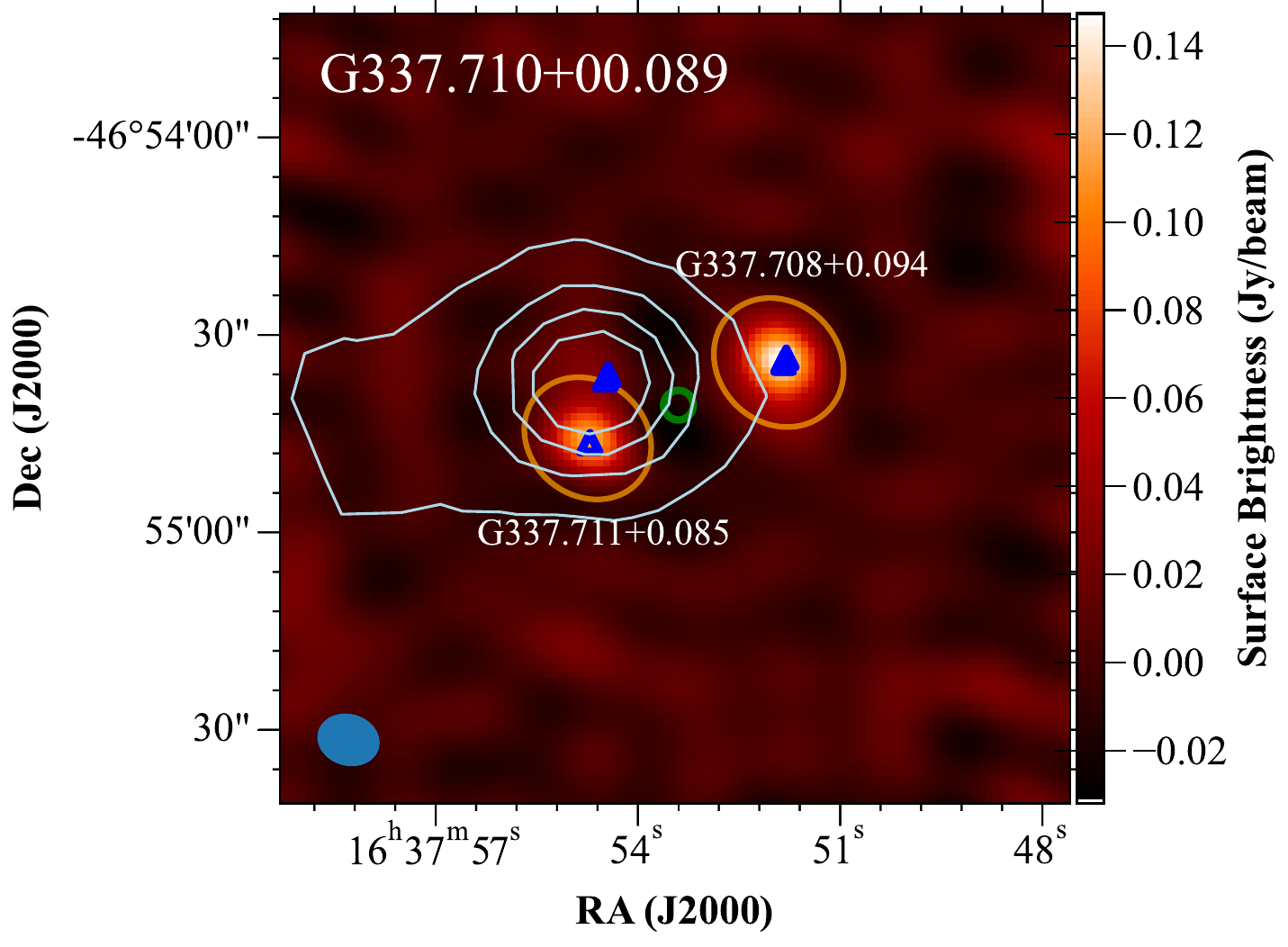}
    \caption{Examples of four 18\,GHz radio maps with different emission types. The top left panel presents a single point source (G345.004$-$0.224), the top right panel shows an example of a single extended radio source (G330.878$-$00.367), the bottom left panel shows the multi-peaked radio source (G318.049+0.086) and the bottom right panel shows an example of a field with two radio sources (G337.708+0.094 \& G337.711+0.085). The orange ellipse shows the resultant fit to the radio emission while the green circles show the position of the methanol maser(s) located in the field. The blue triangles show the position of any low-frequency (5-GHz) radio counterparts. The grey contours trace the 870-$\mu$m dust emission from ATLASGAL  (\citealt{schuller2009}). The filled blue ellipse in the bottom left-hand corner of each image indicates the size and orientation of the synthesised beam. The field name is given in the top left and the radio name is provided if it differs from the field name.} 
    \label{fig:radio_map_examples}
  \end{figure*}
  \setlength{\tabcolsep}{18pt}
\begin{table}
\centering
  \caption{Summary of the observation dates, array configurations used and the number of sources observed at each epoch.}
    \begin{tabular}{cccc}
    \hline
    \multicolumn{1}{c}{{Array config.}}& 
    \multicolumn{1}{c}{{Date}} &
    \multicolumn{1}{c}{{\# of Obs.}} \\   
    \hline
H214  & 27/06/2022 & 84 \\
      & 06/07/2022 & 80 \\
      & 15/07/2022 & 85 \\
      & 22/07/2022 & 88 \\
    \hline
6A    & 01-03/08/2023 & 67 \\

    \hline
    \end{tabular}%
  \label{tab:array_config}%
\end{table}%

\setlength{\tabcolsep}{5pt}
\begin{table}
 
  \caption{Summary of low- and high-resolution observational parameters.}
  \label{tab:obs_parameters}%
  \begin{minipage}{\linewidth}
  \begin{center}
    \begin{tabular}{ccc}
    \hline
    \multicolumn{1}{c}{Parameter} &
    \multicolumn{2}{c}{Array config.} \\
    \cline{2-3}
    
    \multicolumn{1}{c}{} &
    \multicolumn{1}{c}{H214} &
    \multicolumn{1}{c}{6A} \\
    
    \hline
Frequency & 18 \& 24\,GHz & 18 \& 24\,GHz \\
Primary beam (FWHM) & $2\rlap{.}'6$ &  $2\rlap{.}'6$ \\
Synthesised beam$^a$ (FWHM) &  
$\sim 5\rlap{.}''25 \times 6\rlap{.}''89$ & $\sim 0\rlap{.}''36 \times 0\rlap{.}''49$\\
Shortest and longest baseline & 82 \& 247m & 337 \& 5939 \\
Largest angular scale $\theta_{LAS}$  & $\sim$ 40$\arcsec$ & $\sim$ 10$\arcsec$ \\
Total on-source integration time & $\sim$ 10 mins & $\sim$30 mins \\
\hline 
    
    \end{tabular}\\%
    \end{center}
    $^a$ The sizes and shapes of the synthesised beam are dependent on the declination of the source. The values presented here correspond to the average declination (-48\degr) of our sample and are provided as a guide.
  \end{minipage}
\end{table}%

\subsection{Source selection and strategy}
A total of 452 MMB methanol maser sites have been found between $300\degr \leq \ell \leq 355\degr$ (\citealt{green2009}). We exclude any sources located towards the Galactic centre region ($355\degr \leq \ell \leq 5\degr$), as for those accurate kinematic distances are not available, limiting any detailed analysis. We exclude the 57 sources that were observed as part of Paper\,I and 36 MMB sources that do not have an ATLASGAL counterpart. To reduce the number of required observations we identify groups of masers in close proximity to each other that can be observed in a single field of view (i.e., within 2\,arcmin). For 24 cases, two masers are within 30\,arcsec of each other, reducing the total number of pointings needed to 335. 

The low-resolution observations were conducted for all 335 identified fields, while high-resolution observations were focused on 42 compact and optically thick \hii\ regions, identified based on the analysis of the low-resolution data. The observational parameters for both resolutions and frequencies are given in Table\,\ref{tab:field_parameters}. These sources were selected as they met the criteria for HC\,\hii\ region (i.e., optically thick between 5-18\,GHz and compact) candidates, making them good targets for high-resolution follow-up observations. As previously mentioned, both the high- and low-resolution observations were made using the  same correlator set up allowing the resulting data to be combined.

\begin{sidewaystable*}

\setlength{\tabcolsep}{3pt}
\caption{Observational field parameters from all observed fields. The field names and pointing centers are based on the MMB names and positions. Fields marked with a '-' in the "Detection" column were not observed as part of the high-resolution follow-up but were included in the low-resolution observations.}
\begin{tabular}{lcccccccccccc}

    \hline
    \multicolumn{1}{c}{ {Field name}} &
    \multicolumn{1}{c}{RA } &
    \multicolumn{1}{c}{Dec.} &
    \multicolumn{5}{c}{Low Resolution} &
    \multicolumn{5}{c}{High Resolution} \\
    
    \multicolumn{1}{c}{ } &
    \multicolumn{1}{c}{{(J2000)}} &
    \multicolumn{1}{c}{{(J2000)}} &
    \multicolumn{1}{c}{rms} &
    \multicolumn{1}{c}{Detection} &
    \multicolumn{1}{c}{Major }  &
    \multicolumn{1}{c}{Minor}  &
    \multicolumn{1}{c}{Position angle} &
    \multicolumn{1}{c}{rms} &
    \multicolumn{1}{c}{Detection} &
    \multicolumn{1}{c}{Major }  &
    \multicolumn{1}{c}{Minor}  &
    \multicolumn{1}{c}{Position angle} \\
    
    \multicolumn{1}{c}{ } &
    \multicolumn{1}{c}{(h:m:s)} &
    \multicolumn{1}{c}{(d:m:s)} &
    \multicolumn{1}{c}{(mJy)} &
    \multicolumn{1}{c}{ } &
    \multicolumn{1}{c}{ (arcsec)} &
    \multicolumn{1}{c}{ (arcsec)}  &
    \multicolumn{1}{c}{ (\degr)} &
    \multicolumn{1}{c}{(mJy)} &
    \multicolumn{1}{c}{ } &
    \multicolumn{1}{c}{ (arcsec)} &
     \multicolumn{1}{c}{ (arcsec)}  &
      \multicolumn{1}{c}{ (\degr)} \\
    
    \hline
G300.504$-$00.176	&	12:30:03.57	&	$-$62:56:48.8	&	3.1	&	Y	&	10	&	8.8	&	$-$86	&	2.4  	&	$-$	&	7.5	&	6.7	&	79	\\
G300.969+01.148 	&	12:34:53.28	&	$-$61:39:39.9	&	40	&	Y	&	10	&	8.8	&	89     	&	33	    &	Y	&	7.5	&	6.8	&	74	\\
G301.136$-$00.226	&	12:35:35.13	&	$-$63:02:32.6	&	59	&	Y	&	10	&	8.8	&	$-$85	&	58    	&	Y	&	7.5	&	6.8	&	78	\\
G302.032$-$00.061	&	12:43:31.92	&	$-$62:55:06.6	&	56	&	Y	&	10	&	8.9	&	88	    &	41   	&	Y	&	7.6	&	6.9	&	78	\\
G302.034+00.625 	&	12:43:43.44	&	$-$62:13:58.4	&	0.11&	Y	&	9.9	&	9.1	&	$-$78	&	0.14	&	N	&	7.6	&	6.7	&	86	\\
G302.455$-$00.741	&	12:47:08.64	&	$-$63:36:30.2	&	0.29&	N	&	9.9	&	9.3	&	$-$76	&	0.19	&	N	&	7.6	&	6.8	&	85	\\
G304.367$-$00.336	&	13:04:09.81	&	$-$63:10:20.2	&	0.13&	N	&	10	&	9.1	&	$-$89	&	0.15	&	N	&	7.4	&	6.9	&	$-$80	\\
G304.887+00.635 	&	13:08:11.95	&	$-$62:10:21.7	&	0.12&	N	&	9.9	&	9.1	&	$-$82	&	0.17	&	N	&	7.5	&	6.7	&	$-$75	\\
G305.199+00.005 	&	13:11:17.20	&	$-$62:46:45.8	&	51	&	N	&	9.3	&	8.7	&	$-$64	&	13    	&	N	&	7.0	&	6.5	&	$-$57	\\
G305.200+00.019 	&	13:11:16.92	&	$-$62:45:55.0	&	100	&	N	&	9.3	&	8.6	&	$-$59	&	60   	&	N	&	7.0	&	6.5	&	$-$47	\\
G305.208+00.206 	&	13:11:13.70	&	$-$62:34:41.5	&	21	&	Y	&	9.1	&	8.6	&	$-$62	&	9.2   	&	$-$	&	6.9	&	6.6	&	$-$35	\\
G305.248+00.245 	&	13:11:32.47	&	$-$62:32:09.2	&	14	&	N	&	9.2	&	8.6	&	$-$84	&	5.3   	&	N	&	6.8	&	6.5	&	$-$33	\\

    \hline
\end{tabular}    
\footnotesize
\begin{tablenotes}
\item Note: only a small portion of the data is provided here, the full table is only available in
electronic form. \\
\end{tablenotes}
  \label{tab:field_parameters}%

\setlength{\tabcolsep}{4pt}
 \centering 
    \caption{Extracted source parameters from our detected radio sources at low resolution. Column 1 is the radio name, which is a combination of the Galactic longitudes and
    latitudes determined by fitting ellipsoidal Gaussians using {\sc IMFIT}, Columns 2 and 3 are the Right Ascension and Declination of each source, respectively. Columns 4-10 present 18-GHz source parameters and columns 11-17 present 24-GHz source parameters. Col 8$-$10 and 15$-$17 are the deconvolved major, minor and position angle, respectively. Sources with $-$ are those for which sizes could not be deconvolved. Sources with a \textdagger\ indicate those with a poor synthesised beam.}
    \begin{tabular}{lcccccccccccccccc}

    \hline
    \multicolumn{1}{c}{} &
    \multicolumn{1}{c}{RA} &
    \multicolumn{1}{c}{Dec.} &
    \multicolumn{7}{c}{18GHz} &
    \multicolumn{7}{c}{24GHz} \\
    \cline{4-9}
    \cline{11-16}
    \multicolumn{1}{c}{} &
    \multicolumn{1}{c}{} &
    \multicolumn{1}{c}{} &
    \multicolumn{1}{c}{} &
    \multicolumn{1}{c}{} &
    \multicolumn{1}{c}{} &
    \multicolumn{1}{c}{} &
    \multicolumn{1}{c}{} &
    \multicolumn{1}{c}{} &
    \multicolumn{1}{c}{} &
    \multicolumn{1}{c}{} &
    \multicolumn{1}{c}{} \\

    \multicolumn{1}{c}{ } &
    \multicolumn{1}{c}{{(J2000)}} &
    \multicolumn{1}{c}{{(J2000)}} &
    \multicolumn{1}{c}{$f_{\rm peak}$} &
    \multicolumn{1}{c}{$\Delta f_{\rm peak}$} &
    \multicolumn{1}{c}{$f_{\rm int}$} &
    \multicolumn{1}{c}{$\Delta f_{\rm int}$} &
    \multicolumn{1}{c}{Major} &
    \multicolumn{1}{c}{Minor} &
    \multicolumn{1}{c}{PA} &
    \multicolumn{1}{c}{$f_{\rm peak}$} &
    \multicolumn{1}{c}{$\Delta f_{\rm peak}$} &
    \multicolumn{1}{c}{$f_{\rm int}$} &
    \multicolumn{1}{c}{$\Delta f_{\rm int}$} &
    \multicolumn{1}{c}{Major} &
    \multicolumn{1}{c}{Minor} &
    \multicolumn{1}{c}{PA} \\
    
    \multicolumn{1}{c}{Radio name} &
    \multicolumn{1}{c}{(h:m:s)} &
    \multicolumn{1}{c}{(d:m:s)} &
    \multicolumn{1}{c}{(mJy beam$^{-1}$)} &
    \multicolumn{1}{c}{(mJy beam$^{-1}$)} &
    \multicolumn{1}{c}{(mJy)} &
    \multicolumn{1}{c}{(mJy)} &
    \multicolumn{1}{c}{(arcsec)} &
    \multicolumn{1}{c}{(arcsec)}  &
    \multicolumn{1}{c}{(\degr)} &
    \multicolumn{1}{c}{(mJy beam$^{-1}$)} &
    \multicolumn{1}{c}{(mJy beam$^{-1}$)} &
    \multicolumn{1}{c}{(mJy)} &
    \multicolumn{1}{c}{(mJy)} &
    \multicolumn{1}{c}{(arcsec)} &  
    \multicolumn{1}{c}{(arcsec)}  &
    \multicolumn{1}{c}{(\degr)} \\
    
    \hline

G300.504$-$0.173	&	12:30:03.89	&	$-$62:56:39.2	&	36	&	0.85	&	54	&	2.2	&	7.8	&	5.2	&	8.8	    &	24	&	0.47&	47	&	2.2	&	8.0	&	6.1	&	18.6	\\
G300.969+1.148  	&	12:34:53.11	&	$-$61:39:39.9	&	500	&	12   	&	580	&	30	&	4.8	&	2.4	&	$-$44.6	&	490	&	35	&	560	&	79	&	3.7	&	1.6	&	$-$50	\\
G301.136$-$0.225	&	12:35:35.01	&	$-$63:02:30.9	&	870	&	33   	&	1100&	84	&	6.5	&	0.9	&	0.4	    &	1000&	110	&	1400&	300	&	6.8	&	1.5	&	$-$4.4	\\
G302.032$-$0.061	&	12:43:31.84	&	$-$62:55:06.4	&	830	&	15   	&	970	&	28	&	4.7	&	2.8	&	$-$37.6	&	670	&	15	&	840	&	35	&	4.2	&	2.7	&	$-$40.1	\\
G302.033+0.626  	&	12:43:43.70	&	$-$62:13:59.3	&	0.48&	0.052	&	0.45&	0.08&	$-$	&	$-$	&	$-$	    &	0.84&	0.07&	0.78&	0.1	&	$-$	&	$-$	&	$-$	\\
G305.199+0.206  	&	13:11:08.63	&	$-$62:34:42.5	&	110	&	2.0  	&	190	&	11	&	9.4	&	5.7	&	87.1	&	53	&	2.2	&	68	&	6.2	&	4.5	&	2.4	&	87.9	\\
G305.215+0.200  	&	13:11:17.52	&	$-$62:34:59.4	&	130	&	3.3  	&	240	&	19	&	10.2&	6.0	&	77.7	&	46	&	1.2	&	83	&	5.8	&	8.0	&	4.1	&	$-$65.6	\\
G305.359+0.151  	&	13:12:34.52	&	$-$62:37:15.9	&	91	&	1.9   	&	140	&	9.4	&	7.8	&	4.6	&	18.7	&	61	&	1.6	&	89	&	6.4	&	7.3	&	0.8	&	$-$31.5	\\
G305.562+0.013  	&	13:14:26.29	&	$-$62:44:27.9	&	50	&	3.8   	&	54	&	8.0	&	$-$	&	$-$	&	$-$   	&	41	&	2.5	&	45	&	4.5	&	$-$	&	$-$	&	$-$	\\
G305.798$-$0.242	&	13:16:42.60	&	$-$62:58:21.8	&	9.7	&	0.44	&	12	&	1.2	&	$-$	&	$-$	&	$-$    	&	8.5	&	0.64&	11	&	1.4	&	4.2	&	2.6	&	$-$22.2	\\
G305.800$-$0.251	&	13:16:44.19	&	$-$62:58:55.3	&	19	&	0.62	&	29	&	1.4	&	6.9	&	6.2	&	$-$82.8	&	11	&	0.89&	24	&	3.0	&	7.6	&	7.5	&	$-$89.8	\\
G305.886+0.017  	&	13:17:15.05	&	$-$62:42:23.4	&	1.1	&	0.097	&	3.8	&	1.5	&	23.2&	6.8	&	$-$27.8	&	1.3	&	0.1	&	1.7	&	0.2	&	4.2	&	$-$	&	$-$	\\

    \hline
\end{tabular}  

\begin{tablenotes}
\item Note: only a small portion of the data is provided here, the full table is only available in
electronic form. \\
\end{tablenotes}
  \label{tab:low_res_parameters}

\end{sidewaystable*}
\subsection{Data reduction and imaging}

The calibration and data reduction for both datasets were performed using the \miriad\ data analysis tool \citep{miriad}, following standard ATCA procedures. We visually inspected the raw data and performed a flagging procedure to eliminate data affected by radio-frequency interference (RFI) and data significantly affected by imaging artefacts. These artefacts include strong sidelobes contamination from bright sources outside the primary beam, undulations caused by undersampling, and poor synthesized beam quality due to inadequate $uv$ coverage. This flagging procedure, consisting of inspecting plots of visibility data, was performed iteratively until all RFI-affected data were removed. The primary flux and phase calibration were then carried out, and the calibration tables were copied over to the target sources, applying the corrections. The calibrated data were subsequently imaged and CLEANed using the \miriad\ tasks \texttt{INVERT}, \texttt{CLEAN}, and \texttt{RESTOR}. The maps were deconvolved using a robust weighting of 0.5 and a couple of hundred cleaning components, or until the first negative component was encountered.

For the low-resolution data, we imaged a region corresponding to the FWHM size of the primary beam using a pixel size of 1\,arcsec and 126 pixels per side, resulting in an image size of 2.1 $\times$ 2.1 arcmin. For the 42 fields, for which we have high resolution observations, we combine the low-resolution and high-resolution visibiliites to improve the signal to noise ratio, $uv$-coverage, and sensitivity to angular scales. These maps are imaged using a pixel size of 0.2 arcsec, with 120 pixels along each side, which resulted in a image size of $\sim 25 \times 25$\,arcsec. Finally, all images are corrected for the primary beam response.  In Figure\,\ref{fig:rms_hist}, we present histograms of the number of fields as a function of the rms noise for both low- and high-angular resolution datasets. The variation in RMS values arises from the characteristics of the interferometric observations. Our maps are limited by a dynamic range of $\sim$ 80, leading to larger RMS values in instances where a bright point source is present within the field. The rms noise values were estimated from emission-free regions close to the centre of the reduced maps. For a detailed description of the data reduction process, see Papers I and II.

\section{Results}
\label{sec:results}

The final reduced maps were examined for compact, high surface-brightness sources, using a nominal 3\,$\sigma$ detection threshold, where $\sigma$ refers to the image RMS noise level. All detections are visually inspected to confirm the presence of a discrete radio source and to distinguish between genuine sources and possible imaging artefacts from over-resolved structures or bright sidelobes. 

\subsection{Low Resolution Detection Statistics}
\label{sect:low_res_detections}

We have identified 121 discrete radio sources located within 116 of the 335 fields imaged (i.e. $\sim$30\%). The radio emission comes from a single point source in 78 fields, a further 32 fields contain a single extended source. An example of a point and extended radio source are given in the top row of Fig\,\ref{fig:radio_map_examples}. Five fields contain 2 distinct radio sources, while the radio emission in the remaining field (G318.050+00.087) cannot be reliably separated. Examples of these cases are presented in the bottom row of Fig\,\ref{fig:radio_map_examples}.

We use the \texttt{MIRIAD} task \texttt{IMFIT} to determine the flux densities, positions and sizes of the 121 radio detections. This was achieved by carefully drawing a polygon around the detection as an input to \texttt{IMFIT}. \texttt{IMFIT} applies a two-component Gaussian fit to determine source parameters and deconvolve the beam from the source. Following the approach in Papers I and II, we apply the same fitting method to all detections to ensure consistency in our source parameter measurements. We present the radio name, RA and Dec, peak and integrated fluxes and source major, minor axes and position angle for our detections in Table\,\ref{tab:low_res_parameters}. Our principal goal in this study is to identify a sample of \hii\ regions with a particular focus on those that have characteristics similar to those of known HC\,\hii\ regions \citep{Yang2021}.

Observationally, HC\,\hii\ regions are characterised as small, optically thick ionized sources embedded within their natal dust clumps. They are often associated with star formation tracers such as maser emission and molecular outflows. To examine the positional correlation of these star-forming tracers and our radio sources, we created radio maps and plotted the positions of star-forming tracers such as: methanol masers from the MMB survey \citet{caswell2010b,green2010_mmb,caswell2011_mmb}, 870\,\mum\ dust emission from ATLASGAL \citet{schuller2009} and 5-GHz radio sources taken from the literature \citet{Irabor2023,becker1994} (See Fig\,\ref{fig:radio_map_examples}. for examples)

We visually inspect these maps to determine whether our high-frequency radio sources are positionally associated with these star-formation tracers (see discussion of association criteria and Figure\,7 of Paper\,I). Of the 121 high-frequency radio detections, we have identified 80 radio sources that are associated with both 870\,\mum\ dust and methanol maser emission. Comparing these \hii\ region candidates with the classifications of their host clumps made by the ATLASGAL team \citep{urquhart2014_csc,urquhart2018,urquhart2022} we find that 54 are classified as being \hii\ regions. A further 13 are classified as YSOs and four are protostellar in nature. Six clumps have been classified to be photo-dissociation regions (PDRs), while the remaining three clumps are associated with a complex mid-infrared environment that makes a definitive classification uncertain. We have inspected the mid-infrared environments of these three objects and find that they are associated with large-scale structures and extended radio emission. It is likely that these objects are evolved \hii\ regions. Consequently, we exclude these three clumps from further analysis, leaving 77 \hii\ region candidates for further investigation.

\begin{figure*}
    \captionsetup{list=off,format=cont}
    \centering
    \includegraphics[width=0.45\textwidth, height=0.35\textwidth]{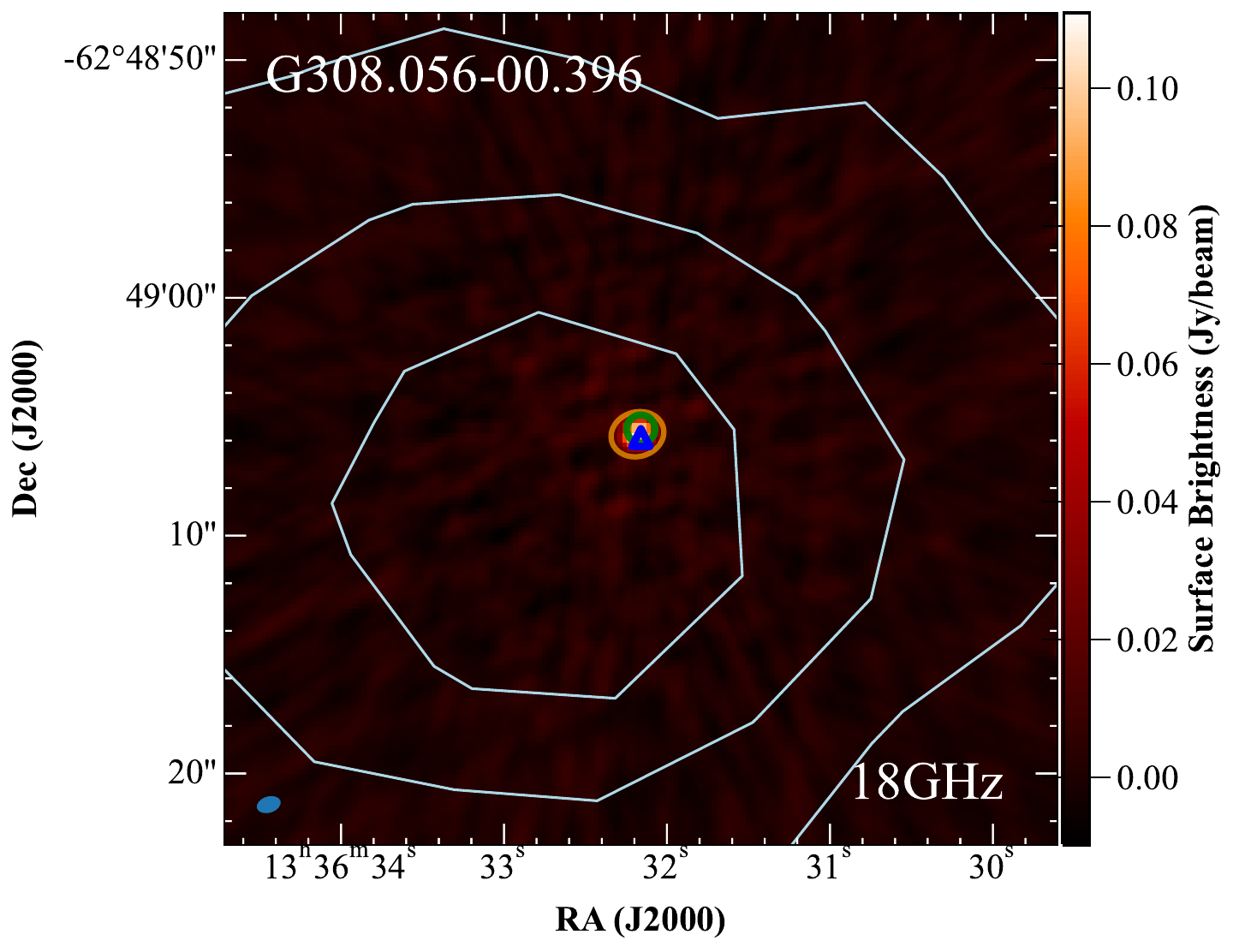}
    \includegraphics[width=0.45\textwidth, height=0.35\textwidth]{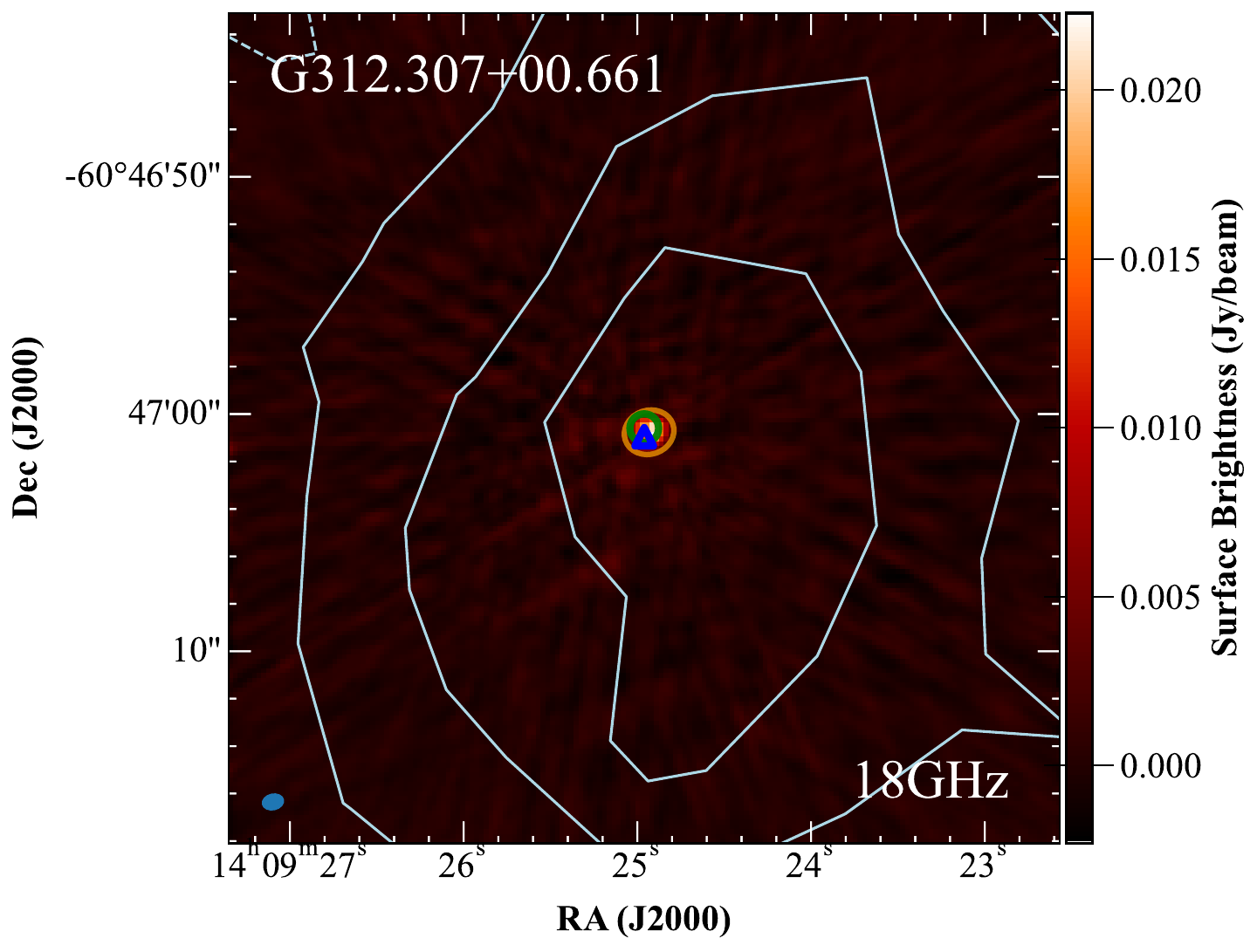}
    \includegraphics[width=0.45\textwidth, height=0.35\textwidth]{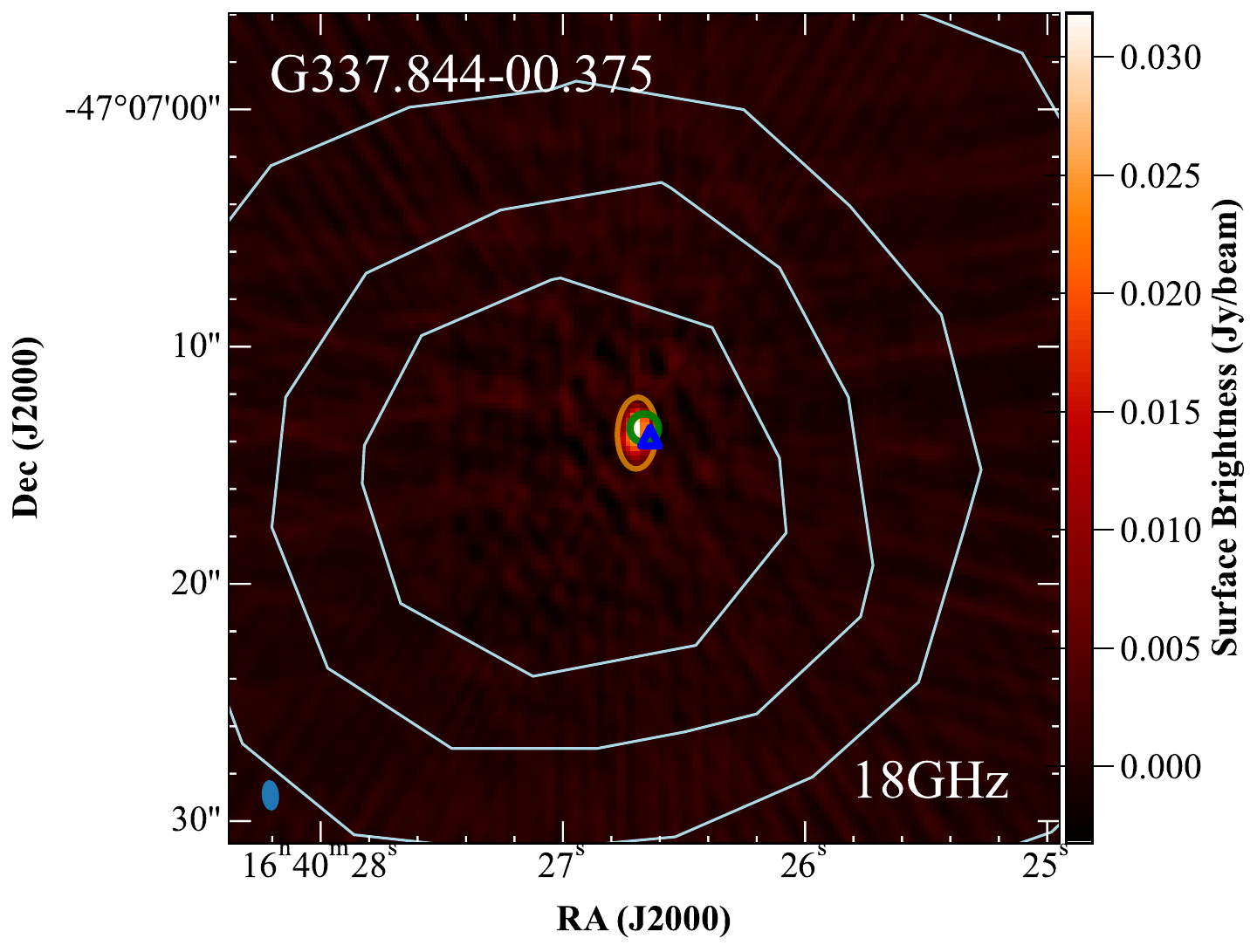}
     \includegraphics[width=0.45\textwidth, height=0.35\textwidth]{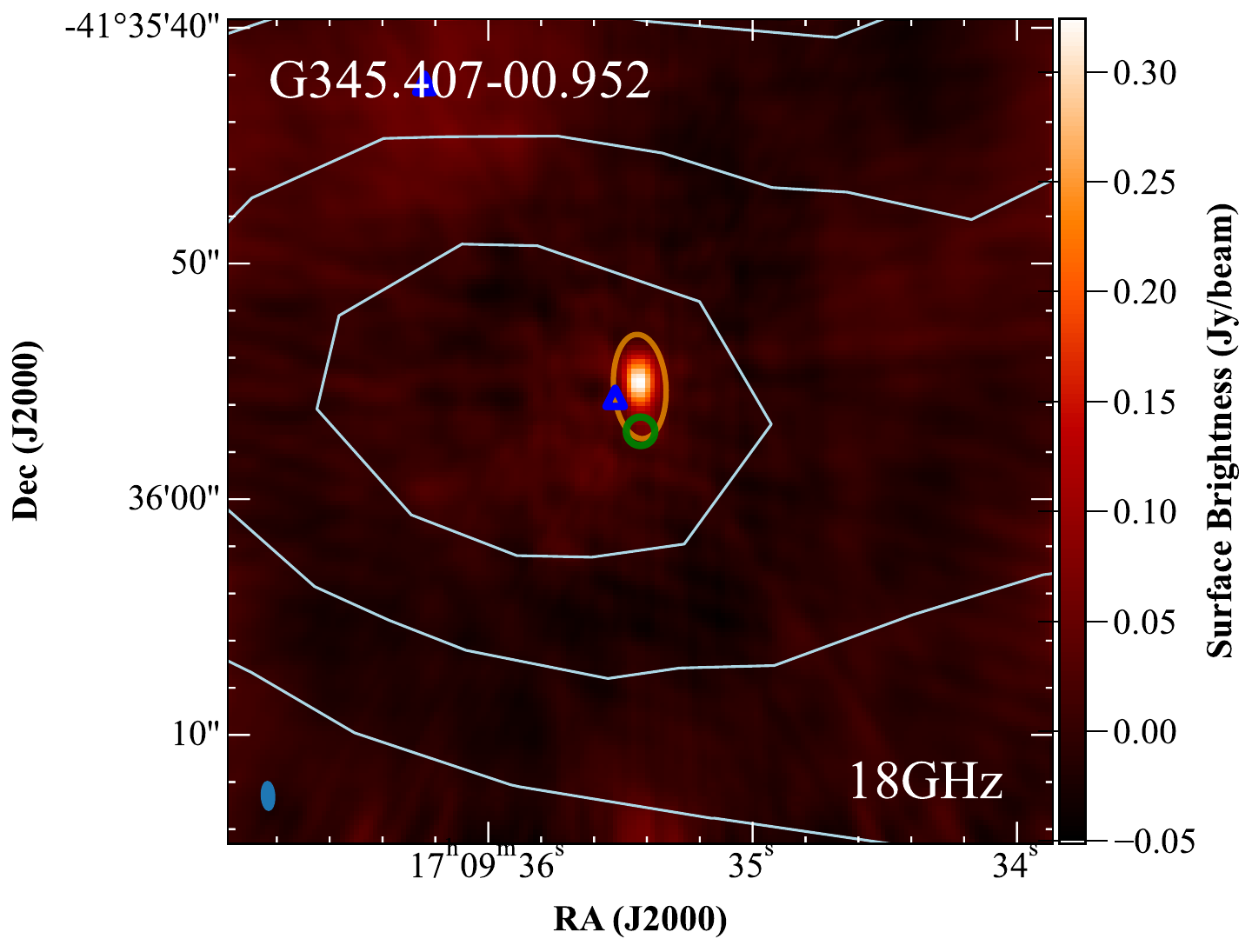}
    
  \caption{Examples of four high resolution 18\,GHz radio detections. The top row presents radio sources G308.056$-$0.396 and G312.307+0.661. These are examples of compact radio sources (y-factor $< 2$). The bottom row presents radio sources G337.844$-$0.375 and G345.407$-$0.952, which are examples of compact but more resolved radio sources (y-factor $< 4$). The symbols are as described in Fig.\,\ref{fig:radio_map_examples}.}
     \label{fig:high_res_examples}
\end{figure*}

Using the available multi-wavelength data from CORNISH-South \citep{Irabor2023} and MAGPIS (for $\ell > 350^\circ$; \citealt{becker1994}), we find that 61 of the 77 radio sources have 5-GHz radio counterparts. Six sources fall outside the nominal latitude range covered by these two 5-GHz surveys, while the remaining 10 lack detectable 5-GHz radio emission. For our 5-GHz radio source sample, we calculate the y-factor, i.e., the ratio of integrated flux density (in Jy) and the peak flux density of brightness (in units of Jy-beam$^{-1}$, $f_{\rm int}/f_{\rm peak}$) and constrain it to $< 2$ to minimize the effects of spatial filtering. This ensures that our sample of 5-GHz radio counterparts are unresolved or compact. We find 17 radio sources have a y-factor $> 2$, and exclude them from the spectral index analysis that follows. This leaves us with a sample of 44 \hii\ regions that are compact and associated with 5-GHz radio emission. These are broken down into 35 CORNISH-South sources and 9 MAGPIS sources.

We expect UC\,\hii\ regions to have turnover frequencies at $\sim$ 5\,GHz, and therefore at higher frequencies, we would expect them to approach a slightly falling, almost flat spectrum ($\alpha$ $\sim$ $-$0.1). In contrast, HC\,\hii\ regions are optically thick at 5 GHz and should have a positive slope (up to $\alpha$ $\sim$ $+2$) between 5 and 18 GHz. We determine the spectral index of these 44 \hii\ region candidates using Eqn.\,1 from Paper\,I. To account for the difference in beam sizes between the 18\,GHz and 5\,GHz observations, we compute the spectral index using the peak flux density at 18\,GHz and the integrated flux at 5\,GHz. Given that all of our 5\,GHz radio counterparts are unresolved, we assume that the peak and integrated intensities are consistent within the measurement uncertainty for these sources. Additionally, we determine a lower limit to the spectral index $\alpha_{\rm min}$ for the 10 radio detections lacking 5-GHz counterparts, by using 3$\sigma$ as the upper limit for the flux density at 5 GHz. In total, we have calculated the spectral index for 54 compact \hii\ regions (44 reliable measurements and 10 lower limits; see Table\,\ref{tab:specindex} for details).

The spectral index analysis has identified 16 radio sources that have a spectral index in the range $-$0.2 < $\alpha$ < 0.3. These objects are considered to be optically thin \hii\ regions. The remaining 38 radio sources have a positive spectral index between 5 and 18\,GHz ($\alpha$ > 0.3) and are classed as optically thick \hii\ regions. For the six sources that fall outside of the 5-GHz survey latitude range we are unable to estimate the optical depth between 5 and 18\,GHz. However, we are able to analyse their emission between 18 and 24\,GHz. If these objects are true HC\,\hii\ regions, they are expected to have a positive spectral index due to their compact size and the optically thick free-free emission that arises from being deeply embedded  dense material in their natal clump. We find that five of the six radio sources have positive spectral indices between 18 and 24\,GHz. For the radio source with a negative spectral index, a visual inspection of its mid-infrared environment shows that the radio emission coincides with extended mid-infrared emission, suggesting it is associated with an evolved \hii\ region. Therefore, we exclude this clump from further analysis. For follow up observations, we have selected the 38 radio sources with a positive spectral index between 5 and 18\,GHz, and an additional 5 sources with a positive spectral index between 18 and 24\,GHz, for follow up observations at higher resolution.

\setlength{\tabcolsep}{6pt}
\begin{sidewaystable*}
\captionsetup{labelfont=bf,format=plain}

\caption{Spectral index and lower limit measurements for the high frequency radio detections that present 5-GHz radio emission. The average offset and size has been taken for sources that contain more than one 5-GHz radio counterpart and is represented with a \textdagger. The 5-GHz integrated flux upper limits are given as 3$\sigma$ thresholds.}
\begin{tabular}{lcccccccccc}
    \hline
    \multicolumn{1}{c}{Radio} &
    \multicolumn{1}{c}{} &
    \multicolumn{1}{c}{$f_{\rm int}$ 5-GHz} &
    \multicolumn{1}{c}{$\Delta f_{\rm int}$ 5-GHz } &
    \multicolumn{1}{c}{Observed 5-GHz size} &
    \multicolumn{1}{c}{$f_{\rm peak}$ 24\,GHz} &
    \multicolumn{1}{c}{$\Delta f_{\rm peak}$ 24\,GHz} &
    \multicolumn{1}{c}{Offset} &
    \multicolumn{1}{c}{} &
    \multicolumn{1}{c}{$\alpha$} &
    \multicolumn{1}{c}{$\Delta \alpha$} \\

    \multicolumn{1}{c}{name} &
    \multicolumn{1}{c}{} &
    \multicolumn{1}{c}{(mJy)} &
    \multicolumn{1}{c}{(mJy)} &
    \multicolumn{1}{c}{(arcsec)} &
    \multicolumn{1}{c}{(mJy beam$^{-1}$)} &
    \multicolumn{1}{c}{(mJy beam$^{-1}$)} &
    \multicolumn{1}{c}{(arcsec)} &
    \multicolumn{1}{c}{} &
    \multicolumn{1}{c}{} &
    \multicolumn{1}{c}{} \\
    \hline\hline
G300.969+1.148	    &		    &	370	    &	18    	&	 $-$   	&	490  	&	35   	&	2.4  	&	     	&	0.19      	&	0.036	\\
G301.136$-$0.225	&		    &	300	    &	11    	&	 $-$   	&	1000	&	110   	&	3.4  	&	     	&    0.77   	&	0.048	\\
G302.033+0.626  	&	<   	&	0.56	&	     	&	     	&	0.84	&	0.07	&	       	&	 >    	&	0.07    	&	     	\\
G305.886+0.017  	&	<   	&	0.46	&	     	&	     	&	1.3  	&	0.1  	&	     	&	 >    	&	0.5     	&		    \\
G308.056$-$0.396	&		    &	110	    &	0.25	&	2.6	    &	190  	&	8.7  	&	0.39	&	      	&	 0.38      	&	0.019	\\
G308.651$-$0.508	&		    &	0.7 	&	0.087	&	2.3  	&	2.9  	&	0.38	&	3.1   	&	     	&	0.89   	    &	0.075	\\
G309.384$-$0.135	&	<   	&	0.35	&	     	&		    &	2.5   	&	0.24	&	      	&	 >    	&	1.0	        &		    \\
G309.920+0.479  	&		    &	350	    &	1.8   	&	2.8  	&	880  	&	49   	&	2.6  	&	     	&	 0.59     	&	0.023	\\
G310.144+0.760  	&		    &	4.1 	&	0.16	&	3.0	    &	7.0     &	0.33	&	2.1   	&	       	&	0.34      	&	0.025	\\
G311.642$-$0.380	&		    &	200	    &	0.47	&	2.7    	&	260  	&	6.1  	&	1.9  	&	     	&	 0.19     	&	0.0094	\\
G312.307+0.661  	&		    &	26	    &	0.17	&	2.9 	&	39   	&	0.63	&	0.66	&	       	&    0.24     	&	0.0071	\\
G318.049+0.086  	&		    &	6.1 	&	0.35	&	2.7   	&	4.6  	&	0.82	&	4.1   	&	     	&  $-$0.18  	&	0.077	\\
G320.233$-$0.284	&		    &	320 	&	1.5  	&	3.3  	&	340  	&	3.5  	&	1.4  	&	     	&	0.034     	&	0.0046	\\

\hline
\end{tabular} 
\begin{tablenotes}
\item Note: only a small portion of the data is provided here, the full table is only available in
electronic form. \\
\end{tablenotes}
\label{tab:specindex}

\setlength{\tabcolsep}{3pt}
\captionsetup{labelfont=bf,format=plain}
\caption{Extracted source parameters from our detected radio sources at high resolution. Column 1 is the radio name, which is a combination of the Galactic longitudes and
latitudes determined by fitting ellipsoidal Gaussians using {\sc IMFIT}, Columns 2 and 3 are the Right Ascension and Declination of each source, respectively. Columns 4-10 present 18-GHz source parameters and columns 11-17 present 24-GHz source parameters. Col 8$-$10 and 15$-$17 are the deconvolved major, minor and position angle, respectively. Sources with $-$ are those for which sizes could not be deconvolved.}
\begin{tabular}{lccccccccccccccccc}

    \hline
    \multicolumn{1}{c}{} &
    \multicolumn{1}{c}{RA} &
    \multicolumn{1}{c}{Dec.} &
    \multicolumn{7}{c}{18GHz} &
    \multicolumn{7}{c}{24GHz} \\
    \cline{4-9}
    \cline{11-16}
    \multicolumn{1}{c}{} &
    \multicolumn{1}{c}{} &
    \multicolumn{1}{c}{} &
    \multicolumn{1}{c}{} &
    \multicolumn{1}{c}{} &
    \multicolumn{1}{c}{} &
    \multicolumn{1}{c}{} &
    \multicolumn{1}{c}{} &
    \multicolumn{1}{c}{} &
    \multicolumn{1}{c}{} &
    \multicolumn{1}{c}{} &
    \multicolumn{1}{c}{} \\
    
    \multicolumn{1}{c}{ } &
    \multicolumn{1}{c}{{(J2000)}} &
    \multicolumn{1}{c}{{(J2000)}} &
    \multicolumn{1}{c}{$f_{\rm peak}$} &
    \multicolumn{1}{c}{$\Delta f_{\rm peak}$} &
    \multicolumn{1}{c}{$f_{\rm int}$} &
    \multicolumn{1}{c}{$\Delta f_{\rm int}$} &
    \multicolumn{1}{c}{Major} &
    \multicolumn{1}{c}{Minor} &
    \multicolumn{1}{c}{PA} &
    \multicolumn{1}{c}{$f_{\rm peak}$} &
    \multicolumn{1}{c}{$\Delta f_{\rm peak}$} &
    \multicolumn{1}{c}{$f_{\rm int}$} &
    \multicolumn{1}{c}{$\Delta f_{\rm int}$} &
    \multicolumn{1}{c}{Major} &
    \multicolumn{1}{c}{Minor} &
    \multicolumn{1}{c}{PA} \\
    
    \multicolumn{1}{c}{Radio name} &
    \multicolumn{1}{c}{(h:m:s)} &
    \multicolumn{1}{c}{(d:m:s)} &
    \multicolumn{1}{c}{(mJy beam$^{-1}$)} &
    \multicolumn{1}{c}{(mJy beam$^{-1}$)} &
    \multicolumn{1}{c}{(mJy)} &
    \multicolumn{1}{c}{(mJy)} &
    \multicolumn{1}{c}{(arcsec)} &
    \multicolumn{1}{c}{(arcsec)}  &
    \multicolumn{1}{c}{(\degr)} &
    \multicolumn{1}{c}{(mJy beam$^{-1}$)} &
    \multicolumn{1}{c}{(mJy beam$^{-1}$)} &
    \multicolumn{1}{c}{(mJy)} &
    \multicolumn{1}{c}{(mJy)} &
    \multicolumn{1}{c}{(arcsec)} &  
    \multicolumn{1}{c}{(arcsec)}  &
    \multicolumn{1}{c}{(\degr)} \\
    
    \hline
G301.136$-$0.225&	12:35:35.09	&	$-$63:02:31.9	&	470	&	31    	&	790	&	76	    &	0.7	&	0.6	&	$-$53	&	510	&	31  	&	1000&	95	&	0.6	&	0.6	&	86.4	\\
G308.056$-$0.396&	13:36:32.18	&	$-$62:49:05.7	&	110	&	5.3   	&	170	&	12     	&	0.7	&	0.4	&	18	    &	85	&	5.1    	&	160	&	14	&	0.7	&	0.5	&	28.3	\\
G309.384$-$0.135&	13:47:23.95	&	$-$62:18:12.3	&	0.94&	0.11	&	2.0	&	0.36	&	1.0	&	0.6	&	74.7	&	0.77&	0.15	&	2.0	&	0.63&	0.8	&	0.5	&	46	\\
G309.920+0.479	&	13:50:41.82	&	$-$61:35:10.5	&	380	&	16     	&	990	&	59     	&	1.0	&	0.9	&	11.5	&	250	&	12  	&	960	&	72	&	1.1	&	0.8	&	18.8	\\
G310.144+0.760	&	13:51:58.32	&	$-$61:15:41.3	&	2.1	&	0.11	&	6.2	&	0.64	&	1.8	&	0.7	&	75     	&	1.4	&	0.13	&	6.4	&	1.4	&	2.0	&	0.6	&	67.5	\\
G312.307+0.661	&	14:09:24.93	&	$-$60:47:00.7	&	21	&	2.0 	&	31	&	4.2 	&	0.6	&	0.4	&	2.6  	&	14	&	1.8 	&	30	&	5.5	&	0.8	&	0.4	&	21.5	\\
G326.448$-$0.748&	15:49:18.65	&	$-$55:16:52.3	&	5.3	&	0.48	&	8.5	&	1.2  	&	0.9	&	0.1	&	$-$11	&	4.2	&	0.48	&	8.7	&	1.5	&	0.9	&	0.3	&	$-$6.0	\\
G328.164+0.586	&	15:52:42.39	&	$-$53:09:52.8	&	2.8	&	0.24	&	5.5	&	0.69	&	1.1	&	0.5	&	$-$18	&	2.4	&	0.22	&	5.2	&	0.71&	1.1	&	0.5	&	$-$18	\\
G328.236$-$0.548&	15:57:58.23	&	$-$53:59:23.5	&	20	&	1.8    	&	35	&	4.9 	&	1.1	&	0.4	&	$-$12.7	&	19	&	1.6   	&	34	&	4.3	&	0.9	&	0.3	&	$-$9.4	\\
G329.183$-$0.314&	16:01:46.95	&	$-$53:11:44.3	&	1.1	&	0.13	&	4.8	&	0.89	&	1.6	&	1.2	&	89.7	&	0.8	&	0.18	&	4.0	&	1.5	&	1.2	&	1.0	&	$-$5.4	\\
G329.272+0.115	&	16:00:21.75	&	$-$52:48:48.1	&	6.6	&	0.57	&	13	&	1.8    	&	1.1	&	0.4	&	$-$13	&	4.8	&	0.52	&	12	&	2.0	&	1.2	&	0.4	&	$-$1.8	\\

    \hline
\end{tabular}  

\begin{tablenotes}
\item Note: only a small portion of the data is provided here, the full table is only available in
electronic form. \\
\end{tablenotes}

  \label{tab:radio_source_param_high_res}%
\end{sidewaystable*}

\begin{figure}
    \centering
      \includegraphics[width=0.49\textwidth]{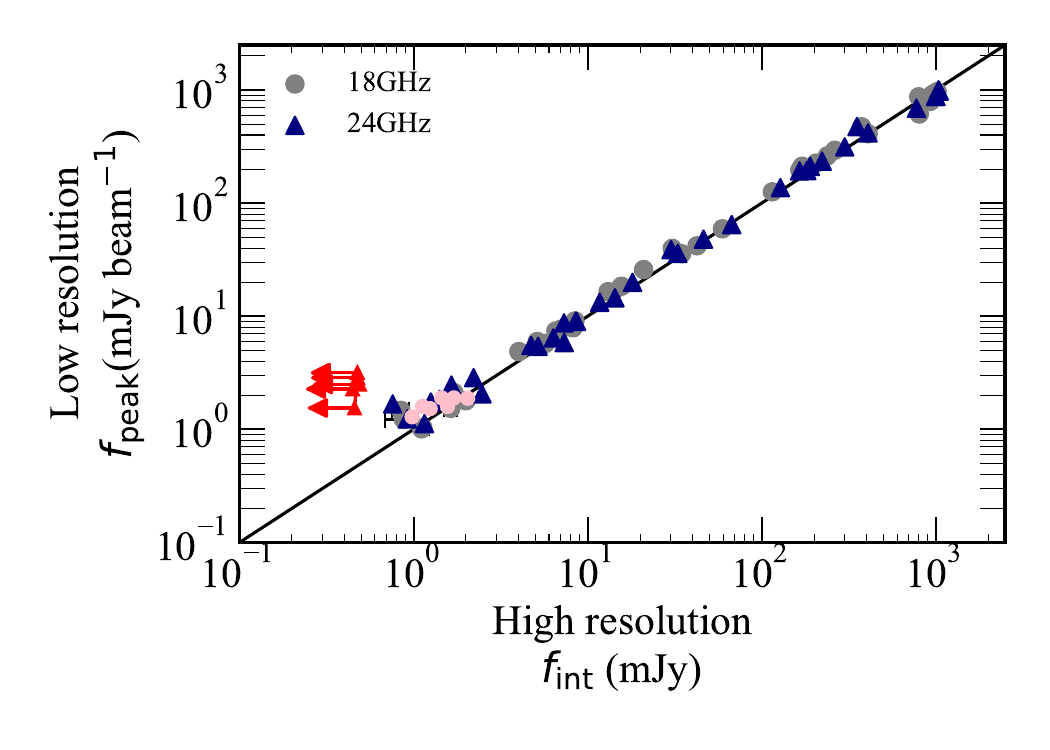}
    \caption{Distribution of integrated flux density from the high resolution study against peak flux densities from the low resolution study. Red upper limits indicate the seven 24\,GHz non-detections, while pink data filled show the fluxes for the corresponding 18\,GHz detections. The solid black line represents the line of equality.}
    \label{fig:int_v_peak_scatter}
\end{figure}

\subsection{High Resolution Detection Statistics}

The high-resolution maps were visually inspected to confirm the presence of radio sources and to identify and exclude any imaging artifacts from over-resolved sources. Radio continuum emission above 3$\sigma$ is detected in 41 out of the 42 observed fields. In total, 43 radio sources were identified at 18 GHz: 39 fields contained a single source, while 2 fields each contained two radio sources. We find that 36 radio sources are detected at 18 and 24\,GHz, leaving 7 radio sources that do not have a 24\,GHz counterpart. The sources detected in both the 18\,GHz and 24\,GHz maps are all compact (y-factor < 2) or well-resolved radio sources (y-factor < 4). We present examples of these radio maps in Fig.\,\ref{fig:high_res_examples}.

\begin{figure*}
    \captionsetup{list=off,format=cont}
    \centering
    \includegraphics[height=0.3\textwidth]{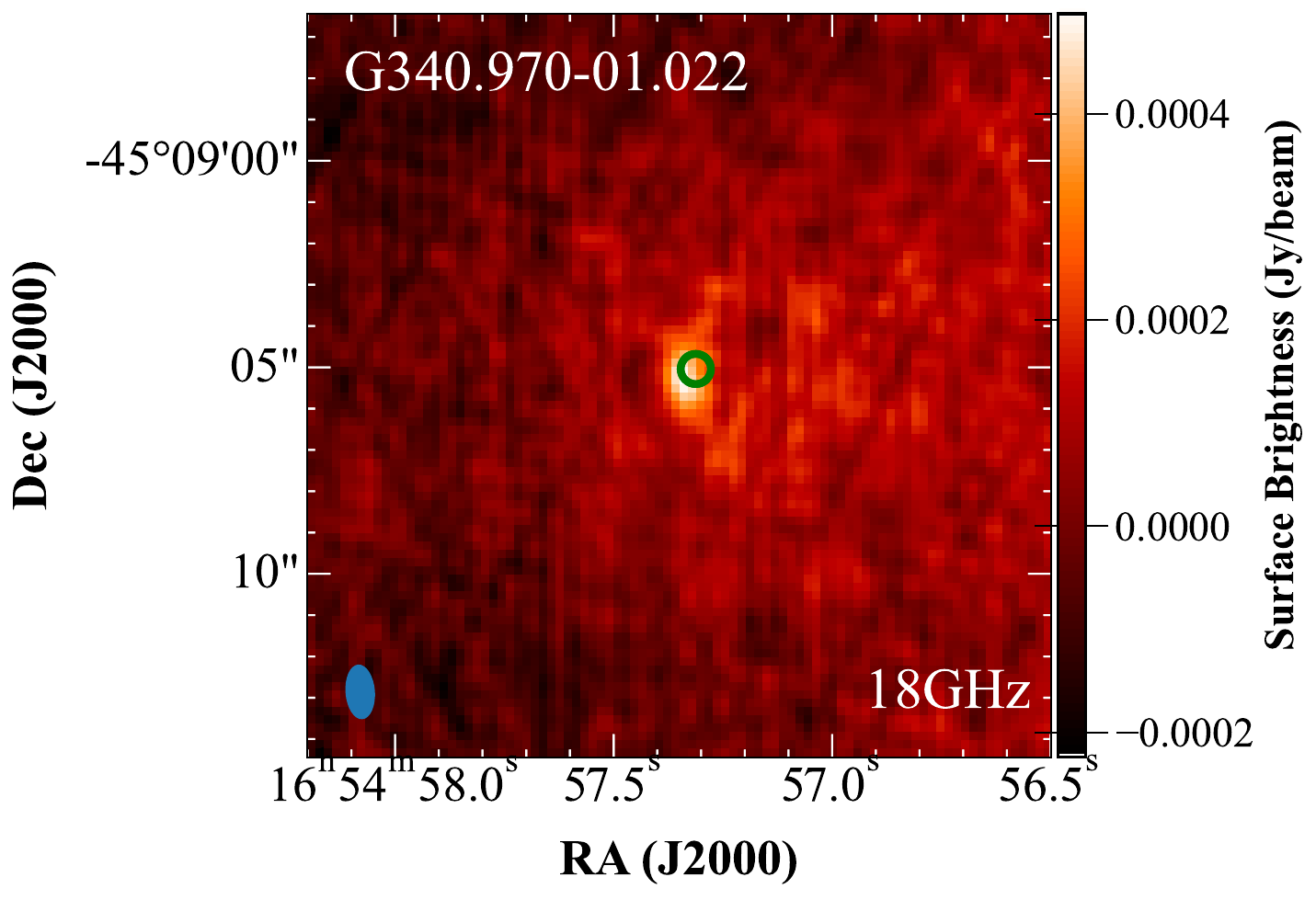}
    \includegraphics[height=0.3\textwidth]{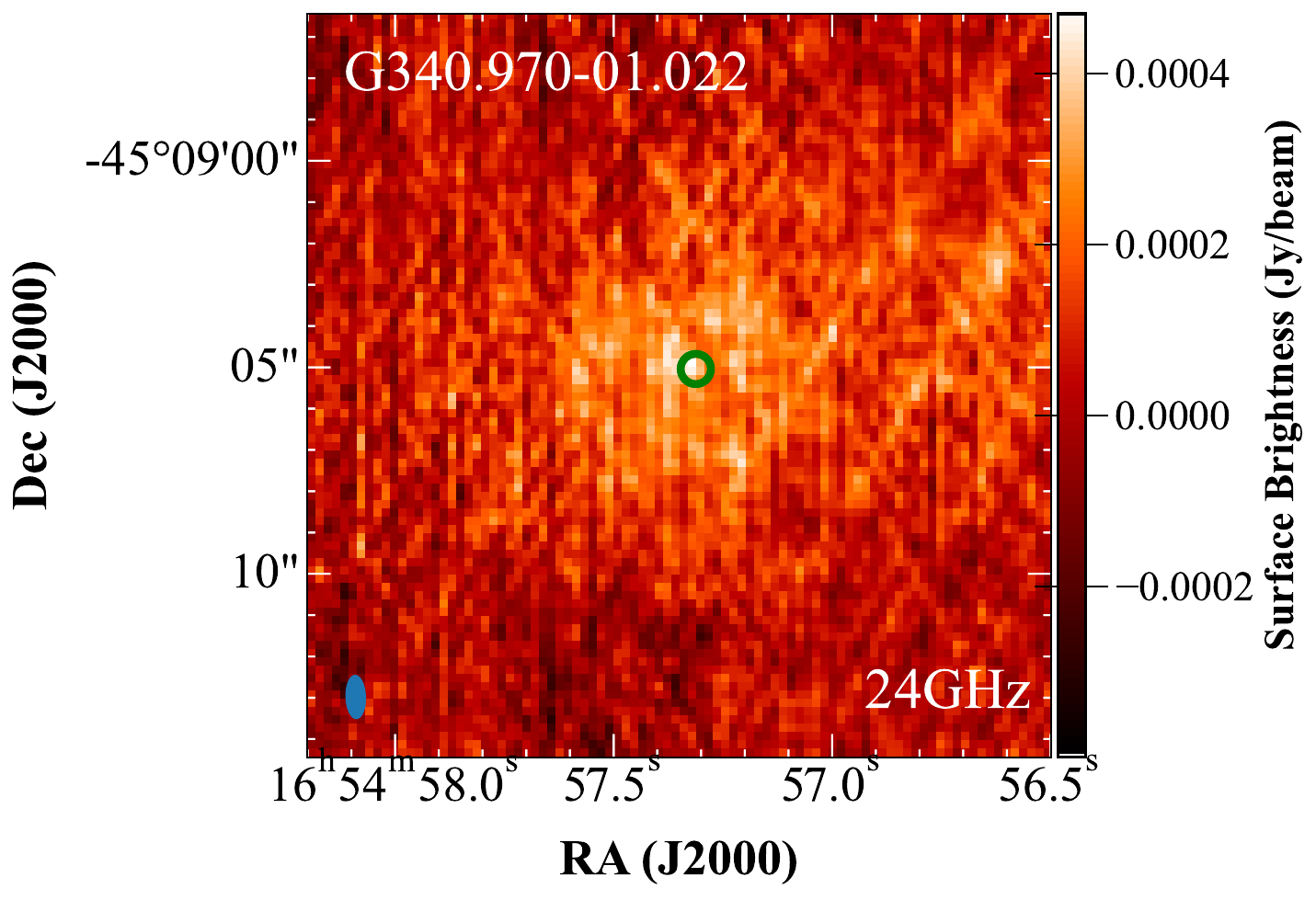}\\ 
    \includegraphics[height=0.3\textwidth]{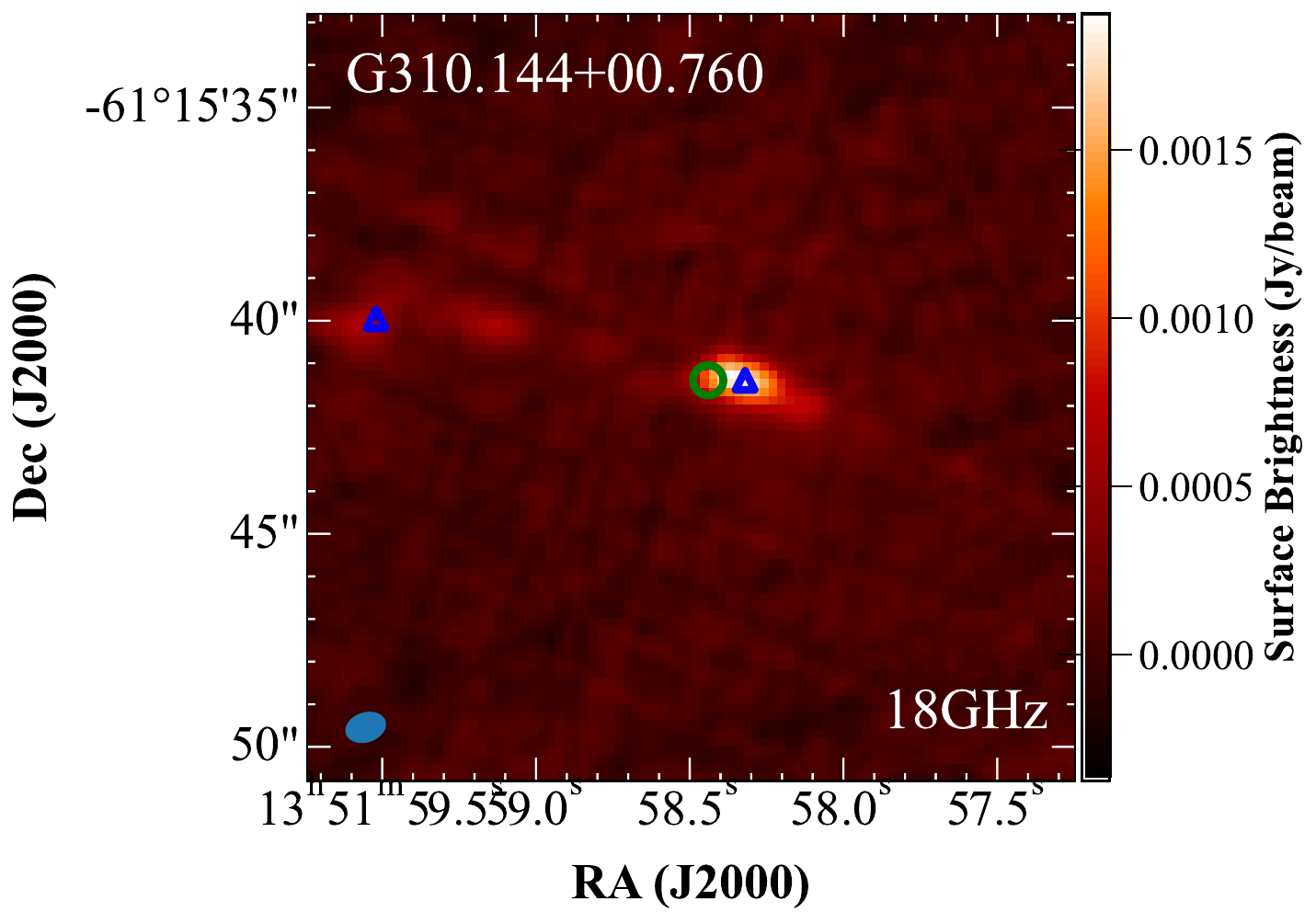}
    \includegraphics[height=0.3\textwidth]{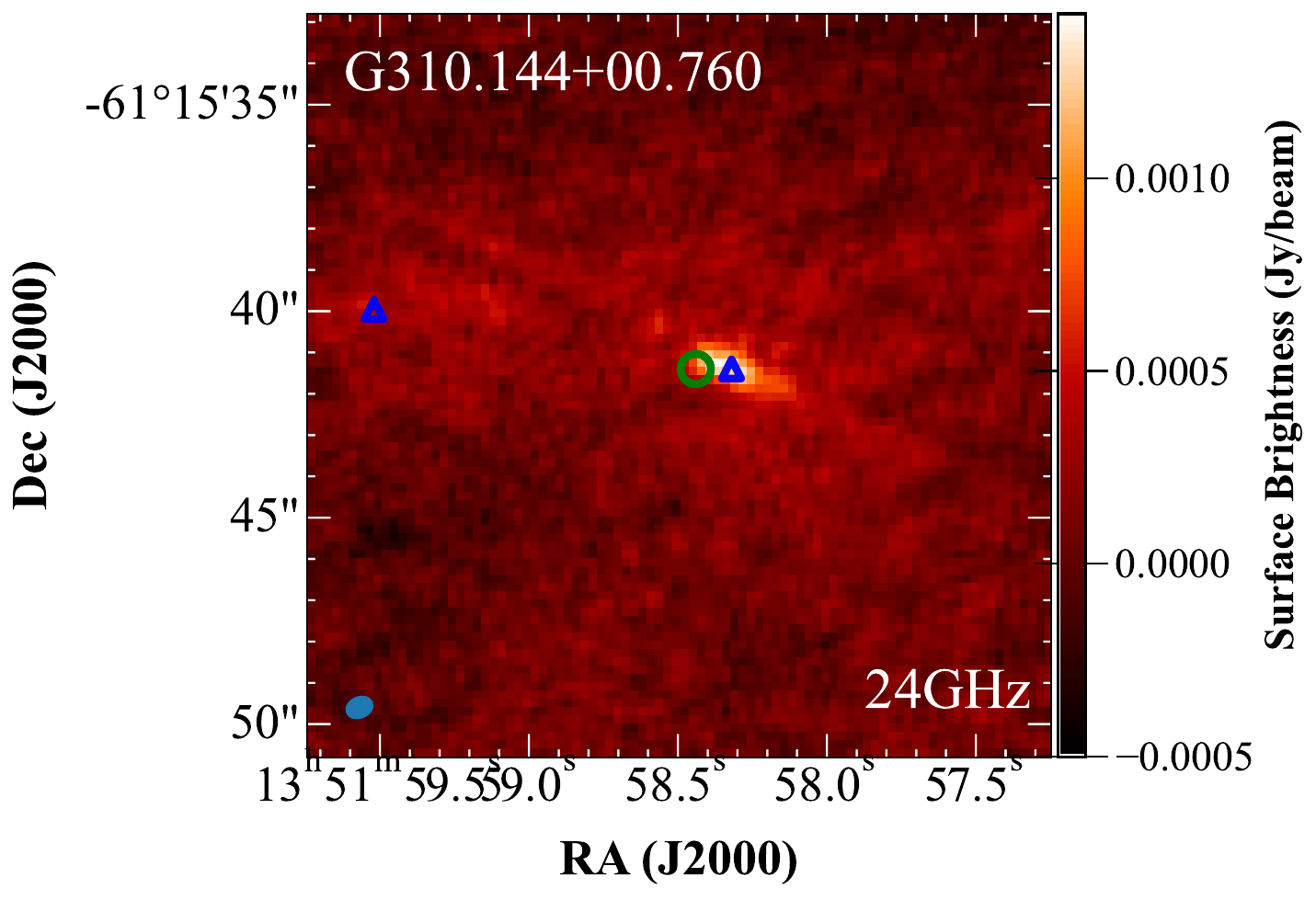}\\
  
  \caption{Examples of two radio fields that have been excluded from further analysis. Upper panel: Radio source G340.970$-$1.021, which is detected at 18\,GHz but not at 24\,GHz. Lower panel: G310.144+0.760, an example of an over-resolved radio source. The symbols and contours are as described in Fig.\,\ref{fig:radio_map_examples}. Full set of excluded fields are presented in Fig.\,\ref{fig:over_resolved_appendix}.}
     \label{fig:high_res_examples_discarded}
\end{figure*}

There are two possible explanations for the 24\,GHz non-detections: i) the 24\,GHz flux is weak and falls below the sensitivity of the observations; ii) the radio emission detected in the low-resolution maps is associated with an extended source that has been filtered out in the high resolution 24\,GHz map. The sensitivity of the high- and low-resolution observations is very similar and so we can rule out the first possibility. To investigate the second possibility, in Figure\,\ref{fig:int_v_peak_scatter}, we present a comparison between the integrated flux densities of sources detected in the higher resolution maps and the peak flux densities of the corresponding sources detected in the lower resolution maps. We determine upper limits for the flux density from the 24\,GHz maps for the 7 non-detections at 24\,GHz (see red upper limits in Fig.\,\ref{fig:int_v_peak_scatter}). Inspection of the plot shows that the high- and low-resolution fluxes agree well for all sources at 18\,GHz and 24\,GHz, and for the 18\,GHz detections that are not detected at 24\,GHz. The significant discrepancy between the upper limits for the 24\,GHz non-detections and the fluxes measured in the low-resolution maps confirms that substantial flux has been filtered out (see upper panel of Fig.\,\ref{fig:high_res_examples_discarded}). This filtering possible explains the one field (G305.887+00.017) where no source was detected at either frequency; in the low-resolution maps, the emission was extended with an SNR $\sim$\,3, and despite being associated with the central MMB maser and dust clump, the weak emission was probably filtered out at higher resolution, leading to the non-detection.

Additionally, we identified another radio source, G310.144+0.760, whose high-resolution flux density is five times lower than its low-resolution flux. Visual inspection of the high-resolution maps reveals that some of the extended structure visible at 18\,GHz is missing in the 24\,GHz map, confirming that flux has been lost due to filtering. The high-resolution maps of this source are shown in the lower panels of Fig.\,\ref{fig:high_res_examples_discarded}. In total, we have identified eight radio sources that are over-resolved at 24\,GHz (maps of these are available in Fig.\,\ref{fig:over_resolved_appendix}), indicating that these objects are not compact and are therefore unlikely to represent HC\,\hii\ regions. Moreover, the lack of detectable 24\,GHz flux density makes it difficult to constrain their physical properties, leading us to exclude these radio sources from further consideration. 

As in the low-resolution analysis, we used the \texttt{MIRIAD} task \texttt{IMFIT} to determine the flux densities, positions, and sizes of the 35 high-frequency radio sources that are detected at both 18 and 24\,GHz. The observational field parameters are presented in Table\,\ref{tab:radio_source_param_high_res}.

\section{Characterising the HC\,\hii\ region candidates}
\label{sec:sedmodelandphysicalproperties}

\setlength{\tabcolsep}{6pt}
\begin{table*}
  \caption{Derived physical parameters for three clumps located at near distances in the literature.}
  \begin{tabular}{lccccccccc}
    \hline
    \multicolumn{1}{c}{}& 
    \multicolumn{1}{c}{} &
    \multicolumn{2}{c}{Distance} &
     \multicolumn{2}{c}{log$_{10}$\,[$M$$_{\rm clump}$]} &
      \multicolumn{2}{c}{log$_{10}$\,[$L$$_{\rm bol}$]} &
    \multicolumn{2}{c}{log$_{10}$\,[\nly]} \\
    \cline{3-4} \cline{5-6} \cline{7-8}  \cline{9-10}

    \multicolumn{1}{c}{Radio}& 
    \multicolumn{1}{c}{Clump} &
    \multicolumn{1}{c}{Near} &
    \multicolumn{1}{c}{Far} &
     \multicolumn{1}{c}{Near} &
    \multicolumn{1}{c}{Far} &
     \multicolumn{1}{c}{Near} &
    \multicolumn{1}{c}{Far} &
     \multicolumn{1}{c}{Near} &
    \multicolumn{1}{c}{Far} \\

    \multicolumn{1}{c}{name}& 
    \multicolumn{1}{c}{name} &
    \multicolumn{1}{c}{(kpc)} &
    \multicolumn{1}{c}{(kpc)} &
    \multicolumn{1}{c}{(\msun)} &
    \multicolumn{1}{c}{(\msun)} &
    \multicolumn{1}{c}{(\lsun)} &
    \multicolumn{1}{c}{(\lsun)} &
    \multicolumn{1}{c}{(photons\,s$^{-1}$)} &
    \multicolumn{1}{c}{(photons\,s$^{-1}$)} \\

    \hline
G308.056$-$0.396 & AGAL308.057$-$00.397 & 0.9 & 9.4 & 1.7 & 3.5 & 2.6 & 4.9  & 46.2 & 48.23 \\
G312.307+0.661 & AGAL312.306+00.661 & 1.0 & 10.6 & 1.0 &  3.3 & 2.5 & 4.7  & 45.6 & 47.6 \\
G346.480+0.132 & AGAL346.481+00.131 & 1.3 & 15.2 & 1.6 & 3.7 & 2.3 & 4.5  & 45.1 & 47.2 \\
    \hline
  \end{tabular}
  \label{tab:nearfar}%
\end{table*}%

In this section, we examine the physical properties of the 35 clumps that host the identified HC\,\hii\ region candidates. We begin by examining the properties of their host clumps, followed by an analysis of the radio characteristics to confirm their nature.

\subsection{Clump properties}

\begin{figure}
    \centering
    \includegraphics[width=0.455\textwidth]{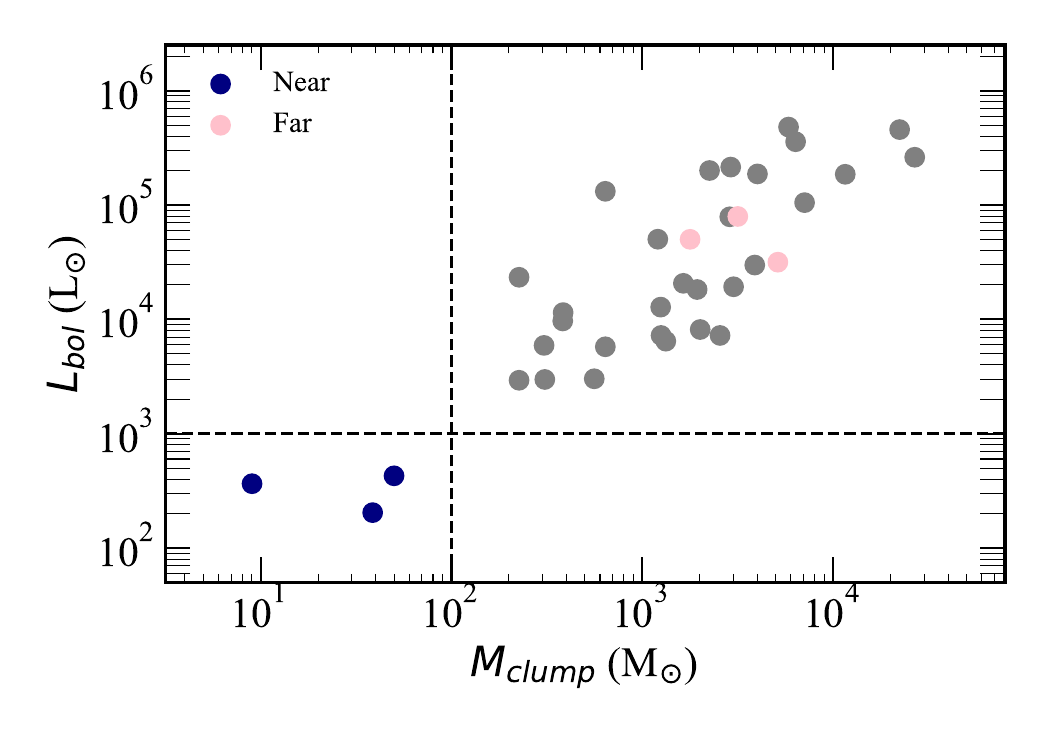}
    
    \caption{Clump masses as a function of bolometric luminosity for our sample of \hii\ regions. For three clumps, the blue points represent the physical parameters obtained adopting the near kinematic distances, while the values for the pink points are derived using the far distances (see main text). The vertical and horizontal dashed lines represents average thresholds for high-mass star forming clumps \citep{urquhart2014_atlas}.}
    \label{fig:masslum}
\end{figure}

Distances and clump properties have been adopted from the literature \citep{urquhart2018,urquhart2022} for 32 of the 35 clumps, while these  data are not available for the three clumps AGAL328.236$-$00.547, AGAL345.003$-$00.224 and AGAL353.409$-$00.361. In Figure\,\ref{fig:masslum}, we present the bolometric luminosities as a function of clump masses for the embedded \hii\ regions. The vertical and horizontal dotted lines represent the typical thresholds for high-mass star-forming clumps (i.e., 10$^3$\,\lsun\ and 100\,\msun; \citealt{urquhart2014_atlas}).
While, as expected, the majority of our clumps are situated in the upper right quadrant, three clumps exhibit bolometric luminosities and masses that are significantly lower than anticipated for high-mass star-forming regions.
\begin{figure}
    \centering
    \includegraphics[width=0.455\textwidth]{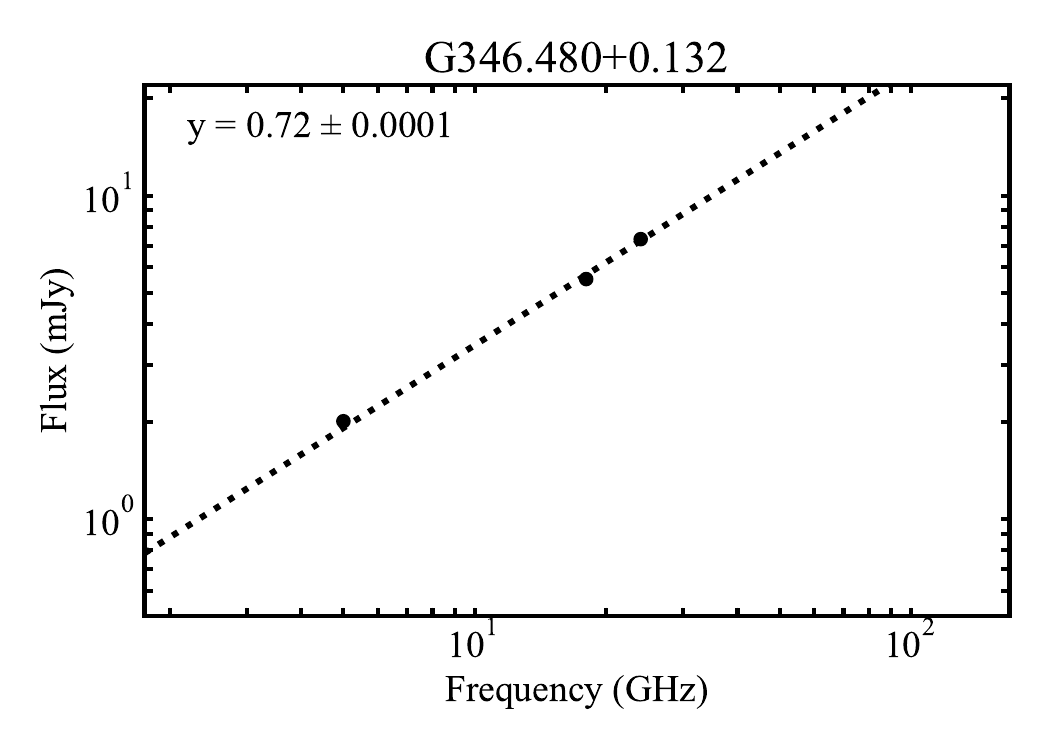}
    \includegraphics[width=0.455\textwidth]{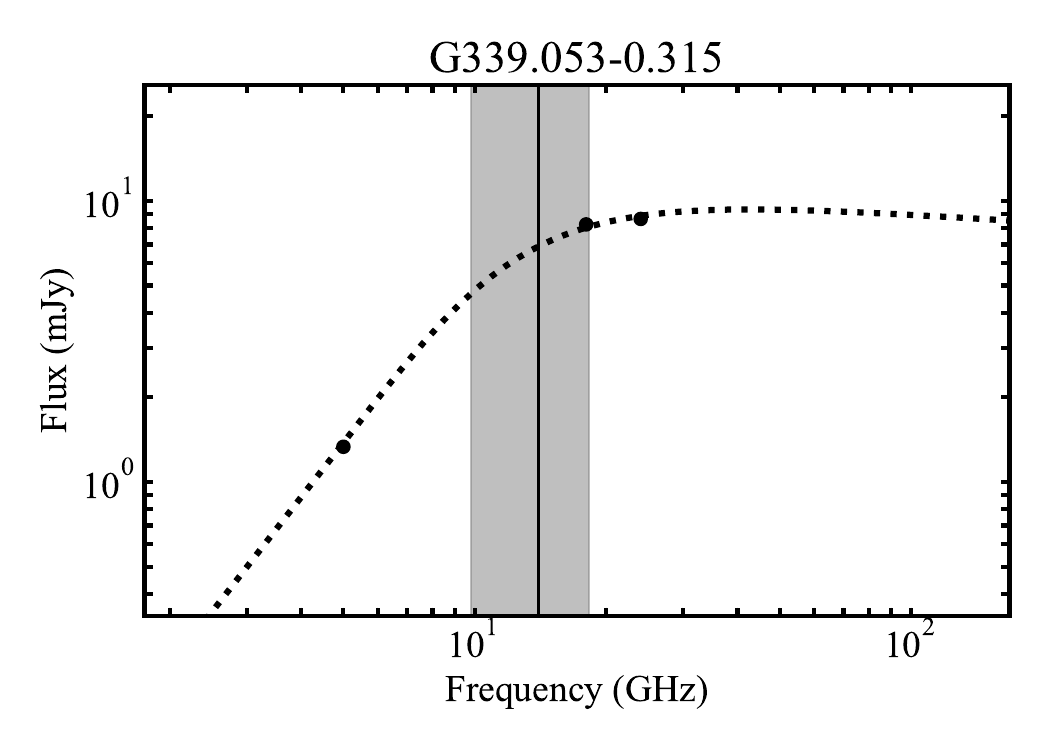} 
    \caption{Two examples of the radio SED models used in our analysis. Upper panel: a single power law fit to the integrated flux density for optically thick radio sources (G346.480+0.132). Lower panel: A simple \hii\ region model applied to radio source G339.053$-$0.315, an example for a source characterised by both optically thick and thin regimes \citep{Yang2021}. The vertical solid line represents the turnover frequency (\vt) and the shaded region provides an estimation of the uncertainty. Full set of radio SEDs are available in Fig.\,\ref{fig:SED_all}.}
    \label{fig:sed_example}
\end{figure}
These three sources are located inside of the solar circle (i.e., $< 8.5$\,kpc from the Galactic Centre) and are subject to the kinematic distance ambiguity. This is because there are two possible distance solutions that are equally spaced either side of the tangent point for each line-of-sight velocity; these are referred to as the near and far distances. In the literature, all three host clumps have been placed at the near distance ($< 1.5$\,kpc) \citep{urquhart2022}. 
However, the clump properties are not consistent with high-mass star formation, indicating that the far distance is more appropriate. We have, therefore, determined the far distance using the \citet{reid2016} model and recalculated the clump masses and luminosities accordingly (see the pink circles shown in Fig.\,\ref{fig:masslum} for updated values). Table\,\ref{tab:nearfar}, presents the near and far distances for these clumps alongside their corresponding bolometric luminosities, clump masses and Lyman continuum fluxes. 

\subsection{\hii\ region properties}

In this section, we determine the physical characteristics (i.e., diameters, emission measure, electron densities, Lyman-continuum flux and turnover frequency) of the 35 radio detections. To extend the frequency coverage of the radio data, we again utilize the 5-GHz flux densities extracted from the CORNISH-South and MAGPIS surveys (see Sect.\,\ref{sect:low_res_detections}). The 5-GHz data is available for 32 sources (24 reliable measurements and 8 lower limits), for which we examine the radio wavelength spectral energy distributions (SEDs). We find that 28 sources are well-fitted by the simple \hii\ region model (see next paragraph for details). The remaining four sources, which are optically thick, are fitted with a single power-law. Three of these sources, G350.011$-$1.342, G352.623$-$1.076, and G352.630$-$1.067, lie outside the coverage of the 5-GHz surveys and thus lack corresponding 5-GHz detections. For these three sources, we derive the spectral index between 18 and 24\,GHz, obtaining values of 0.38, 0.74, and 0.97, respectively. The latter two values align with those reported for HC\,\hii\ regions in the literature \citep{murphy2010,Yang2019}, while the lower spectral index of 0.38 for G350.011$-$1.342 suggests its turnover frequency may lie within or near the observed range. Given that these sources are based on only two data points, additional high-frequency observations are necessary to accurately determine their true optical depth. In Figure \ref{fig:sed_example}, we provide examples of both fitting models used in this analysis.

\setlength{\tabcolsep}{4pt}
\begin{table*}
\caption{Derived physical parameters for 35 \hii\ regions. The sources whose properties were not extracted via the SED fitting are denoted by a \textdagger.}
\begin{tabular}{lccccccccc}

    \hline
    \multicolumn{1}{c}{Radio name} &
    \multicolumn{1}{c}{Distance} &
    \multicolumn{1}{c}{log$_{10}$\,[$L$$_{\rm bol}$]} &
    \multicolumn{1}{c}{Diameter} &
    \multicolumn{1}{c}{\nelectron} &
    \multicolumn{1}{c}{$EM$} &
    \multicolumn{1}{c}{log$_{10}$\,[\nly]} &
    \multicolumn{1}{c}{\vt} &
    \multicolumn{1}{c}{Spectral} &
     \multicolumn{1}{c}{Reference} \\
    
    \multicolumn{1}{c}{ } &
    \multicolumn{1}{c}{(pc)} &
    \multicolumn{1}{c}{(\lsun)} &
    \multicolumn{1}{c}{{(pc)}} &
    \multicolumn{1}{c}{{(10$^4$\,cm$^{-3}$)}} &
    \multicolumn{1}{c}{(10$^7$\,pc\,cm$^{-6}$)} &
    \multicolumn{1}{c}{(photons\,s$^{-1}$)} &
    \multicolumn{1}{c}{GHz} &
    \multicolumn{1}{c}{type}    &
    \multicolumn{1}{c}{} \\
    
    \hline
G301.136$-$0.225\textdagger	&	4.31	&	5.33	&	0.0133	&	39.93	&	28.37	&	48.25	&	$-$    	&	O8.5	&	HCHII	\\
G308.056$-$0.396          	&	9.40	&	4.90	&	0.0457	&	5.22	&	12.44	&	48.23	&	5.82	&	O8.5	&	Transition	\\
G309.384$-$0.135          	&	5.29	&	3.91	&	0.0013	&	64.13	&	54.69	&	45.75	&	11.77	&	B1    	&	HCHII	\\
G309.920+0.479          	&	5.35	&	5.30	&	0.0412	&	8.67	&	3.10	&	48.51	&	8.99	&	O7.5	&	Transition	\\
G312.307+0.661          	&	10.60	&	4.70	&	0.0326	&	4.14	&	5.60	&	47.6	&	3.98	&	B0   	&	Transition	\\
G326.448$-$0.748          	&	3.75	&	3.99	&	0.0032	&	25.71	&	21.22	&	46.15	&	7.50	&	B1  	&	HCHII	\\
G328.164+0.586          	&	7.08	&	4.32	&	0.004	&	27.25	&	29.93	&	46.49	&	8.83	&	B0.5	&	HCHII	\\
G328.236$-$0.548          	&	11.66	&	$-$	    &	0.018	&	12.06	&	26.21	&	47.75	&	8.29	&	B0     	&	HCHII	\\
G329.183$-$0.314          	&	11.29	&	5.27	&	0.0057	&	22.98	&	30.22	&	46.79	&	8.88	&	B0.5	&	HCHII	\\
G329.272+0.115          	&	4.57	&	4.06	&	0.0027	&	54.93	&	80.31	&	46.51	&	14.13	&	B0.5	&	HCHII	\\
G329.610+0.114          	&	3.82	&	3.48	&	0.001	&	54.06	&	30.09	&	45.29	&	8.86	&	B1  	&	HCHII	\\
G333.387+0.032          	&	10.69	&	4.28	&	0.011	&	10.13	&	11.32	&	46.98	&	5.56	&	B0.5	&	Transition	\\
G335.789+0.174          	&	3.35	&	4.26	&	0.0009	&	49.78	&	21.86	&	45.05	&	7.61	&	B2     	&	HCHII	\\
G337.097$-$0.929          	&	3.01	&	3.47	&	0.0007	&	82.19	&	49.58	&	45.21	&	11.23	&	B1	    &	HCHII	\\
G337.705$-$0.053          	&	12.19	&	5.42	&	0.0492	&	8.13	&	3.25	&	48.67	&	9.20	&	O7  	&	UCHII	\\
G337.844$-$0.375          	&	2.98	&	4.37	&	0.0044	&	30.41	&	40.25	&	46.65	&	10.17	&	B0.5	&	HCHII	\\
G339.053$-$0.315          	&	7.15	&	3.76	&	0.0034	&	48.04	&	78.99	&	46.7	&	14.02	&	B0.5	&	HCHII	\\
G339.282+0.136          	&	4.63	&	3.60	&	0.0024	&	35.42	&	30.14	&	46.03	&	8.87	&	B1     	&	HCHII	\\
G340.249$-$0.372          	&	3.64	&	4.47	&	0.0154	&	11.46	&	20.27	&	47.51	&	7.34	&	B0     	&	HCHII	\\
G343.757$-$0.163\textdagger	&	2.50	&	3.81	&	0.0125	&	1.19	&	0.08	&	45.09	&	$-$  	&	B2  	&	Transition	\\
G343.929+0.125\textdagger	&	16.97	&	4.90	&	0.0444	&	5.01	&	1.78	&	48.31	&	$-$   	&	O8.5	&	Transition	\\
G345.004$-$0.224          	&	2.73	&	$-$ 	&	0.0147	&	13.29	&	26.01	&	47.56	&	8.26	&	B0     	&	HCHII	\\
G345.407$-$0.952          	&	1.37	&	5.12	&	0.0112	&	15.36	&	26.46	&	47.32	&	8.33	&	B0     	&	HCHII	\\
G346.480+0.132\textdagger	&	15.20	&	4.50	&	0.1024	&	4.14	&	0.28	&	47.2	&	$-$    	&	B0     	&	UCHII	\\
G347.628+0.148          	&	9.57	&	5.68	&	0.0518	&	5.22	&	1.41	&	48.41	&	6.18	&	O8	    &	UCHII	\\
G350.011$-$1.342\textdagger	&	3.11	&	4.70	&	0.0054	&	8.54	&	0.56	&	45.88	&	$-$  	&	B1  	&	Transition	\\
G350.015+0.433          	&	12.67	&	5.55	&	0.0703	&	3.81	&	1.02	&	48.51	&	5.29	&	O7.5	&	UCHII	\\
G350.340+0.141          	&	10.91	&	4.11	&	0.0148	&	9.88	&	14.49	&	47.34	&	6.26	&	B0     	&	Transition	\\
G350.343+0.116          	&	10.81	&	3.86	&	0.0122	&	15.54	&	29.36	&	47.46	&	8.75	&	B0    	&	HCHII	\\
G351.383$-$0.181          	&	10.09	&	3.86	&	0.002	&	53.91	&	57.7	&	46.13	&	12.08	&	B1     	&	HCHII	\\
G352.517$-$0.155          	&	10.8	&	5.27	&	0.0208	&	18.14	&	68.38	&	48.2	&	13.09	&	O8.5	&	HCHII	\\
G352.623$-$1.076\textdagger	&	1.33	&	3.47	&	0.0048	&	1.98	&	0.08	&	44.26	&	$-$  	&	B3  	&	Transition	\\
G352.630$-$1.067\textdagger	&	1.33	&	3.77	&	0.0028	&	6.38	&	0.15	&	44.62	&	$-$  	&	B3  	&	Transition	\\
G353.410$-$0.360          	&	3.55	&	$-$ 	&	0.0196	&	16.23	&	51.75	&	47.75	&	10.39	&	B0    	&	HCHII	\\
G354.724+0.300          	&	22.02	&	5.02	&	0.1716	&	2.48	&	1.06	&	49.33	&	5.38	&	O5.5	&	UCHII	\\
   \hline
\end{tabular}    
  \label{tab:phys_parameters}%
\end{table*}%

The simple \hii\ region model is constructed using the standard uniform electron density model for a bound ionized \hii\ region given by \citet{mezger1967}. Using the model provided by \citet{Yang2021}, the radio SED for each source has two free parameters: the electron density (\nelectron) and the diameter (diam). We obtain the best estimate for the two parameters by fitting the available radio continuum emission over a range of frequencies. The uncertainties of the integrated flux densities are obtained from \texttt{IMFIT} (see Table\,\ref{tab:radio_source_param_high_res}) and are taken into consideration during the fitting process. The emission measure can be calculated using $EM$ = \nelectron$^2$ $\times$ diam. The model has two components and comprises both optically thick and optically thin ionized gas. The output yields estimates for properties such as electron density (\nelectron) and diameter, specifically for the optically thin ionized gas between 5 and 24\,GHz. For these sources, we also determine the turnover frequency, \vt, at which the emission changes from the optically thick to the thin regimes, which is defined as the frequency for which the optical depth $\tau = 1$. Using the formula provided by \citet{wilson2013} assuming a homogeneous, optically thin \hii\ region, the turnover frequency can be expressed as a function of the emission measure and electron temperature \Telectron:

\begin{equation}
    \bigg(\frac{\nu_{\textup{t}}}{\mathrm{GHz}}\bigg) = 0.3045\,\bigg(\frac{{T_{\textup{e}}}}{\mathrm{10^4K}}\bigg)^{-0.643}\,\bigg(\frac{EM}{\mathrm{cm^{-6}\,pc}}\bigg)^{0.476},
    \label{tof}
\end{equation} 

\noindent where \Telectron\ is assumed to be 10\,000\,K and $EM$ is provided by the \hii\ region model. The typical error of \vt\ is $\sim$ 30\,per\,cent considering a 20\,per\,cent error in the flux density and 10\,per\,cent in the diameter measurement. The distribution of \vt\ in our sample ranges from 4 to 15\,GHz and has a mean of $\sim$ 9\,GHz.

To calculate the number of Lyman continuum photons emitted per second from the \hii\ regions, we use \citep{panagia1973,carpenter1990,urquhart2013_cornish}:

\begin{equation}
    \bigg(\frac{N_\textup{{Ly}}}{\mathrm{s^{-1}}}\bigg)= 8.9\times10^{40} \,\bigg(\frac{S_{\textup{$\nu$}}}{\mathrm{Jy}}\bigg)\,\bigg(\frac{d}{\mathrm{pc}}\bigg)^2\,\bigg(\frac{\nu}{\mathrm{GHz}}\bigg)^{0.1},
    \label{flux}
\end{equation} 

\noindent where $S_{\textup{$\nu$}}$ is the optically thin integrated flux density at 24\,GHz, $d$ is the heliocentric distance to the source and $\nu$ is 24\,GHz. The estimated uncertainty in the derived \nly\ flux is $\sim$20 per\,cent considering the error in the distance and flux densities  (e.g., \citealt{urquhart2013_cornish,sanchez_monge2013}).

We derive the physical parameters for the seven \hii\ regions that remain optically thick at 24\,GHz using equations from the literature (see Section 3.3 of \citealt{Patel2024}) and use Eqn.\,\ref{flux} to estimate the Lyman continuum photon flux. Due to their optically thick nature, the 24\,GHz flux is likely to be underestimated, and consequently, the electron density, emission measure and Lyman flux must be considered to be lower limits. The SED and derived physical properties for all 35 sources are given in Table\,\ref{tab:phys_parameters}.

\subsection{Distinguishing between \hii\ regions types}

\begin{figure}
    \centering
    \includegraphics[width=0.49\textwidth]{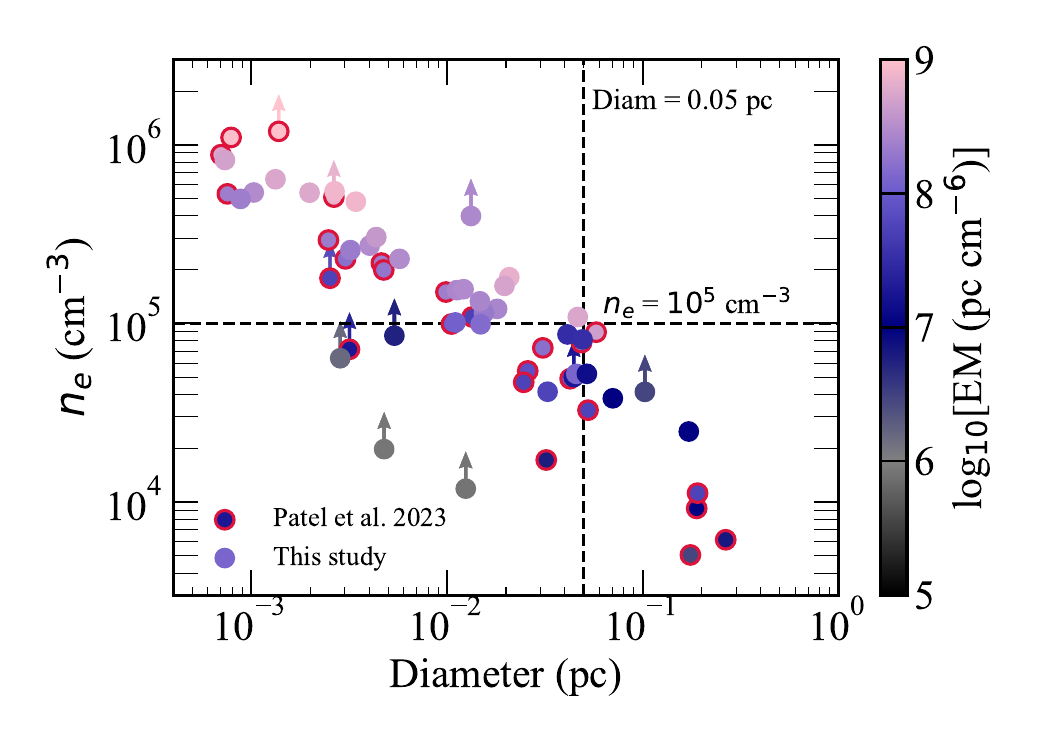}
    \caption{Distribution of the physical properties of 61 sources identified in this work (35) and Paper\,II (26). The colour of the data points represents the emission measure of the radio source (see colour bar on the right for corresponding values). Filled circles with upward-pointing arrows denote upper limits for \nelectron\ and EM, indicating optically thick radio sources. The region in the upper left part of the plot bounded by the vertical and horizontal dotted lines shows the region parameter space where HC\,\hii\ regions are found (see Table\,\ref{tab:criteria}).}
    \label{fig:phys_parameters_dist}
\end{figure}

We use the physical parameters derived in the previous section to distinguish between HC\,\hii\ and UC\,\hii\ regions. We summarize the general criteria for HC and UC\,\hii\ regions and the intermediate objects between these two classes in Table\,\ref{tab:criteria}. In Figure\,\ref{fig:phys_parameters_dist}, we show the distribution of the electron density, diameter and emission measure for the 35 HC \hii\ region candidates (lower limits for the electron density are given for optically thick sources). On this plot we, indicate a region of parameter space (vertical and horizontal dashed lines) where we expect to find HC\,\hii\ regions (i.e., \nelectron\ $>$ $10^5$\,$\mathrm{cm^{-3}}$ and diam $<$ 0.05\,pc) . 

\setlength{\tabcolsep}{6pt}
\begin{table}
  \caption{Quantitative criteria of the physical parameters for HC\,\hii\ regions, intermediate objects between the two stages and UC\,\hii\ regions from the literature.}
  \begin{tabular}{ccccc}
    \hline
    \multicolumn{1}{c}{Parameters}& 
    \multicolumn{1}{c}{Diameter} &
    \multicolumn{1}{c}{\nelectron} &
    \multicolumn{1}{c}{EM} \\

    \multicolumn{1}{c}{}& 
    \multicolumn{1}{c}{(pc)} &
    \multicolumn{1}{c}{(cm$^{-3}$)} &
    \multicolumn{1}{c}{(pc\,cm$^{-6}$)} \\

    \hline
HC\,\hii\ region	        &	$<$ 0.05 	     &	$>$ $10^5$ 	   &  $>$ $10^8$ \\
Intermediate objects	    &	0.05 $-$ 0.1	 &	$10^4-10^5$        &  $10^7-10^8$	\\
UC\,\hii\ region	        &	$>$ 0.1	     &	$<$ $10^4$       &  $<$ $10^7$	\\

    \hline
  \end{tabular}
  \label{tab:criteria}%
\end{table}%

Based on the criteria given in Table\,\ref{tab:criteria} and inspection of Fig.\,\ref{fig:phys_parameters_dist} we have found 20 sources that satisfy the HC\,\hii\ regions' electron density and size criteria, thus confirming their classification as HC\,\hii\ regions. A further 12 sources satisfy the size criterion for HC\,\hii\ regions but have lower electron densities. Among these, five are optically thick, indicating that their electron densities are lower limits. Three of these are in close proximity to the electron density threshold and we therefore consider them to also be confirmed as HC\,\hii\ regions. The characteristics of the remaining two optically thick sources fall significantly below the threshold and their nature is not certain. The 7 optically thin sources in this region of the plot exhibit parameters between the HC\,\hii\ region and UC\,\hii\ region thresholds, and are classified as intermediate objects. The remaining three sources are located in the bottom-right quadrant and are consistent with a classification as UC\,\hii\ region.

\hii\ regions and radio jets can be often mistaken for each other due to their similar properties. Non-relativistic and non-magnetic radio jets produce free-free emission primarily dominated by thermal bremsstrahlung emission and have spectral indices between $-0.1 \leq \alpha \leq 1.6$, values that are very similar to those of compact and optically thick \hii\ regions \citep{purser2016}. Fortunately, radio jets are predominately associated with significantly lower bolometric to Lyman photon fluxes ratios making them potentially easy to distinguish. Observationally, radio jets and \hii\ regions have very different morphologies. Radio jets appear as elongated structures frequently containing knots (see \citealt{purser2016} for more detail) whereas, HC\ and UC\,\hii\ regions usually have simple, isolated, often unresolved structures.

\begin{figure}
    \centering
    \includegraphics[width=0.49\textwidth]{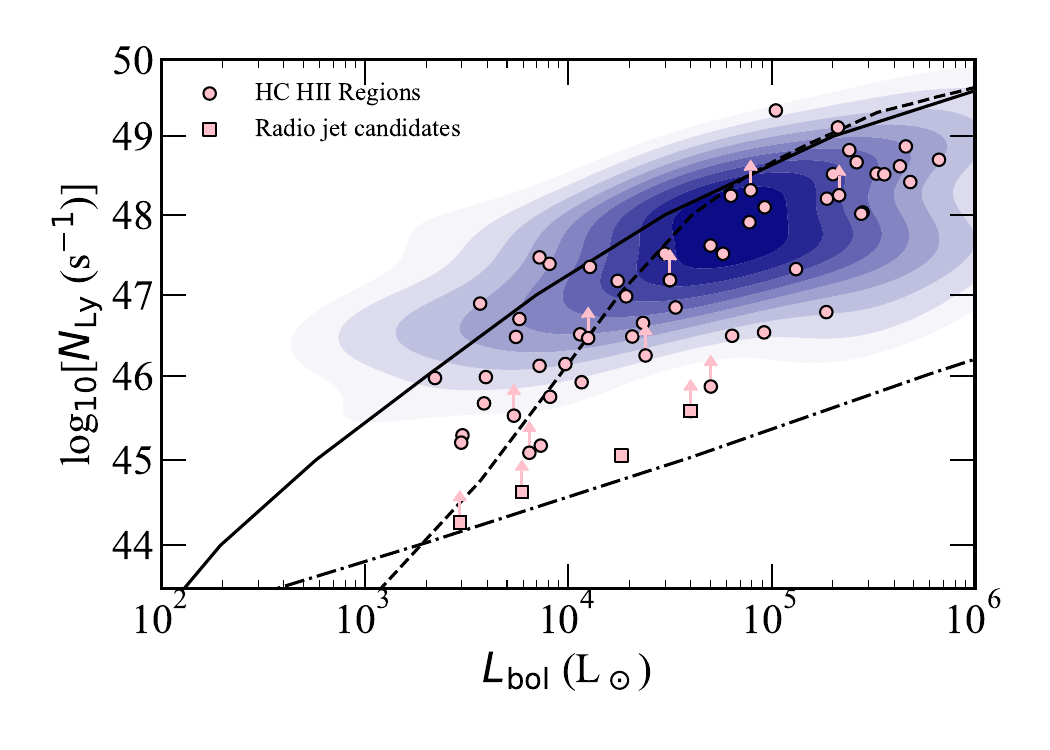}
    \caption{Lyman continuum photon flux vs. bolometric luminosity for the entire \hii\ region sample. blue contours highlight previously identified UC and HC\,\hii\ regions, while pink squares indicate radio-jet candidates, with optically thick jets represented as upper limits. The dashed line illustrates the relationship for OB ZAMS star models \citep{martins2005,davies2011} and the dot-dashed line represents a fit to radio jets found in \citet{purser2016} and \citet{purser2021}. The solid black line corresponds to the Lyman continuum flux expected from a blackbody with the same radius and temperature as a ZAMS star (see \citet{sanchez_monge2013} for more detail).}
    \label{fig:Lyman_bol}
\end{figure}

To avoid any contamination of our sample of compact \hii\ regions from jets, we compare the bolometric luminosity with the Lyman continuum photon flux in Fig.\,\ref{fig:Lyman_bol}. Note that three clumps lack bolometric luminosity measurements and are thus not included in this plot. We include the luminosity–Lyman photon flux relationship (dashed line) for OB ZAMS stars from a table of values given in \citet{davies2011}. The dot-dashed line represents the power-law relation for the radio jets identified by \citet{Anglada1995}. The grey and pink upper limit square represents the HC\,\hii/jet-like candidate from this work and Paper\,II. The blue contours represent the number density distribution of the known HC\,\hii\ and UC\,\hii\ regions \citep{Kalcheva2018,Yang2019,Yang2021}

The majority of the sources in our sample aligns well with the expected OB ZAMS line in Fig.\,\ref{fig:Lyman_bol} and their location in the plot is consistent with the previously identified \hii\ regions, reinforcing their classification as HC\,\hii\ and UC\,\hii\ regions. It is noteworthy that the Lyman continuum photon flux exhibits a two orders of magnitude scatter among our \hii\ regions. This variation is likely due to dust absorption, as highlighted by \citet{Yang2021}, who found that approximately 67 per cent of Lyman continuum photons are absorbed by dust within compact \hii\ regions. The degree of dust absorption tends to increase in more compact and younger \hii\ regions, thereby contributing significantly to the observed scatter. Such effects must be considered when interpreting Lyman continuum photon flux in star-forming regions.

By targeting sites of methanol masers we have been able to identify \hii\ regions that surround lower-luminosity stars. However, this is likely attributed to the fact that the \hii\ regions reported in the literature were identified from a lower-frequency survey that is sensitive primarily to the brightest \hii\ regions, as they are optically thick at these frequencies. Whereas, sites of methanol masers are not intrinsically bright at higher resolutions and high-frequency radio observations are more sensitive to lower-luminosity high-mass stars.

The dot-dashed line in Fig.\,\ref{fig:Lyman_bol} represents the expected fit for radio jets found in \citet{Anglada1995}. There is, therefore, a clear distinction between bright \hii\ regions (\lbol $> 10^{4}$\,\lsun) and radio jets; however, the two models converge and intersect around $\sim 2000$\,\lsun, making it difficult to distinguish between jets and \hii\ regions for objects within the 10$^{3}$ to 10$^{4}$\,\lsun\ range. Three of our radio sources lie close to this jet line. Two of these, G352.623$-$1.076 and G352.630$-$1.067, are located in a region of the parameter space where the parameters for jets and \hii\ regions converge. However, both of these sources are optically thick and so their Lyman flux is underestimated, making a definitive classification uncertain.

A recent study by \citet{Chen2019} reports a strong spatial correlation between the radio source G352.630$-$1.067 and a 6.7\,GHz methanol maser (G352.624$-$01.077; \citealt{green2010_mmb}). The maser emission is observed to be distributed along a linear structure oriented from southwest to northeast.

\subsection{Combining the SCOTCH sample}
\begin{figure}
    \centering
    \includegraphics[width=0.455\textwidth]{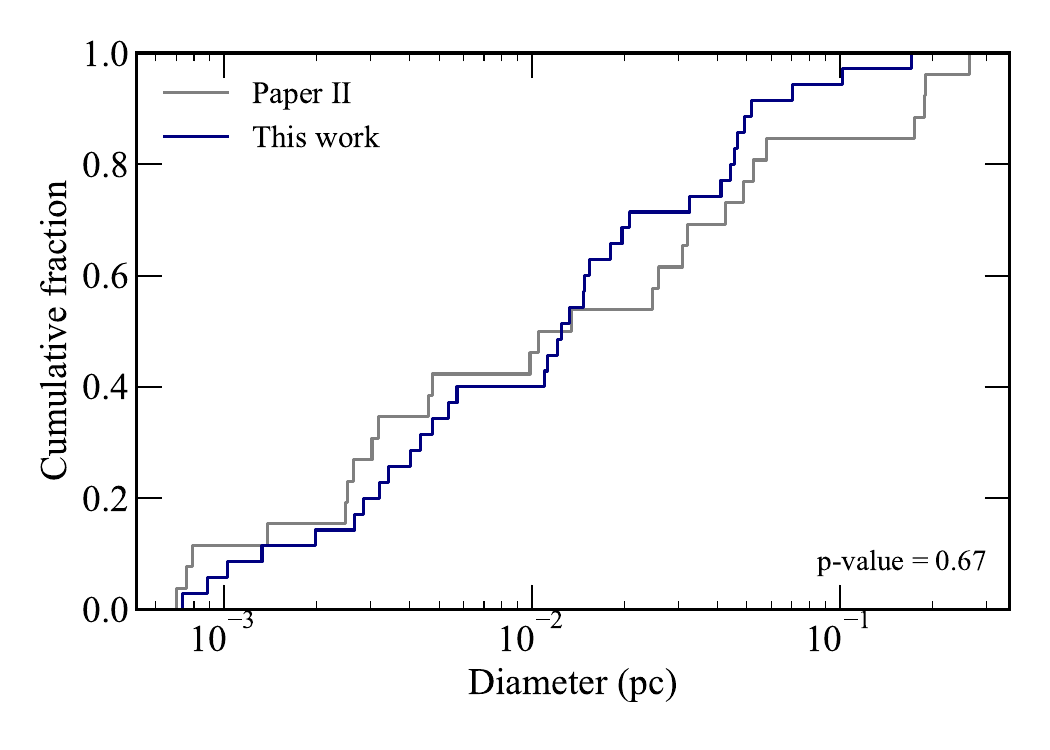}
    \includegraphics[width=0.455\textwidth]{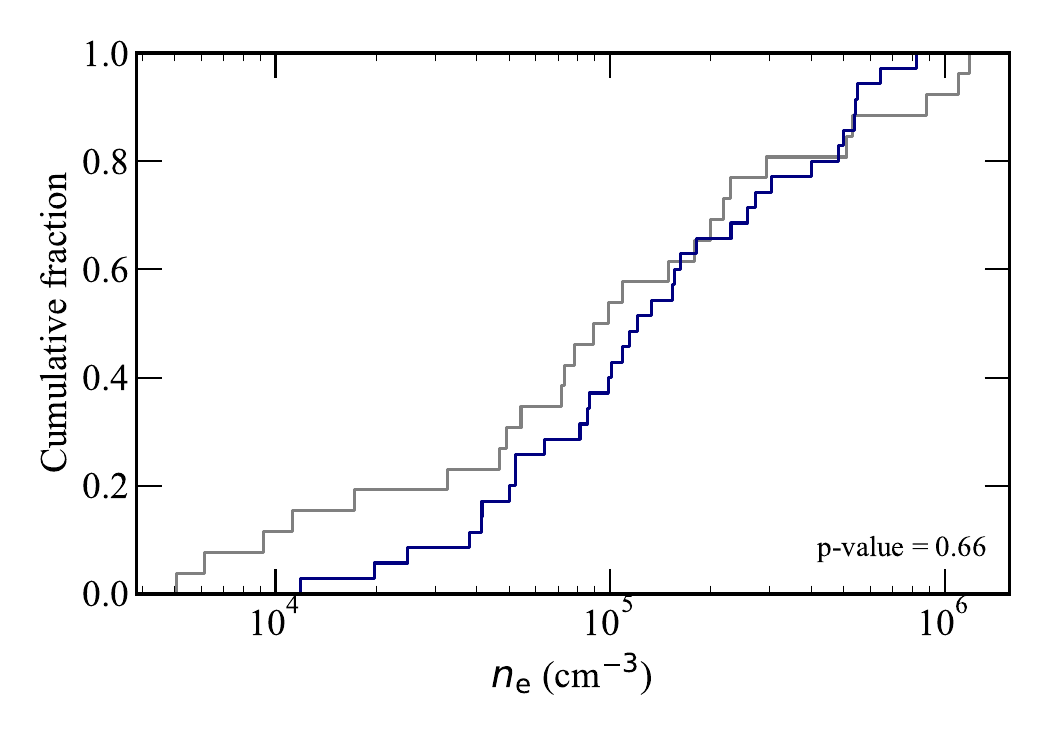} \\
    \caption{Cumulative distribution functions for diameter (top panel) and \nelectron\ (bottom panel) for the sample of HC\,\hii\ regions identified in Paper II and in this work.}
    \label{fig:cdf_diam_ne}
\end{figure}
In Figure\,\ref{fig:phys_parameters_dist}, we present physical parameters for a total of 61 sources, 31 of them discussed in this work, along with the 26 \hii\ regions identified in Paper\,II, which include 13 HC\,\hii\ regions, 6 intermediate objects, 6 UC\,\hii\ regions, and one radio jet candidate. A comparison of the distributions of physical parameters of both samples reveals similar trends, which is expected, given that both sets of sources originate from the same underlying population.
To ensure the comparability of the two datasets, we focused on the key physical properties, diameter and electron density (\nelectron). The cumulative distribution functions for these parameters are shown in Figure\,\ref{fig:cdf_diam_ne}, showing strong agreement between the two samples. We performed a two-sample Kolmogorov-Smirnov (KS) test to determine whether the electron densities and diameters in both samples are drawn from the same population. The KS-test returned p-values of 0.67 for diameter and 0.66 for \nelectron, indicating no statistically significant differences between the two samples. This statistical consistency suggests that the two datasets can be reliably combined for further analysis. Given that the emission measure is proportional to both the electron density and diameter, it is also expected to show consistency across both samples.
By combining these 26 sources with the 35 newly identified \hii\ regions from this study, we form a unified sample of 61 \hii\ regions. This combined sample consists of 33 HC\,\hii\ regions, 15 intermediate objects, 9 UC\,\hii\ regions, and 4 radio jet candidates. Notably, 11 sources remain optically thick at 24\,GHz, likely indicating that these are the youngest \hii\ regions, as their high optical depths suggest they are still deeply embedded within their natal environments and have yet to expand significantly. This combined dataset will be referred to as the SCOTCH sample throughout the remainder of this paper.

\subsection{Physical properties of the SCOTCH sample}

\begin{figure}
    \centering
    \includegraphics[width=0.455\textwidth]{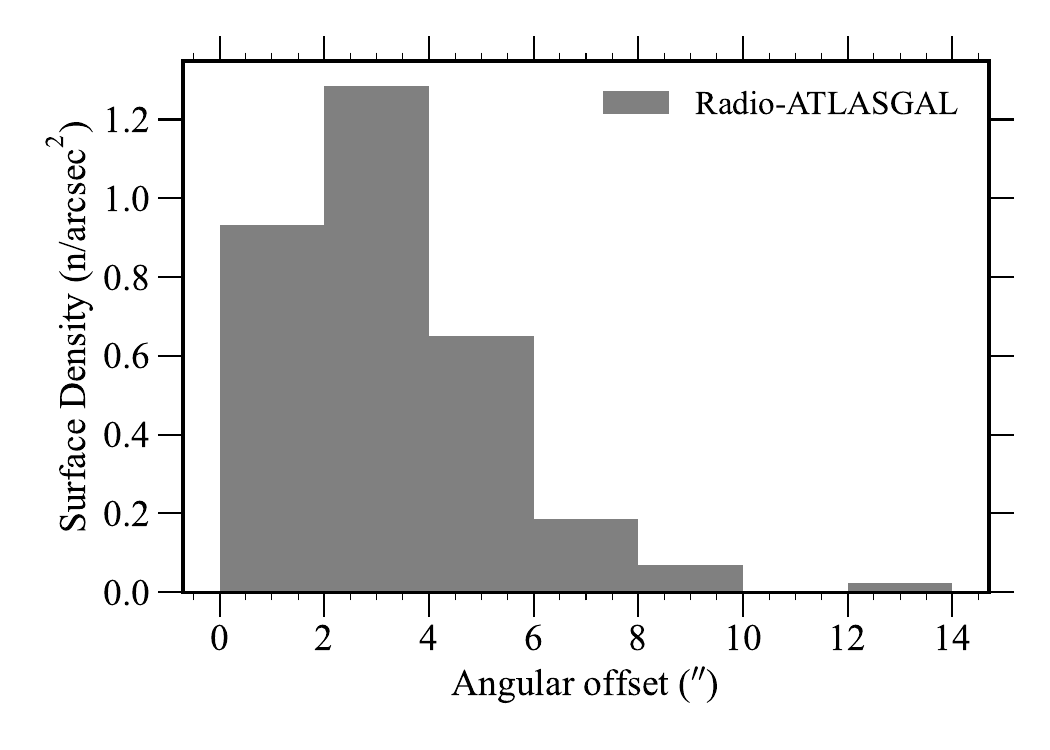}
    \includegraphics[width=0.455\textwidth]{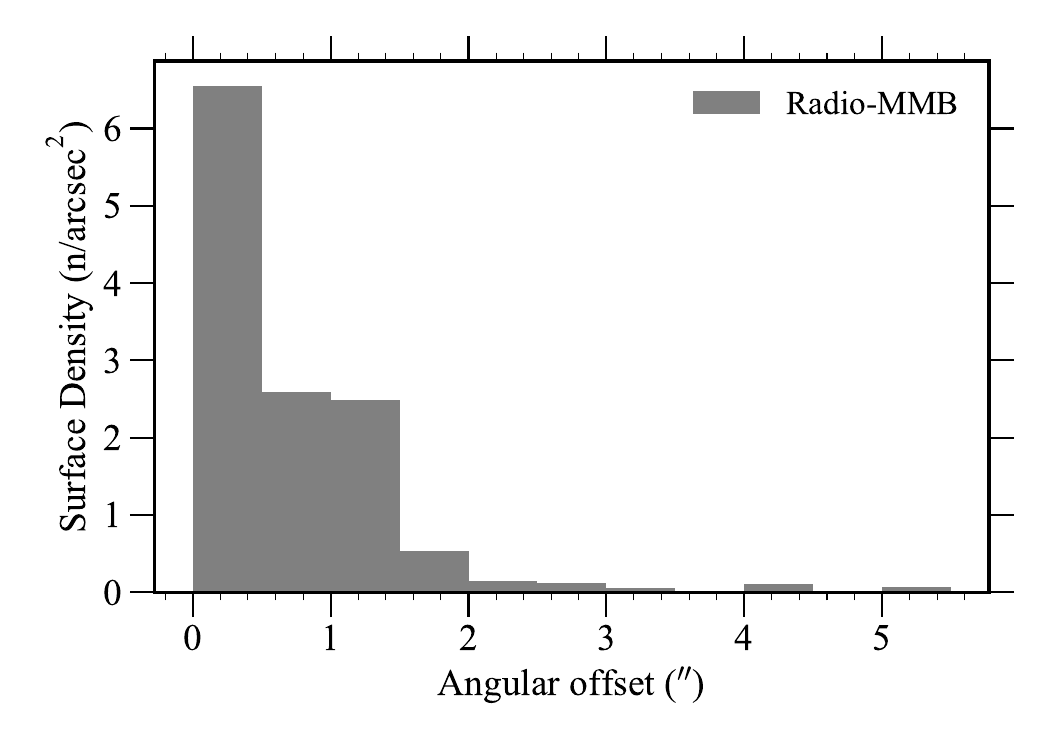} \\
    \caption{Histograms of the angular separation between the detected radio sources and their nearest ATLASGAL dust clump (top panel), MMB maser (bottom panel). The histograms are binned using a value of 2\,arcsec and 0.5\,arcsec, respectively.
    \label{fig:offset_histograms}}
\end{figure}

\begin{figure*}
    \centering
    \includegraphics[width=0.455\textwidth]{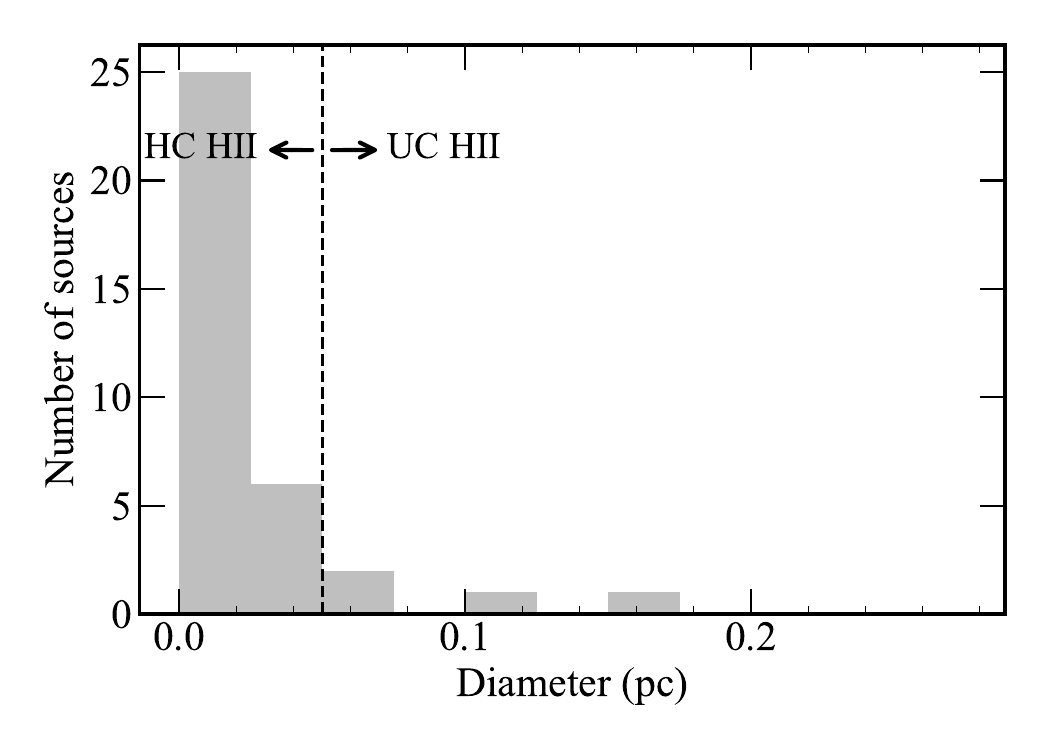}
    \includegraphics[width=0.455\textwidth]{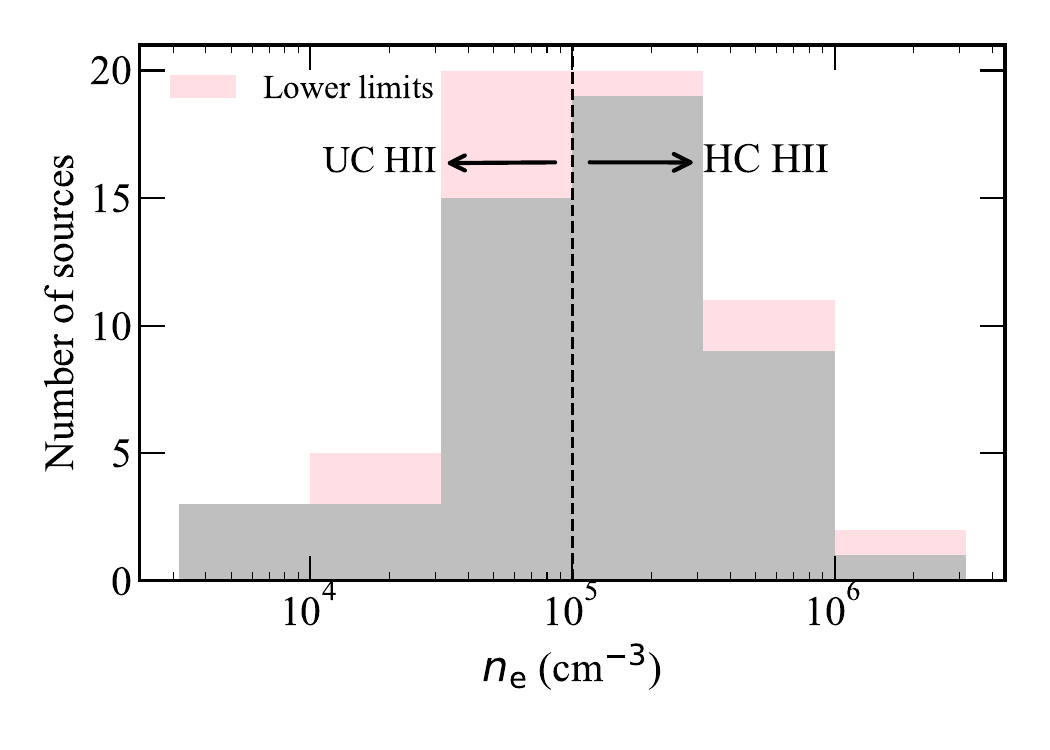} \\
    \includegraphics[width=0.455\textwidth]{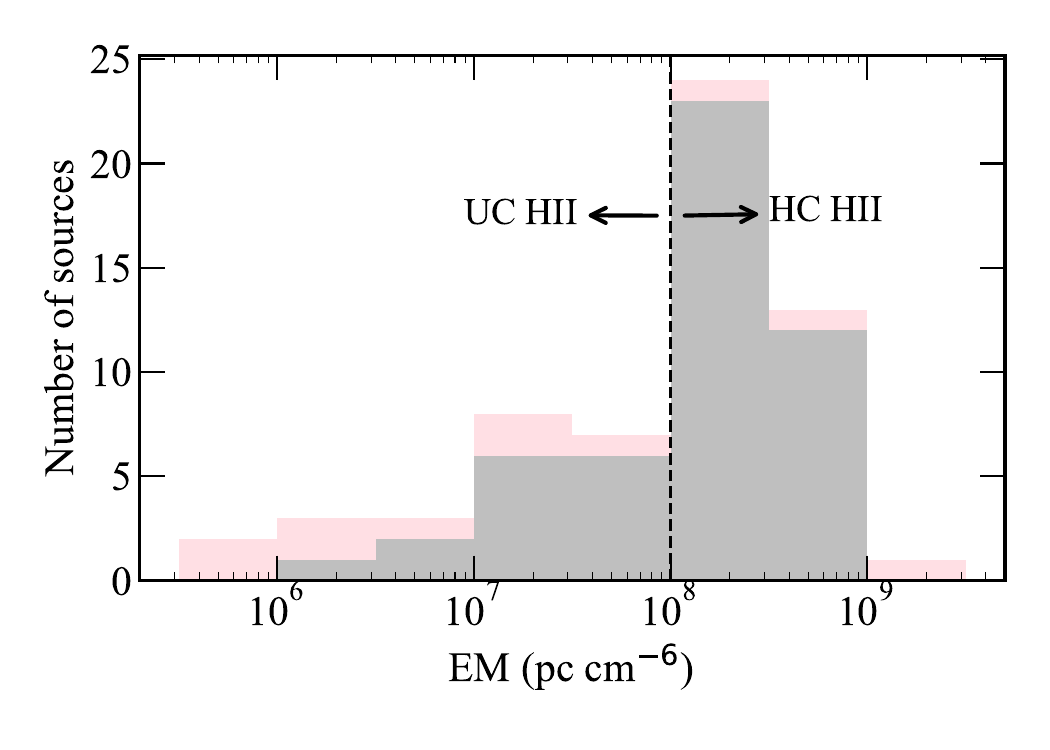}
    \includegraphics[width=0.455\textwidth]{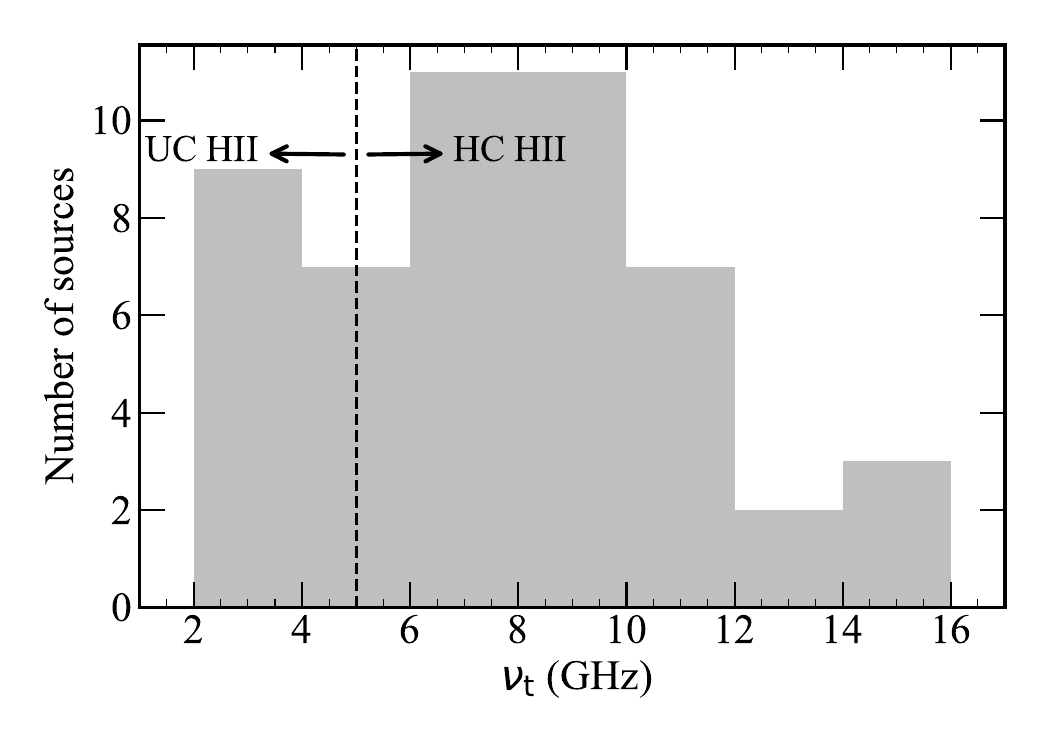}
    
    \caption{Histograms presenting the distribution of source parameters for the full sample of SCOTCH \hii\ regions. The diameter (upper left panel), electron density (\nelectron; upper right panel) emission measure (lower left panel) and turnover frequency (\vt; lower right panel) are shown. All histograms include all  61 \hii\ regions identified with the exception of the turnover frequency plot as this excludes optically thick regions. The pink fraction of the distribution represents values that are lower limits. The bin sizes are 0.025\,pc, 0.5\,dex, 0.5\,dex and 2\,GHz for the diameter, \nelectron, emission measure and turnover frequency, respectively. The black vertical line and arrows emphasize the region of the parameter space in which we expect to find HC\,\hii\ regions.}
    
    \label{fig:histogram_diam_ne_em}
\end{figure*}

\begin{figure}
    \centering
    \includegraphics[width=0.455\textwidth]{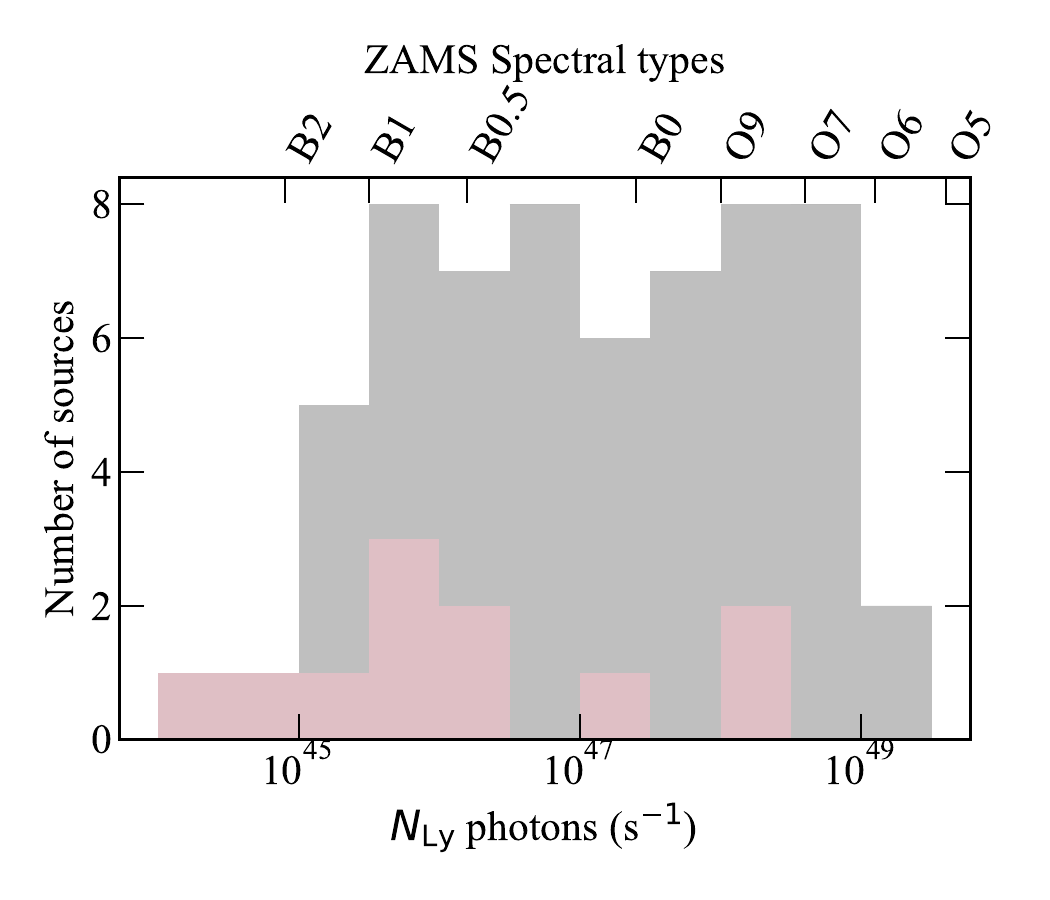} 
    \caption{Histogram of the Lyman continuum photon  flux for the \hii\ regions identified throughout SCOTCH. The bin size is 0.5\,dex. The pink fraction of the distribution represents values that are lower limits.}
    \label{fig:turnoverandlyman}
\end{figure}

In this section, we investigate the physical properties and star formation tracers associated with the SCOTCH sample. It is expected that HC\,\hii\ regions are deeply embedded and detected towards the centre of dense clumps and often are in close proximity to methanol masers. Therefore, we match the SCOTCH sample to ATLASGAL clumps and their observed MMB methanol masers.

Figure\,\ref{fig:offset_histograms}, presents histograms illustrating the distribution of angular offsets between the positions of methanol masers and the peaks of dust emission. These plots demonstrate that the compact \hii\ regions identified in SCOTCH are all embedded towards the centre of their natal clumps and are tightly correlated with methanol masers. 
Furthermore, out of the 476 methanol masers observed, 57 are associated with high-frequency radio continuum emission within 5\,arcsec, accounting for approximately 12\,per\,cent of the total sample. Of these, 47 masers are located within 2\,arcsec (representing $\sim$ 83\,per\,cent of the detections). This association rate is comparable to the 12\,per\,cent reported by \citet{nguyen2022} in their analysis of methanol masers from the Global View of Star Formation in the Milky Way (GLOSTAR) survey \citep{brunthaler2021}, which focused on the first quadrant of the Galactic plane. The remaining 88\,per\,cent of methanol masers, which are not currently associated with high-frequency radio emission, are likely in an earlier stage of their evolution.

To differentiate between the classes of \hii\ regions, we present the distributions of the diameter, electron density, emission measure and turnover frequency in Fig.\,\ref{fig:histogram_diam_ne_em}. The emission measure and electron density of the optically thick radio sources are given as lower limits and are represented by the pink region of the histogram. In Figure\,\ref{fig:turnoverandlyman}, we present the Lyman photon flux distribution for the sample.  The vertical dotted line in Fig.\,\ref{fig:histogram_diam_ne_em} represents the typical threshold for HC\,\hii\ regions (see Table\,\ref{tab:criteria}) and demonstrates that our sample contains parameters consistent with those of HC\,\hii\ regions.

The calculated Lyman photon fluxes range from 10$^{44.2}$ to 10$^{49.4}$ s$^{-1}$ and are consistent with a ZAMS star between B3 and O5 \citep{panagia1973}, assuming that a single ZAMS star is the primary source of the ionizing photons and there is no absorption from dust in the ionization-bounded \hii\ region (the corresponding spectral types are given on the top of Fig.\,\ref{fig:turnoverandlyman}). The estimated uncertainty in the \nly\ is $\sim$ 20 per\,cent, which takes into consideration the errors associated with the flux densities and distances (e.g., \citealt{urquhart2013_cornish,sanchez_monge2013}). The spectral types of stars can be over or under-estimated if multiple stars are responsible for the ionisation or if there is dust absorption present in the \hii\ region \citep{wood1989b,Yang2021}. For UC\,\hii\ regions, the dust absorption fraction varies from $\sim$ 50$-$90\,per\,cent, which can decrease the flux density by a factor of two or more. This implies that the observed flux may be lower than the expected flux and that the spectral types we derive may be earlier than estimated \citep{wood1989b, kurtz1994, Yang2021}. Consequently, the Lyman continuum fluxes reported here should be regarded as lower limits. Nevertheless, all of the radio sources exhibit sufficient Lyman fluxes to indicate the presence of an embedded high-mass star.

\begin{figure}
    \centering
    \includegraphics[width=0.49\textwidth]{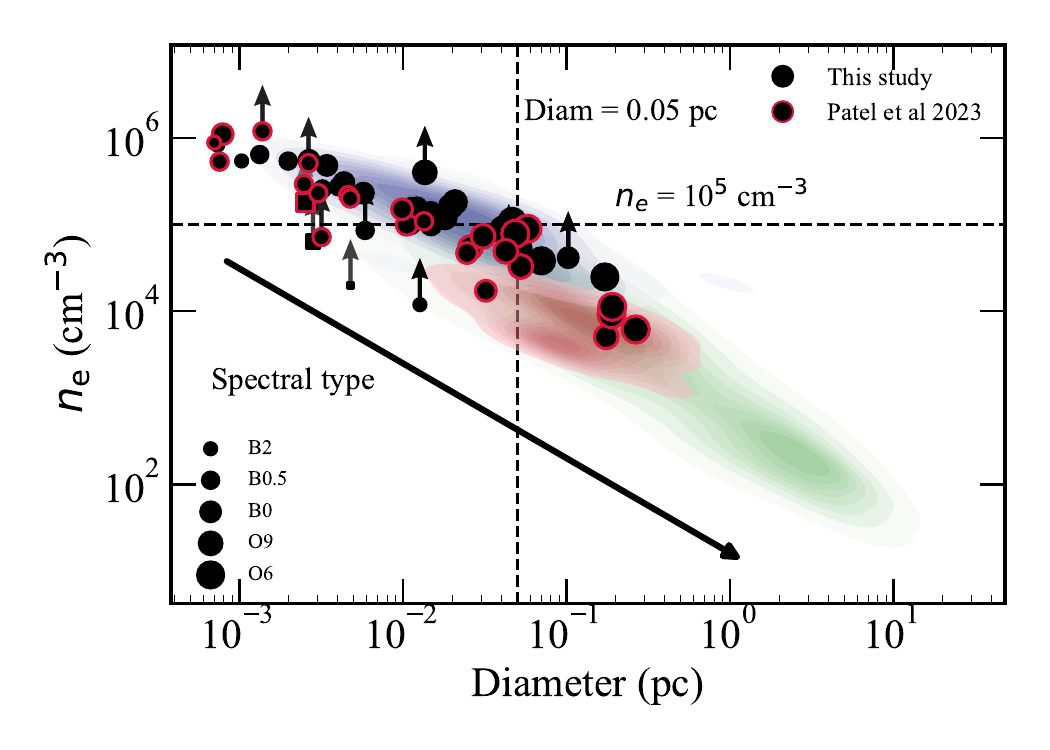}
    \caption{Distribution of the electron density as a function of diameter for the 61 objects identified in SCOTCH. The size of the data points denote the approximate ZAMS spectral type. The square data points represent the jet-like candidates. For the  optically thick sources the  entries for electron density and spectral type are lower limits.  The blue coloured area represents a kernel density estimation for the previously identified HC\,\hii\ and intermediate \hii\ regions \citep{,Yang2019,Yang2021}. The area coloured in pink  represent the UC\,\hii\ regions as identified in \citet{Kalcheva2018} and the green area present the compact regions in \citet{Khan2024}. The black solid arrow shows the typical evolutionary trend a compact \hii\ region is likely to follow as these objects expand (See Fig.\,11 of Paper II).}
    \label{fig:global_parameters}
\end{figure}

In the following section, we discuss in detail the similarities and differences between the compact \hii\ regions identified in this series of studies and the established population of \hii\ regions from the literature \citep{Kalcheva2018,Yang2019,Yang2021}.

\subsection{Comparison with UC \hii\ regions}
\label{sec:globalproperties}

\begin{figure}
    \centering
    \includegraphics[width=0.49\textwidth]{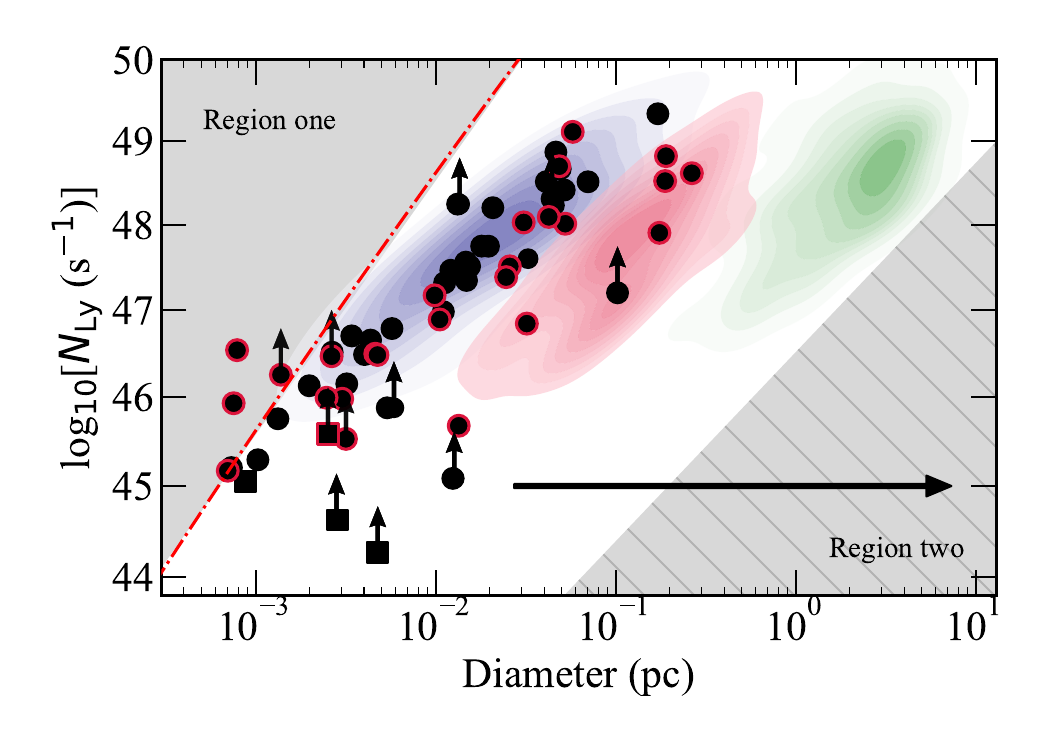}
      \caption{Distribution of the Lyman continuum flux as a function of diameter for the 61 objects identified in SCOTCH. The square data points represent the jet-like candidates. The coloured areas are as described in Fig.\,\ref{fig:global_parameters}. The black arrow represents the implied evolutionary trend for \hii\ regions. The red dot-dashed line represents the Strömgren expansion of \hii\ regions, assuming a constant electron density of $10^6$ cm$^{-3}$ and a recombination coefficient $\alpha = 2.7 \times 10^{-13}$ (cm$^{-3}$).}
    \label{fig:diamlymanKDE}
\end{figure}

The full SCOTCH survey has identified 33 HC\,\hii\ regions, 15 intermediate objects, 9 UC\,\hii\ regions and 4 radio jet candidates. In Figure\,\ref{fig:global_parameters}, we show the distribution of SCOTCH sources as a function of diameter, electron density and spectral type. We include the distributions of similar HC\,\hii\ regions and intermediate \hii\ regions identified by 
\citet{Yang2019} and \citet{Yang2021}; blue filled contours and  UC\,\hii\ regions and compact \hii\ regions identified in \citet{Kalcheva2018}; pink filled contours and \citet{Khan2024}; green filled contours, respectively (see Table\,\ref{tab:criteria} for \hii\ region definitions).

The distribution of the data for the SCOTCH sources overlaps with those of both previously identified HC and UC\,\hii\ and intermediate \hii\ regions, creating a continuum of \hii\ regions that exhibit a strong correlation between their sizes and electron densities. 
However, we note that our \hii\ region sample generally consists of smaller and denser objects that tend to surround early B-type stars. The overlap with the UC\,\hii\ regions is quite small, indicating that SCOTCH has primarily identified a sample of young \hii\ regions including some of the youngest yet discovered. There are a few outliers that do not overlap with either the HC and UC\,\hii\ region envelopes, however, these are optically thick and their electron densities lower limits. 

To investigate the evolution of \hii\ regions, it is essential to understand the relationship between their size and the number of ionizing photons. In Figure\,\ref{fig:diamlymanKDE}, we show the Lyman continuum flux as a function of diameter for all compact \hii\ regions identified in the SCOTCH survey, alongside the HC\,\hii\ regions and intermediate objects identified by \citet[][blue filled contours]{Yang2019,Yang2021}, the UC\,\hii\ regions identified by \citet[][red filled contours]{Kalcheva2018} and the compact \hii\ regions identified by \citet[][green filled contours]{Khan2024}. For optically thick sources upper limits for \nly\ are plotted. 

The distributions of all three \hii\ region samples form diagonal, elongated bands in the figure, indicating a trend of increasing diameter with photon flux. However, the Lyman photon flux across the three \hii\ region stages remains similar, ranging from approximately 10$^{46}$ to 10$^{49}$\,s$^{-1}$. The fact that the distributions for these stages overlap indicates evolution of the \hii\ regions is from left to right, as shown by the black arrow in Fig.\,\ref{fig:diamlymanKDE}. This leads us to conclude that the Lyman photon flux remains constant throughout the formation of the HC\,\hii\ region and during its expansion from 0.05\,pc to 5\,pc; this is consistent with the findings of \citet{Yang2021} and Paper\,II, but here we have extended the analysis to include compact \hii\ regions. This would also suggest that the final mass of the star driving the \hii\ region is determined in the \hchii\ region stage. 

The overall distribution of the individual \hii\ region stages presents a distinctive pattern that requires further explanation. If we take the distribution of the \hchii\, region sample in Fig.\,\ref{fig:diamlymanKDE}, we note a paucity of sources with small sizes and high Lyman photon fluxes (the grey region or region one in Fig.\,\ref{fig:diamlymanKDE}) and a lack of large sizes with low Lyman photon fluxes (grey hatched region or region two in Fig.\,\ref{fig:diamlymanKDE}); if the evolution of \hii\ regions proceeds from the left to the right then these areas of parameter space should not be empty. The sparseness of sources in the blue region  of Fig.\,\ref{fig:diamlymanKDE} is likely to be the result of the enormous number of ionizing photons that drives the rapid expansion of \hii\ regions around late O-type stars, making it extremely unlikely to detect an \hchii\ region around an O-type star (also see discussion of \hii\ region sizes around O-type and B-type stars in \citealt{urquhart2013_cornish}). Region one is broadly in agreement with the dot-dashed red line which represents the initial Strömgren expansion of \hii\ regions.
Although only a few sources fall within this  region, it is likely that these are situated in complex natal environments.

The lack of sources in region two of Fig.\,\ref{fig:diamlymanKDE} is the result of the fixed number of Lyman flux and expanding \hii\ region, which results in a decrease of the surface brightness. This affects all \hii\ regions as they expand. However, the lower Lyman photon flux emitted by B-type stars causes the radio emission from the ionized bubbles to drop below the survey's sensitivity limits, resulting in a lack of large \hii\ regions around early B-type stars. The combination of rapid expansion for O-type stars and limitations in telescope sensitivity to larger \hii\ region around early B-type stars result in the diagonal morphologies seen for the three stages shown in Fig.\,\ref{fig:diamlymanKDE}. We note a complete lack of any compact \hii\ regions associated with B-type stars and this is consistent their surface brightness being below the telescope sensitivity.

Comparing the distribution of the SCOTCH \hii\ regions to the distribution of  \hii\ region in other  stages we, find that six are coincident with the \uchii\ region distribution (filled red contours) but the majority are tightly correlated with the \hchii\ region distribution. Furthermore, the distribution of SCOTCH sources shown in Fig.\,\ref{fig:diamlymanKDE} reveals that we have extended the observed trend for \hchii\ regions to lower Lyman photon fluxes and smaller sizes. We note three significant outliers in both the literature \hii\ region distribution and our extended \hchii\ region distribution; however, all of these outliers are optically thick, with two identified as potential jet candidates.

\section{Summary and Conclusions}
\label{sec:conclusions}

We present the results of multi-resolution, high-frequency radio continuum  observations with the ATCA towards 335 methanol masers located within the fourth quadrant of the Galactic plane. The observations were conducted using the H214 and 6A antenna configurations, which provide both large and small scale angular resolutions of 20\,arcsec and 0.5\,arcsec, respectively. The main aim of this work is to identify new HC\,\hii\ regions located within the target region, for which previously no high-frequency observations were available. We target methanol masers as these are known to be excellent tracers of young and embedded high-mass stars \citep{Breen2013_6_7}.

From the 335 fields observed we find 121 reliable radio detections above 3$\sigma$ in the low-resolution maps. Using archival 5-GHz data (e.g., CORNISH-South and MAGPIS) we identify 42 HC\,\hii\ region candidates (i.e., very compact and/or optically thick objects); these were subsequently followed up at high-resolution to confirm their nature. We construct SEDs between 5$-$24\,GHz and use these to derive physical properties including size, electron density, \nelectron, emission measure, EM, Lyman continuum photon flux, \nly and turnover frequency, \vt, assuming an ionization-bounded \hii\ region with a uniform electron density model. In the present study we have identified 20 HC\,\hii\ regions, 9 intermediate objects, 3 UC\,\hii\ regions and 3 radio jet candidates. 

In Papers I and II of this series we had presented the results of similar high- and low-resolution observations towards an initial sample of 141 methanol masers. We detected 68 discrete radio sources in the low-resolution maps, 39 of which were followed up at high resolution. This led to the classification of 13 HC\,\hii\ regions, 6 intermediate objects, and 6 UC\,\hii\ regions. Combined, this series of papers presents high-frequency radio observations towards 476 MMB methanol masers, has identified 189 radio sources, 134 of which are directly associated with the methanol masers targeted, corresponding to an association rate of $\sim$30\,per\,cent. This has resulted in the identification of 33 HC\,\hii\ regions, 15 intermediate objects, 9 UC\,\hii\ regions, and 4 radio jet candidates. This sample includes 11 objects that are optically thick at 24\,GHz; these are categorised as 4 HC\,\hii\ regions, 4 intermediate objects, 1 UC\,\hii\ region, and 2 radio jet candidates. \\

\noindent Our main findings are as follows:

\begin{enumerate}[label=(\roman*),leftmargin=1\parindent]

     \item The compact \hii\ regions identified by SCOTCH are deeply embedded within dusty clumps and are positionally associated with methanol masers and 5-GHz radio emission, all of which are key tracers of the final stages of massive star formation.\\
    
    \item We have identified 48 new HC\,\hii\ regions and intermediate objects. Combining these with the 28 HC\,\hii\ regions and intermediate objects previously reported in the literature, we have increased the sample of \hchii\ regions by a factor of $\sim$3. Furthermore, the HC\,\hii\ regions identified in these works are generally smaller, denser, and surround lower-luminosity stars compared to previously known HC\,\hii\ regions.\\

    \item Of the 11 optically thick sources, 2 are classified as radio jet candidates,
    three are mid-infrared dark, while the remainder are associated with an infrared point source within a protocluster. All of these objects are embedded towards the centres of dense high-mass clumps and show signs of active star formation, indicating they are likely to be the youngest \hii\ regions in our sample.\\
    
    \item A comparison of the HC\,\hii\ regions identified in this work with previously identified HC\,\hii\ regions shows that their physical parameters are consistent. Comparing these with the larger population of \uchii\ regions and compact \hii\ regions, we find that they form a continuum covering three orders of magnitude in size and electron density. Therefore, the evolution of \hii\ regions appears to be a continuous process rather than consisting of definitive evolutionary stages. \\

    \item Our survey of 476 methanol masers from the MMB catalogue resulted in the identification of 48 HC\,\hii\ regions and intermediate objects, corresponding to a detection rate of approximately 10\,per\,cent. With 498 masers remaining unobserved at high frequencies, extending this study, particularly in the first and second quadrants of the Galactic plane, could potentially identify an additional 50 HC\,\hii\ regions. This would significantly increase the known population of HC\,\hii\ regions, offering a more complete view of their physical properties and their role in the early stages of massive star formation.

\end{enumerate}

\section*{Acknowledgements}

AYY acknowledges the support from the National Key R\&D Program of China No. 2023YFC2206403, and from National Natural Science Foundation of China (NSFC) grants No. 12303031 and No. 11988101. 
MAT gratefully acknowledges the support of the Science \& Technology Facilities Council through grant award ST/W00125X/1. We are deeply saddened by the passing of our co-author, Prof. Dr. Karl Menten, during the final stages of preparing this paper. His contributions to this work were invaluable, and he will be greatly missed.

\section*{Data Availability}

The data underlying this article are available in the article and in its online supplementary material. The full version of Tables\,3,4,5 and 6 is available on CDS. Copies of the SEDs and three colour composite maps are available in the Appendix.



\bibliographystyle{mnras}
\bibliography{bibtex_ash_new}



\newpage
\appendix

\section{Excluded radio maps}

  \begin{figure*}
  \centering
  \includegraphics[width=0.49\textwidth, height=0.35\textwidth]{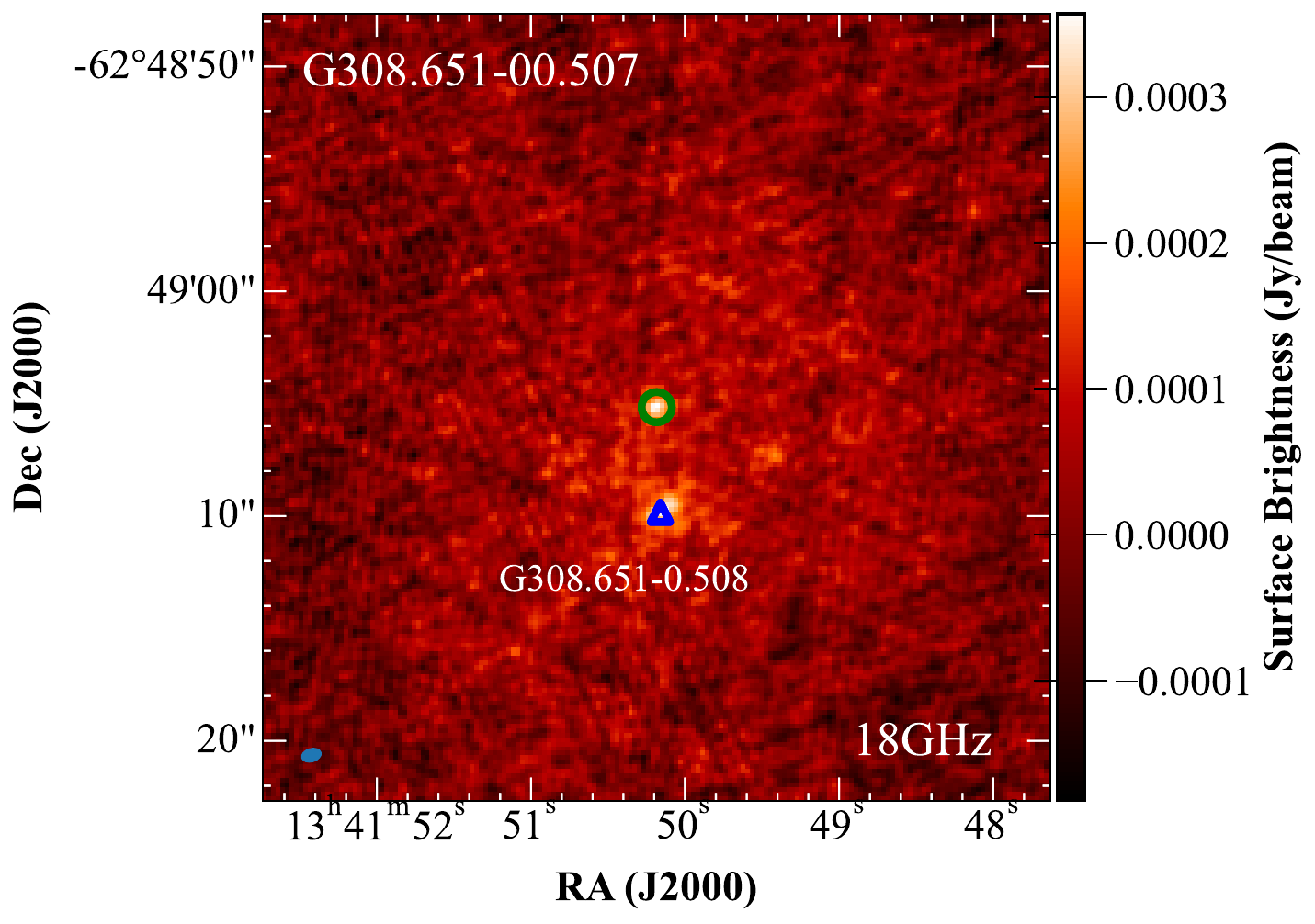}
  \includegraphics[width=0.49\textwidth, height=0.35\textwidth]{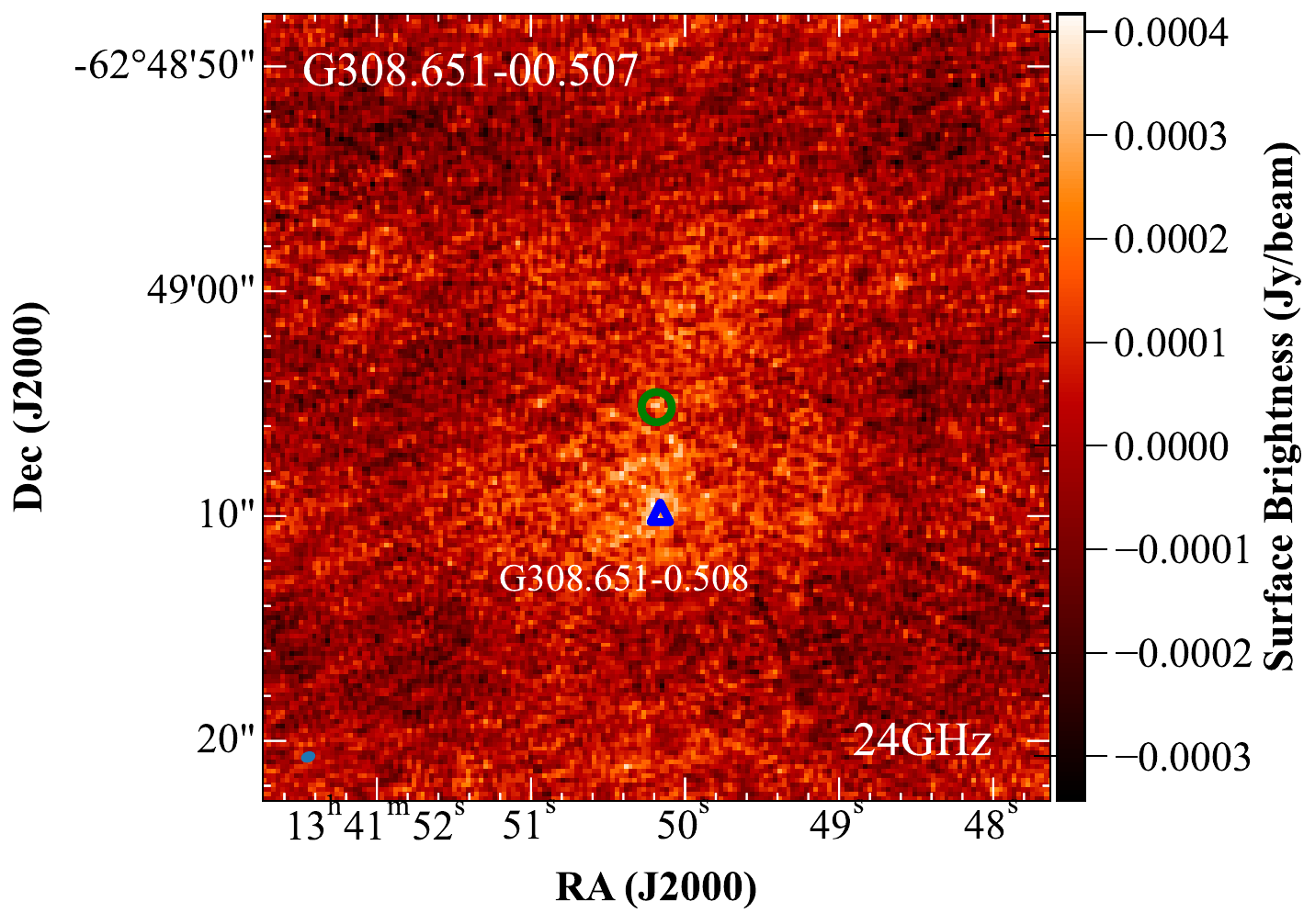}
  \includegraphics[width=0.49\textwidth, height=0.35\textwidth]{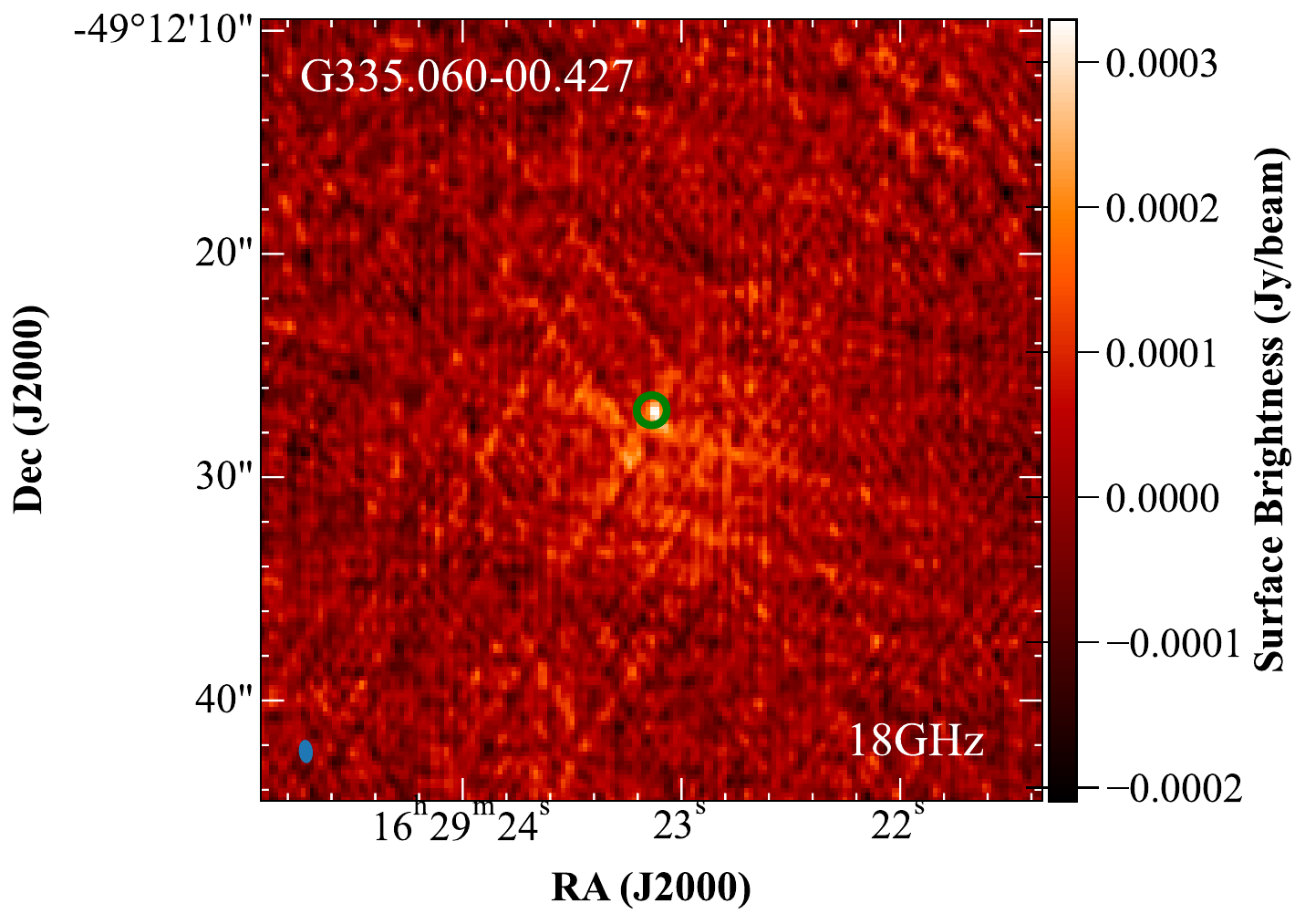}
  \includegraphics[width=0.49\textwidth, height=0.35\textwidth]{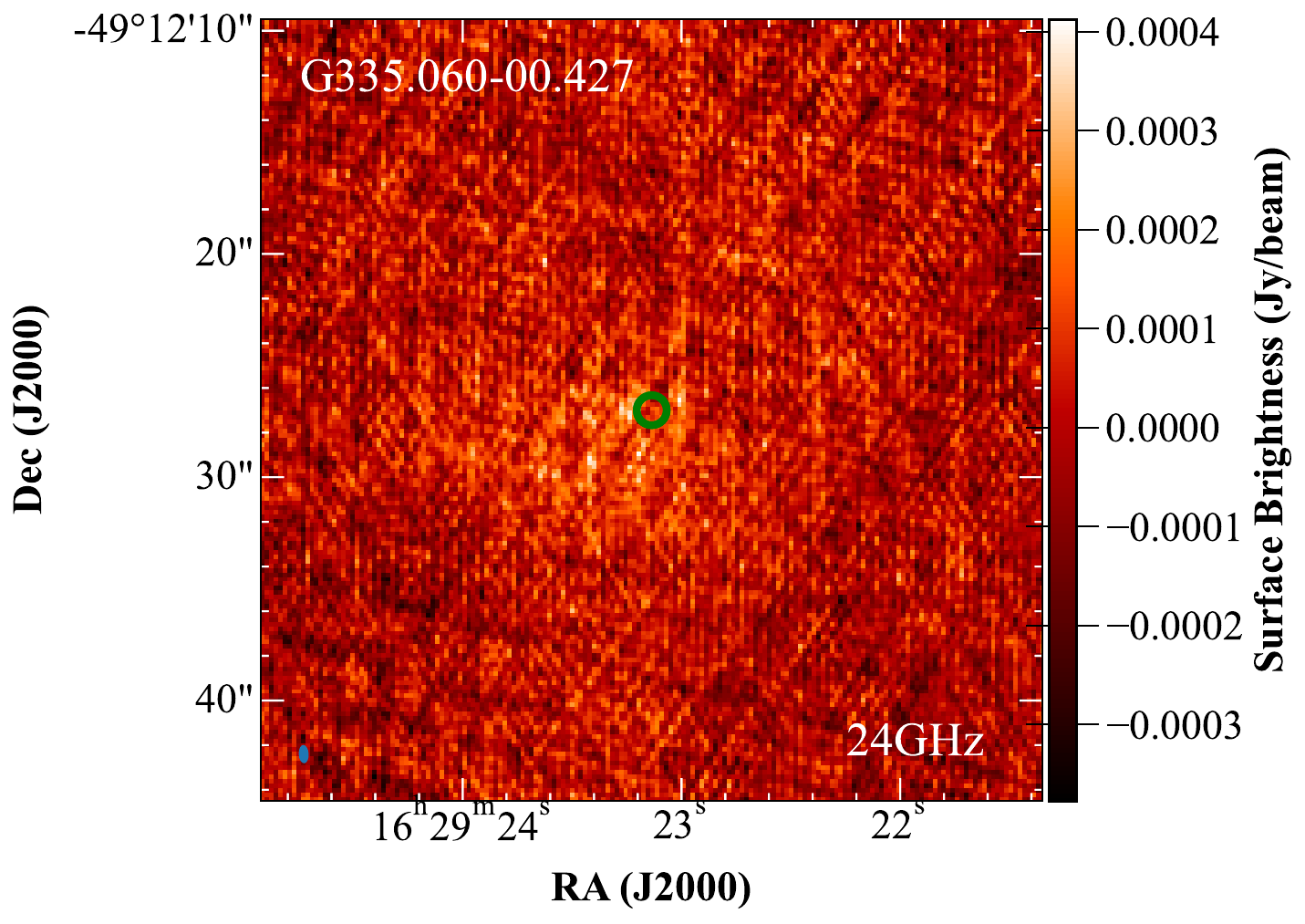}
  \includegraphics[width=0.49\textwidth, height=0.35\textwidth]{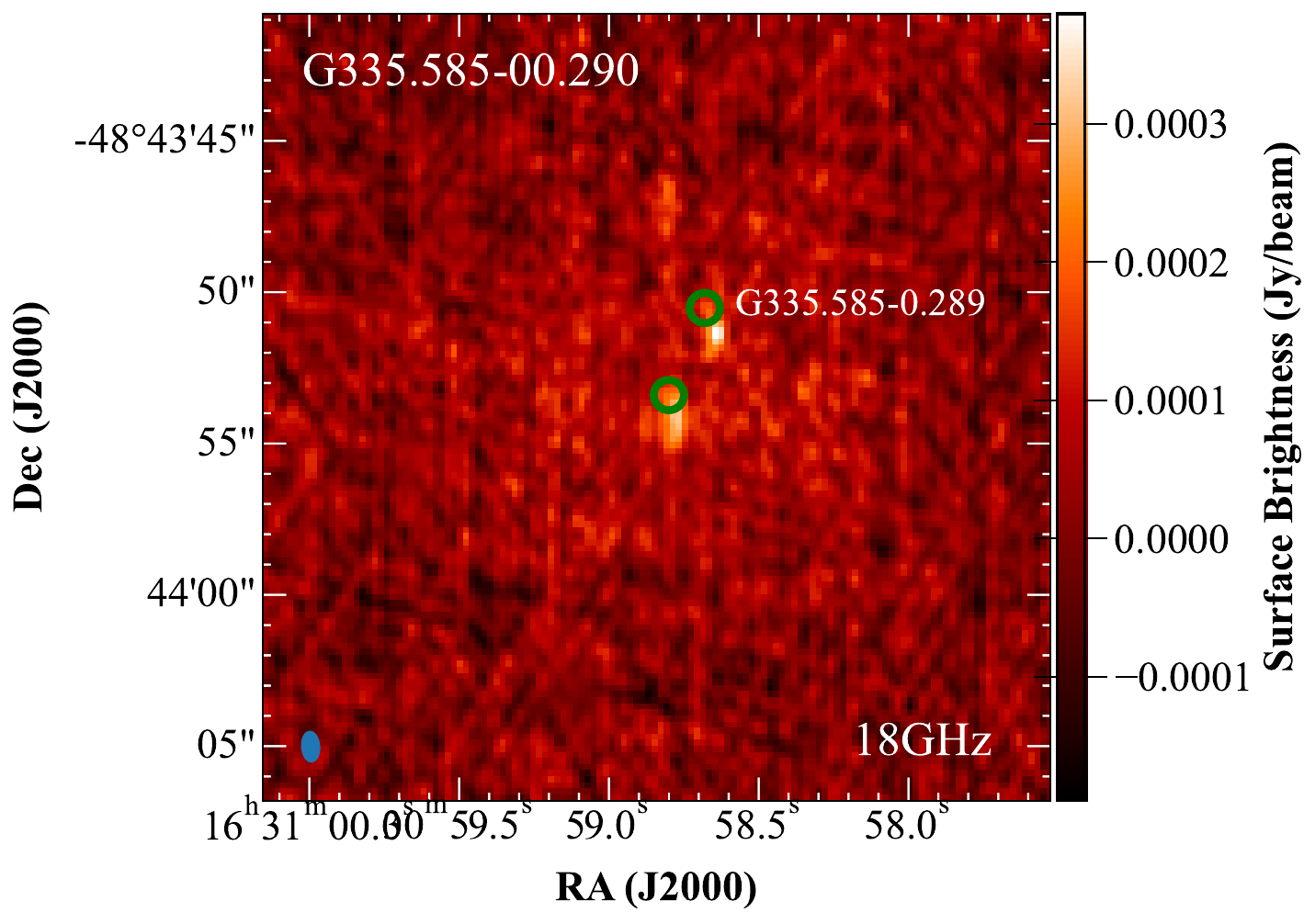}
  \includegraphics[width=0.49\textwidth, height=0.35\textwidth]{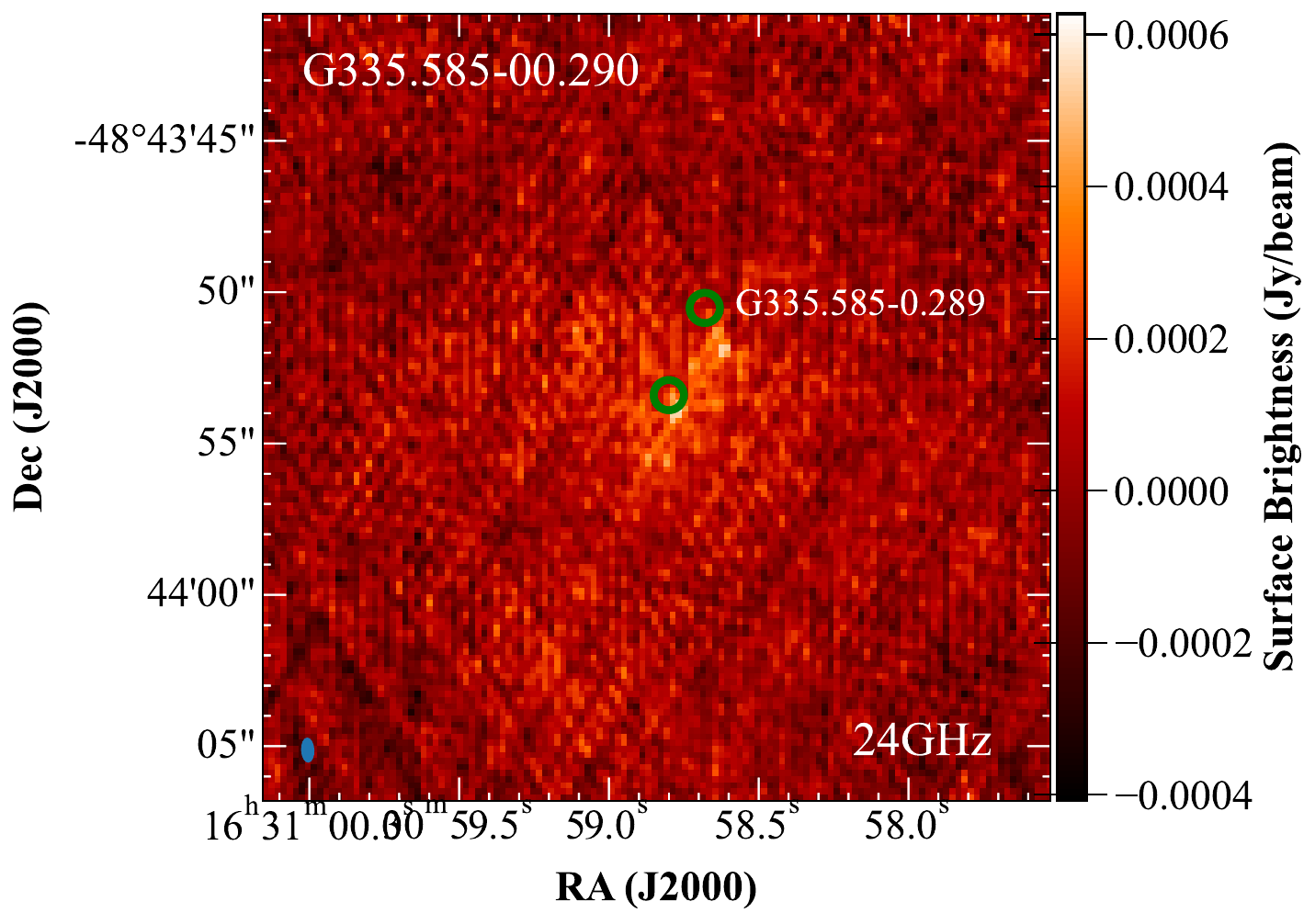}
  \caption{Examples of radio fields that have been excluded in this study. The green circles show the position of the methanol maser(s) located in the field. The blue triangles show the position of any low-frequency (5-GHz) radio counterparts. The filled blue ellipse in the bottom left-hand corner of each image indicates the size and orientation of the synthesised beam. The field name is given in the top left and the radio name is provided if it differs from the field name.}
   \label{fig:over_resolved_appendix}
   \end{figure*}
    \begin{figure*}
    \ContinuedFloat
    \captionsetup{list=off,format=cont}
    \centering
    \includegraphics[width=0.49\textwidth, height=0.35\textwidth]{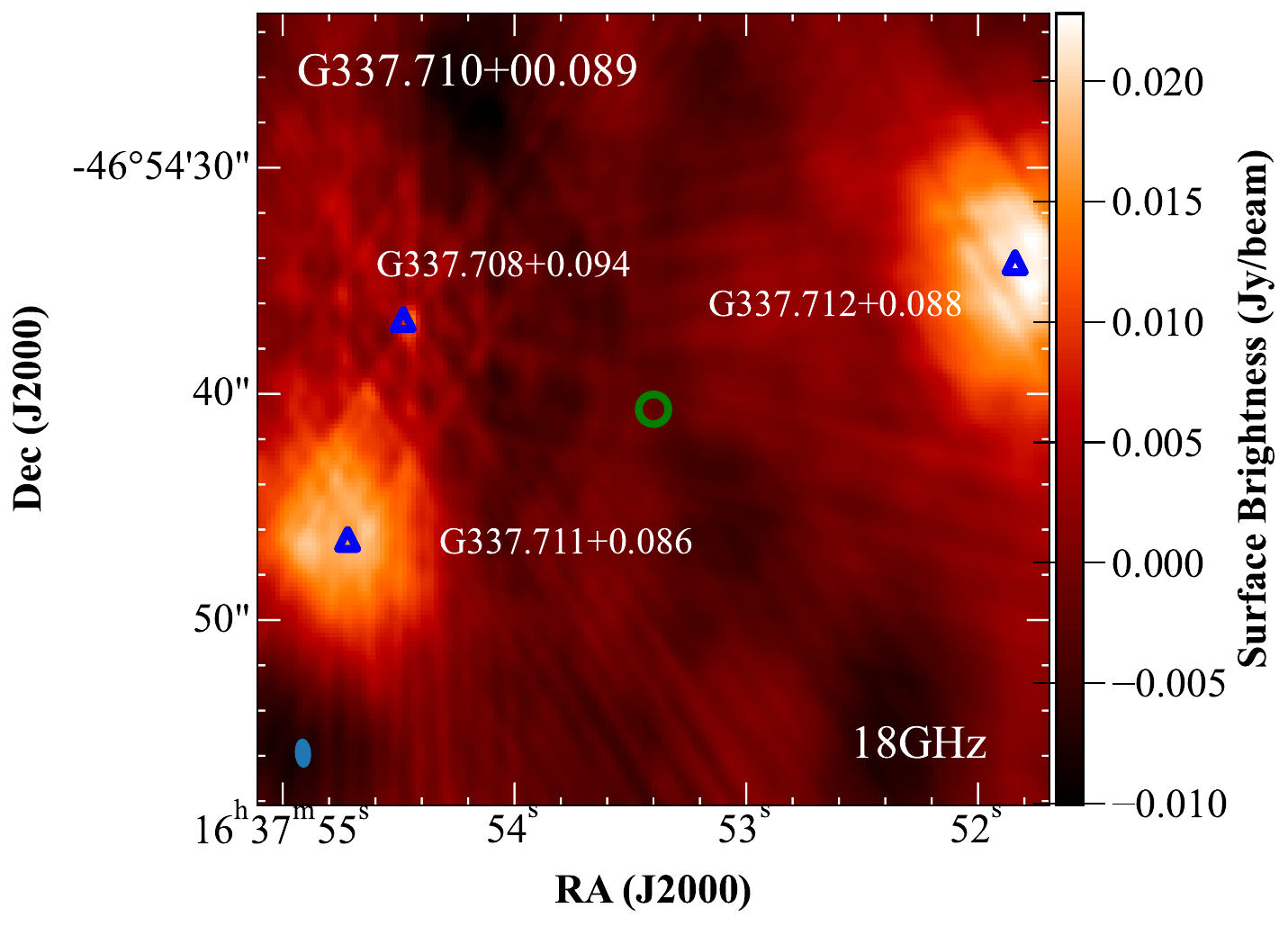}
  \includegraphics[width=0.49\textwidth, height=0.35\textwidth]{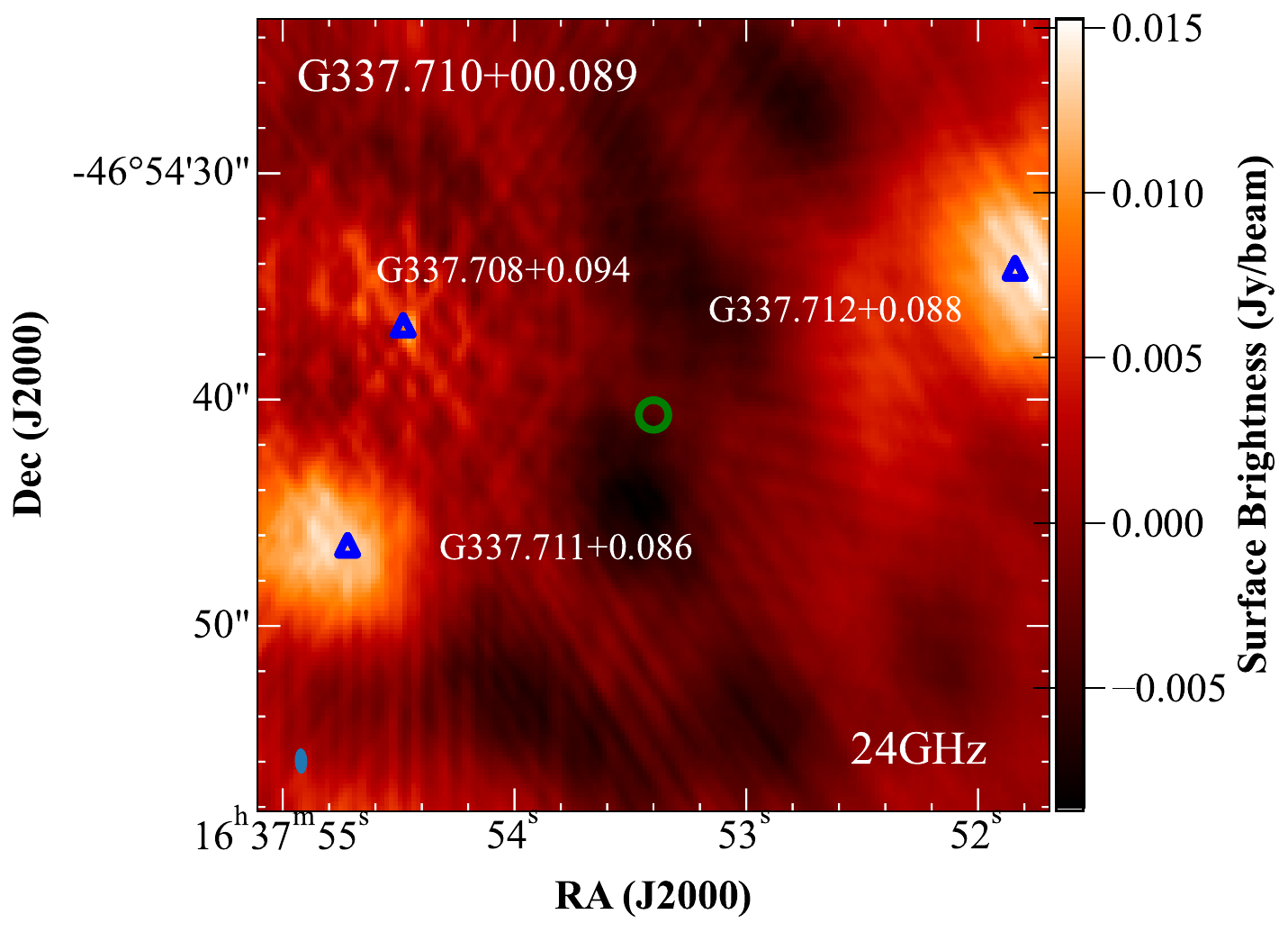}
  \includegraphics[width=0.49\textwidth, height=0.35\textwidth]{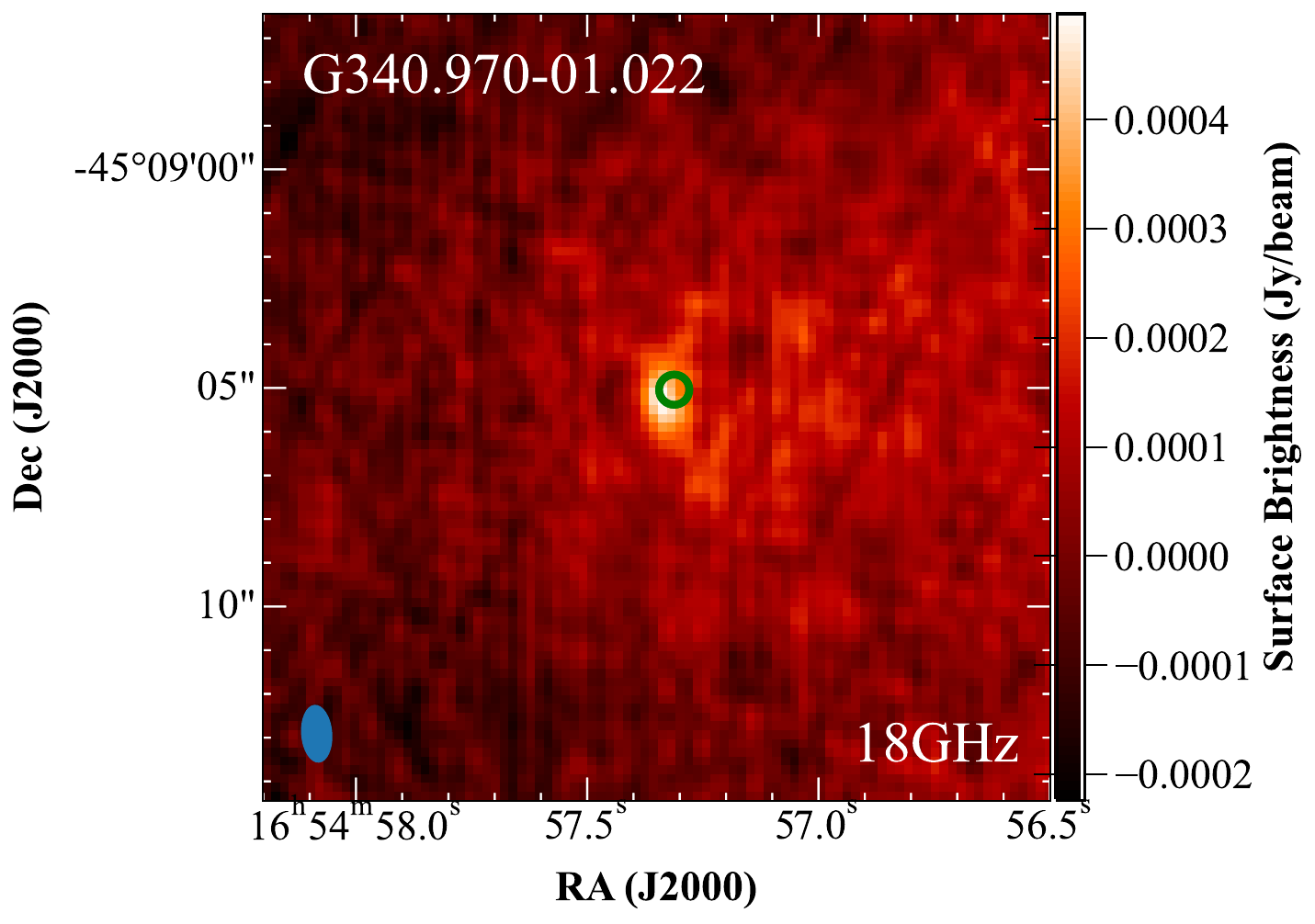}
  \includegraphics[width=0.49\textwidth, height=0.35\textwidth]{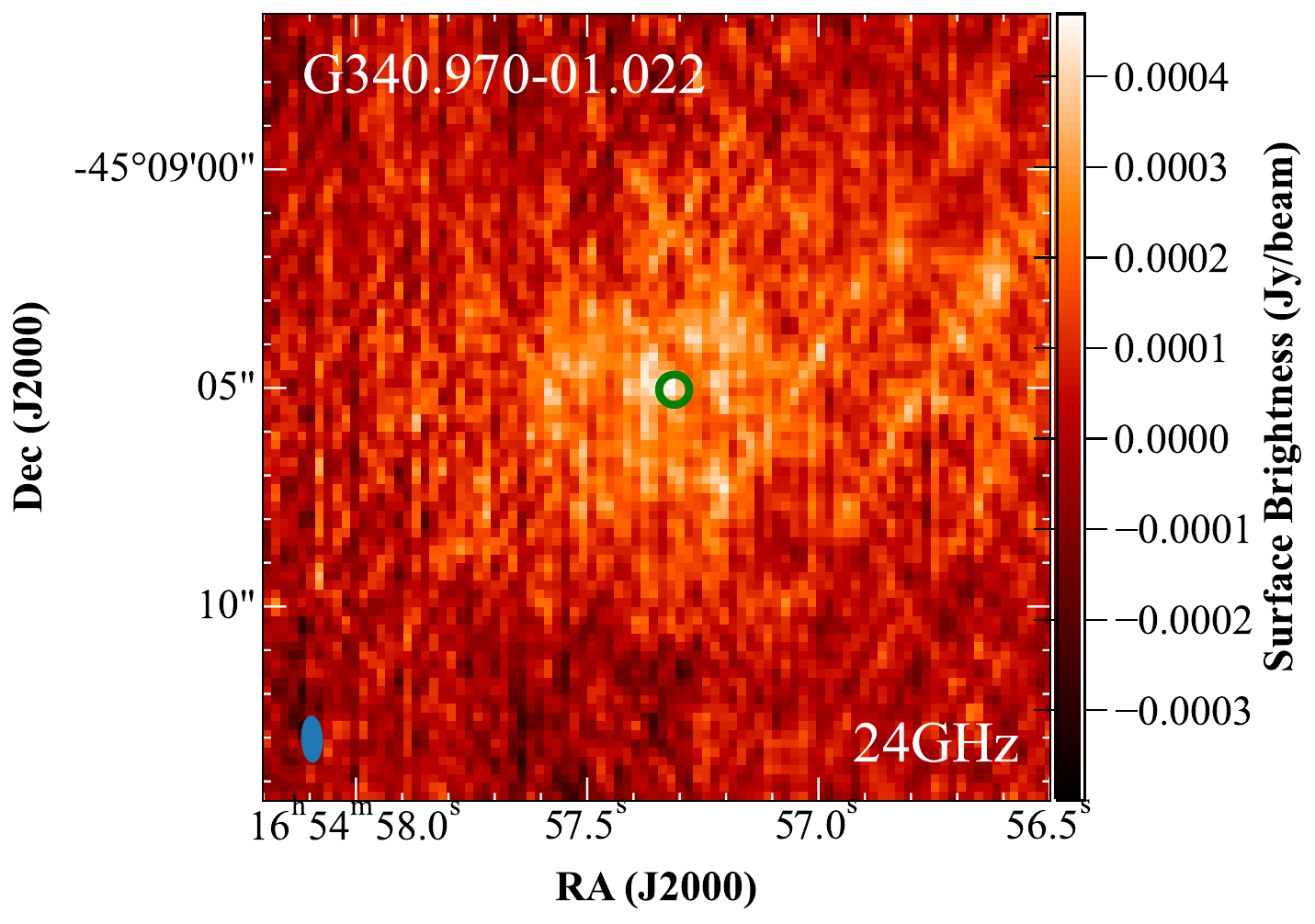}
  \includegraphics[width=0.49\textwidth, height=0.35\textwidth]{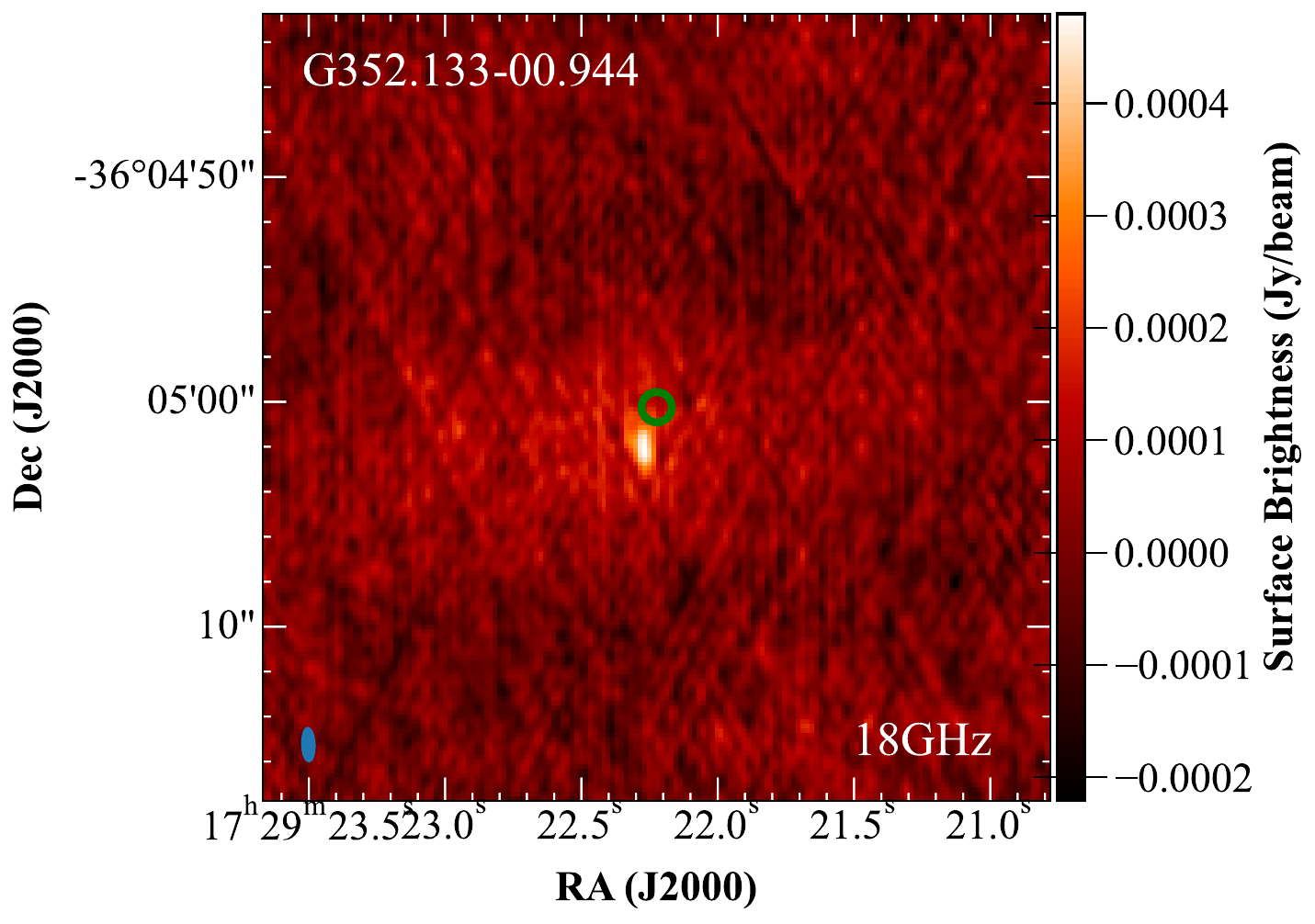}
  \includegraphics[width=0.49\textwidth, height=0.35\textwidth]{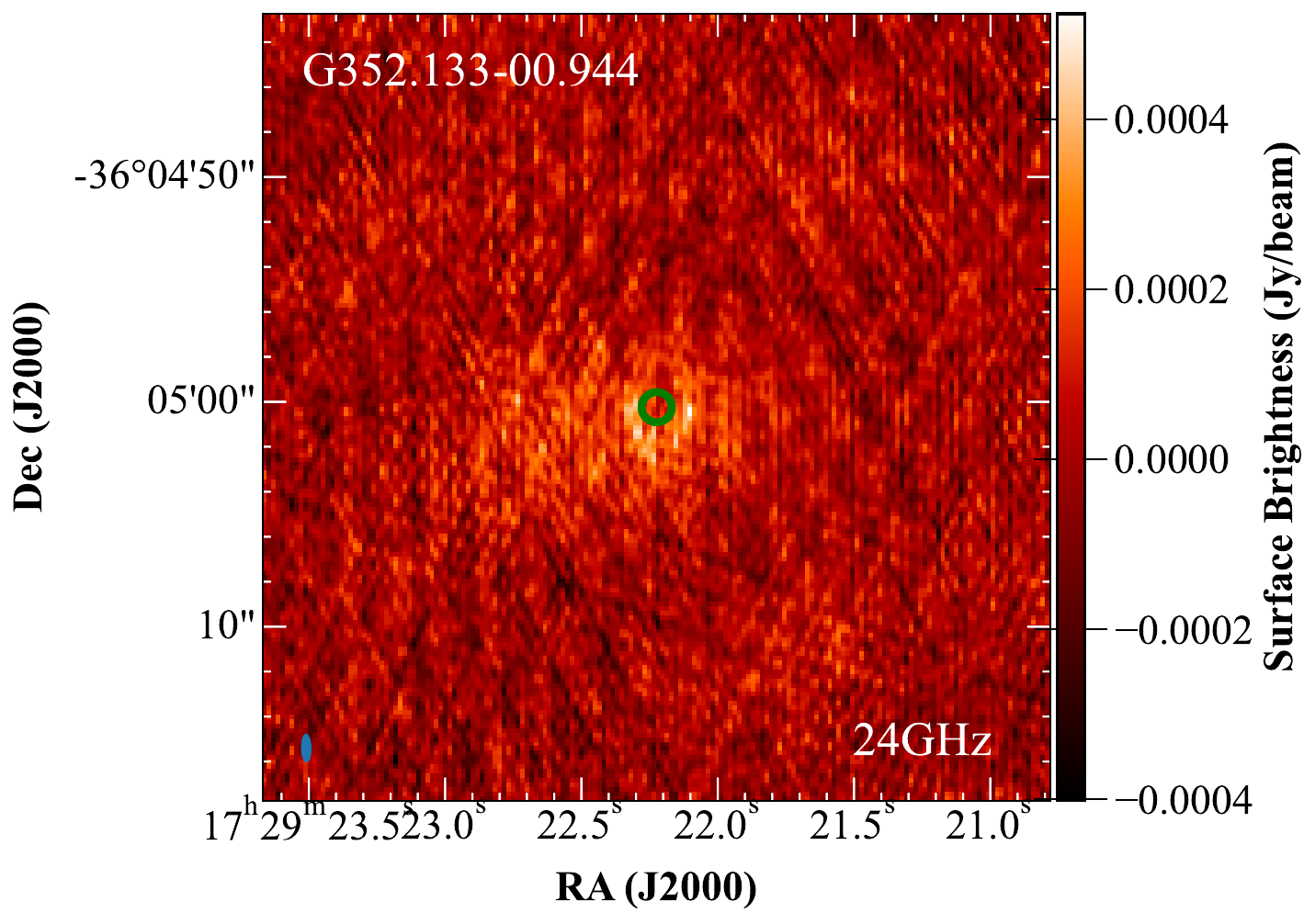}
  \caption{Cont.}
\end{figure*}

\section{Radio SEDs}

\begin{figure*}
    \centering
    \includegraphics[width=0.33\textwidth]{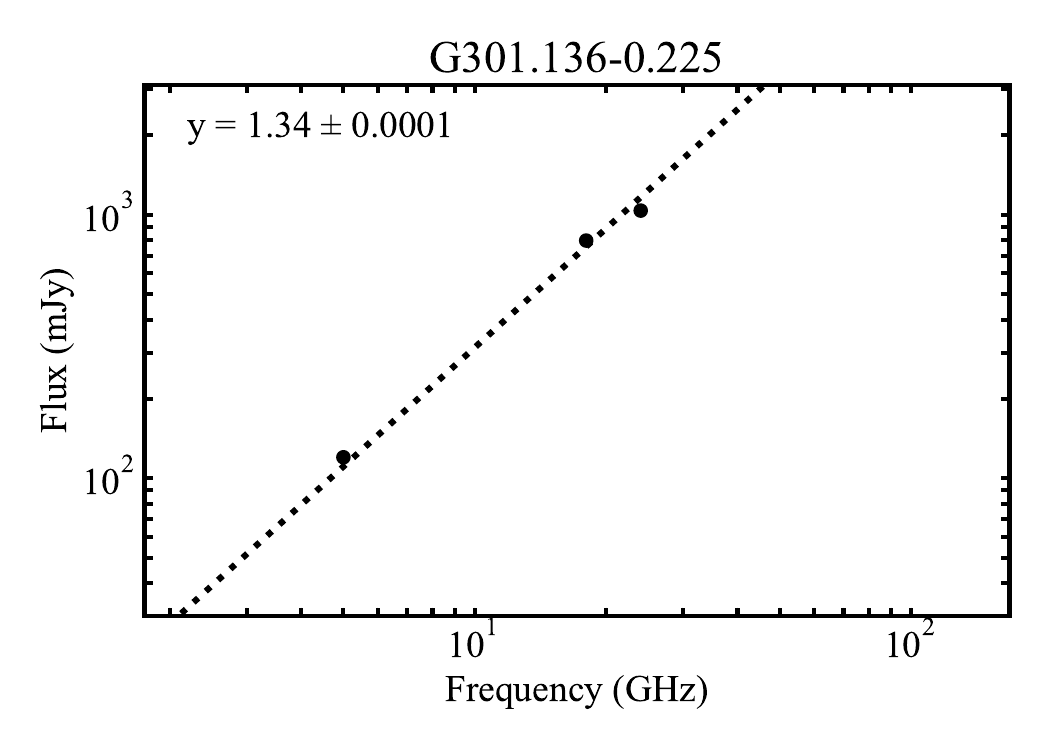}
    \includegraphics[width=0.33\textwidth]{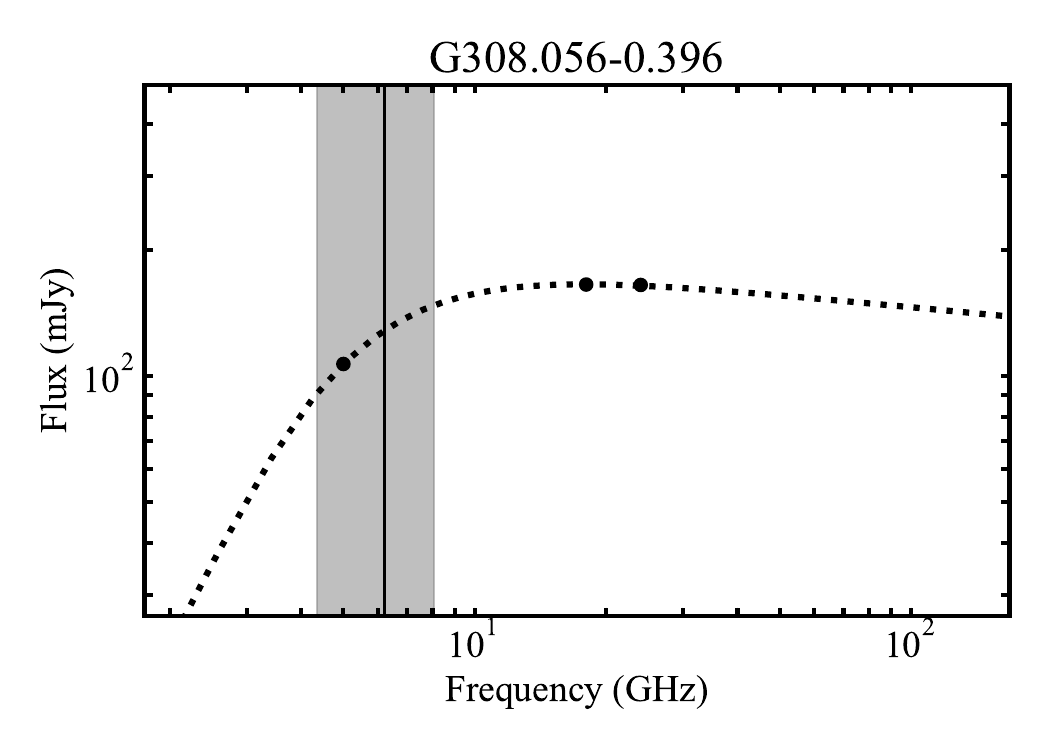}
    \includegraphics[width=0.33\textwidth]{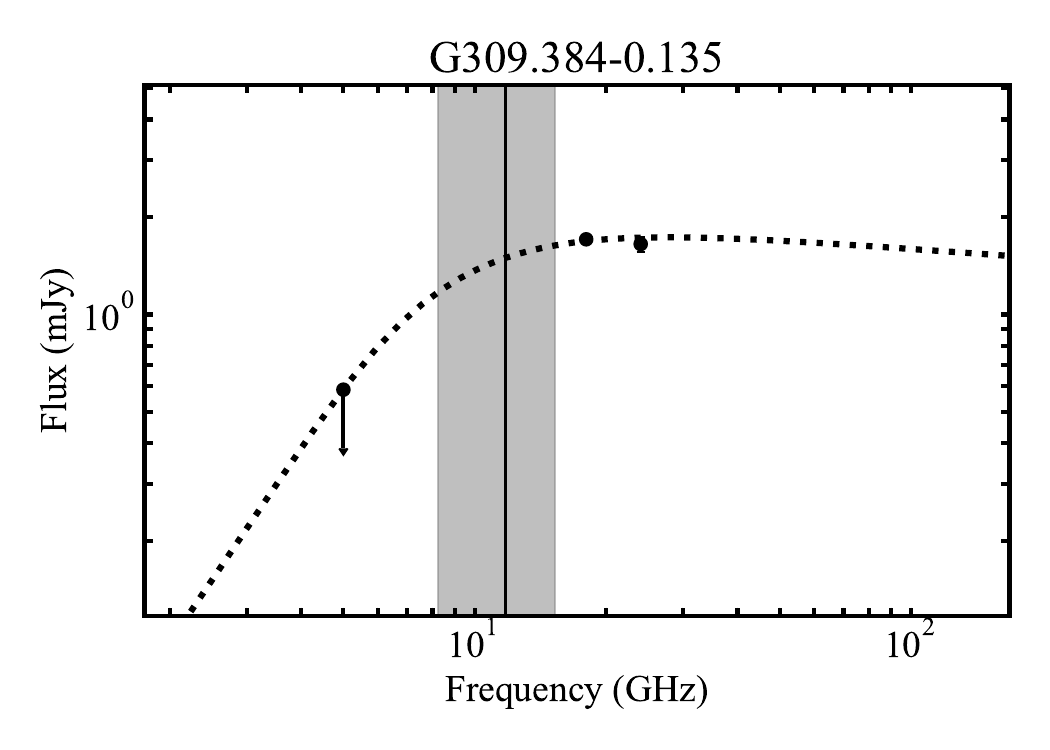}
    \includegraphics[width=0.33\textwidth]{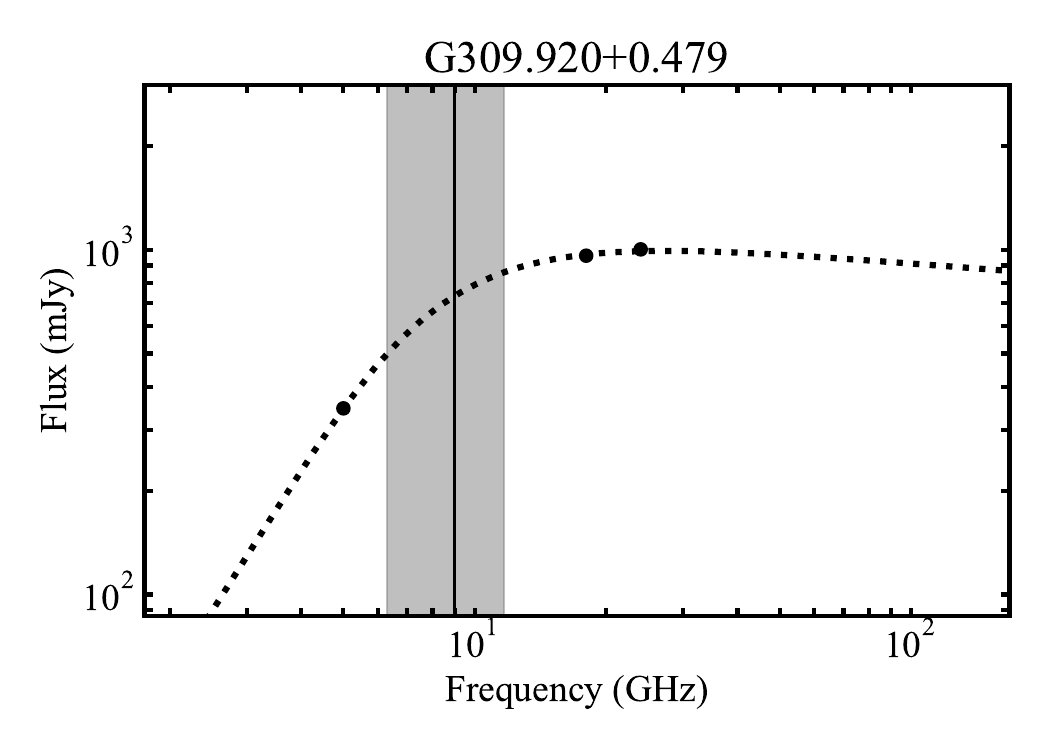}
    \includegraphics[width=0.33\textwidth]{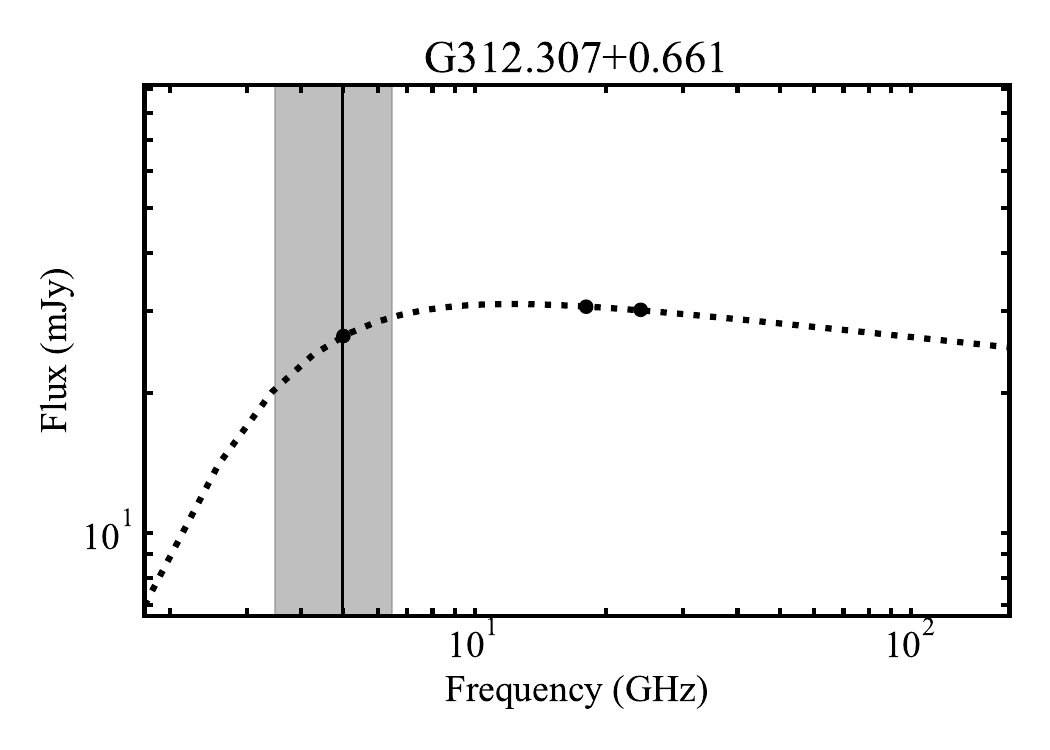}
    \includegraphics[width=0.33\textwidth]{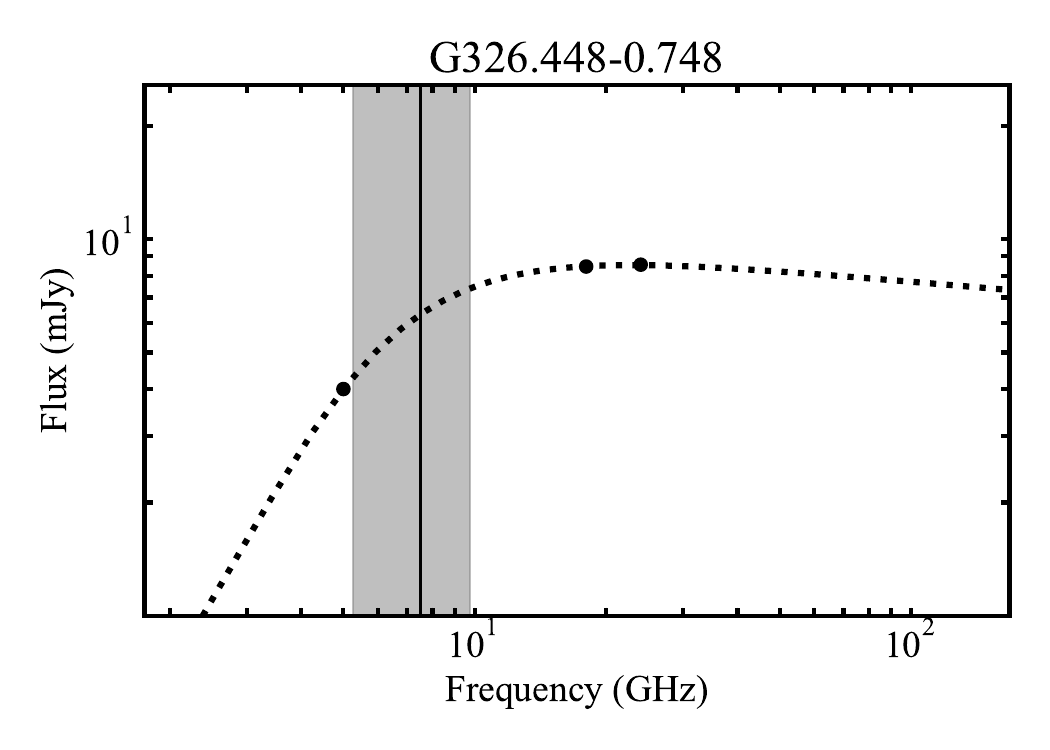}
    \includegraphics[width=0.33\textwidth]{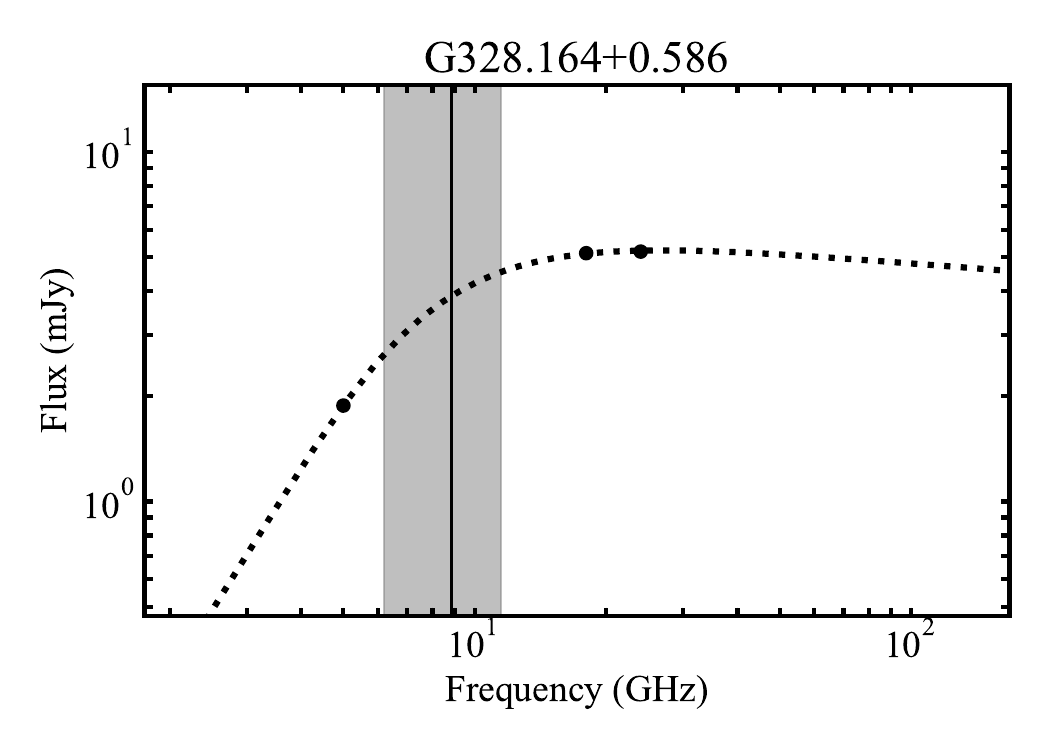}
    \includegraphics[width=0.33\textwidth]{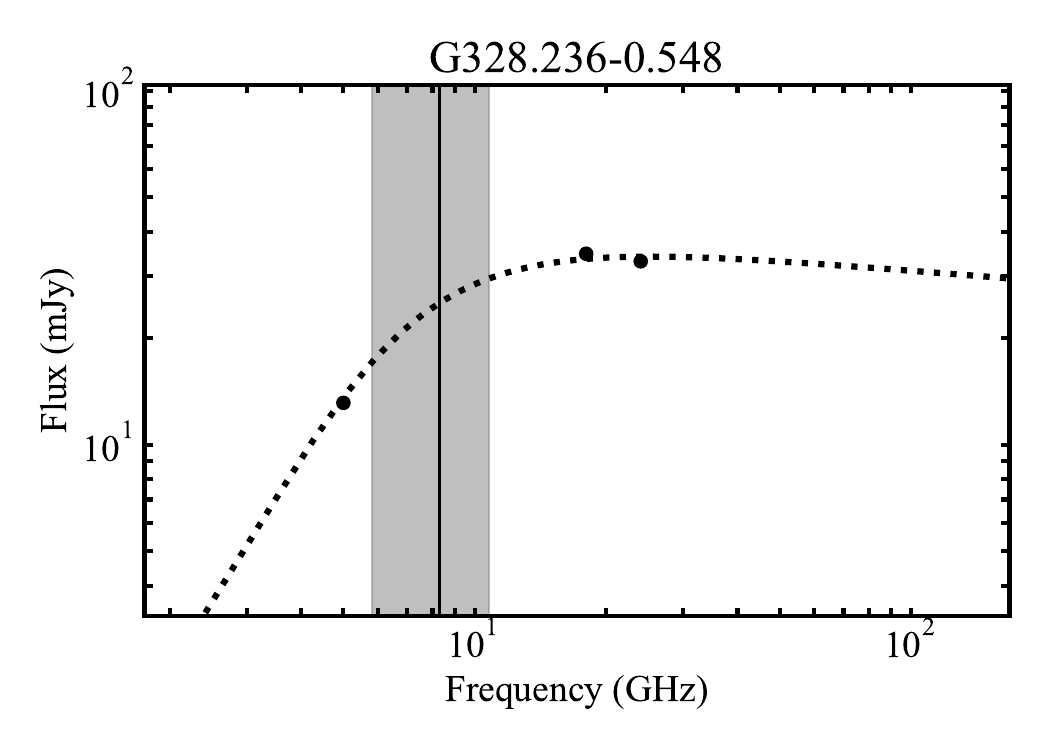}
    \includegraphics[width=0.33\textwidth]{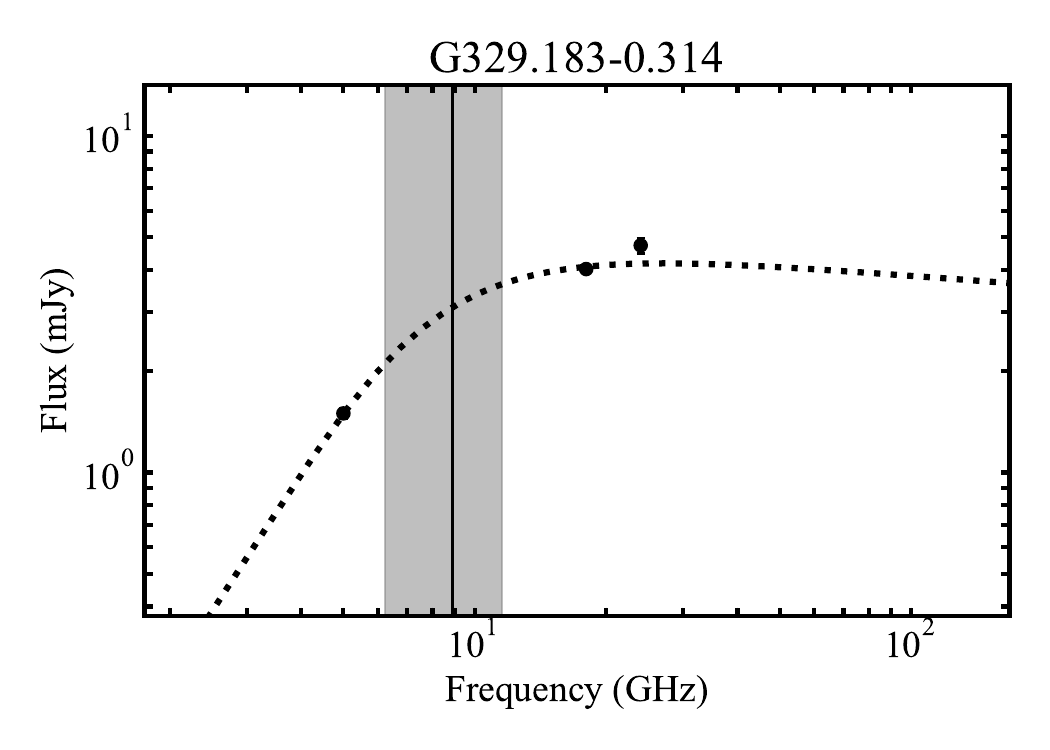}
    \includegraphics[width=0.33\textwidth]{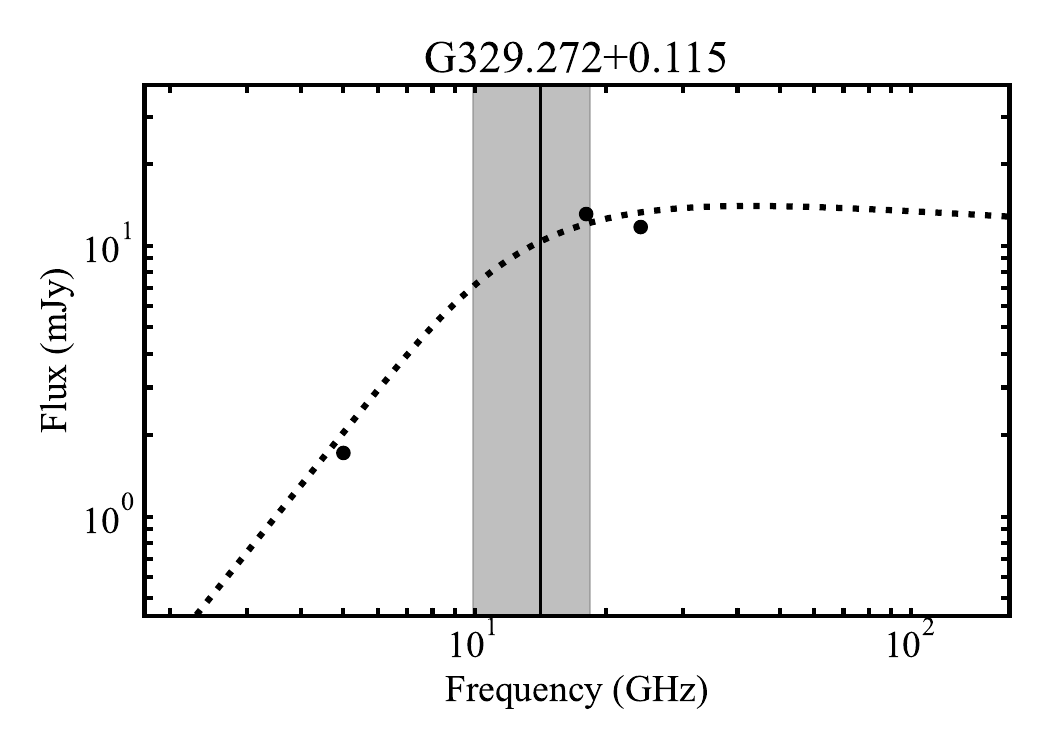}
    \includegraphics[width=0.33\textwidth]{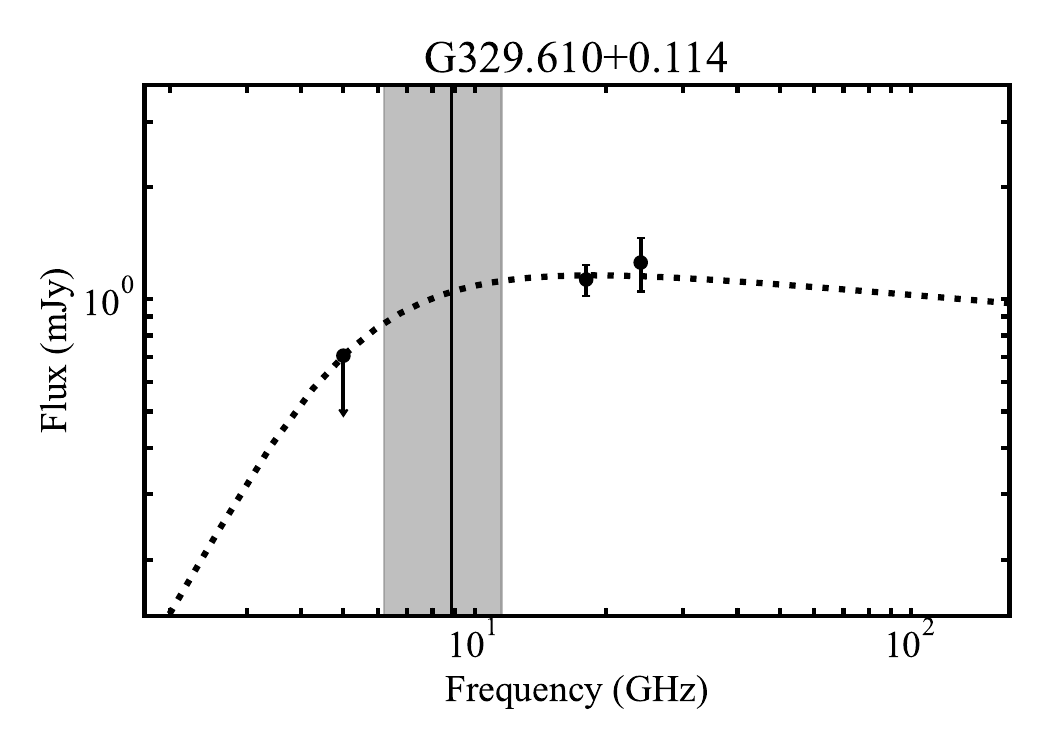}
    \includegraphics[width=0.33\textwidth]{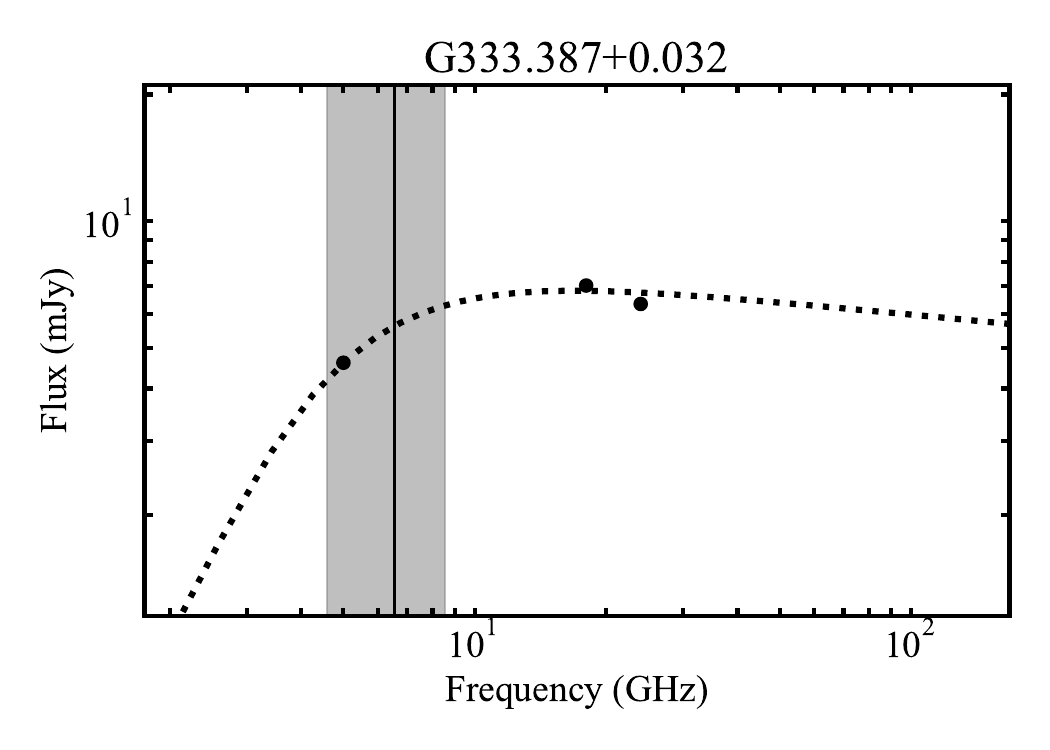}
    \includegraphics[width=0.33\textwidth]{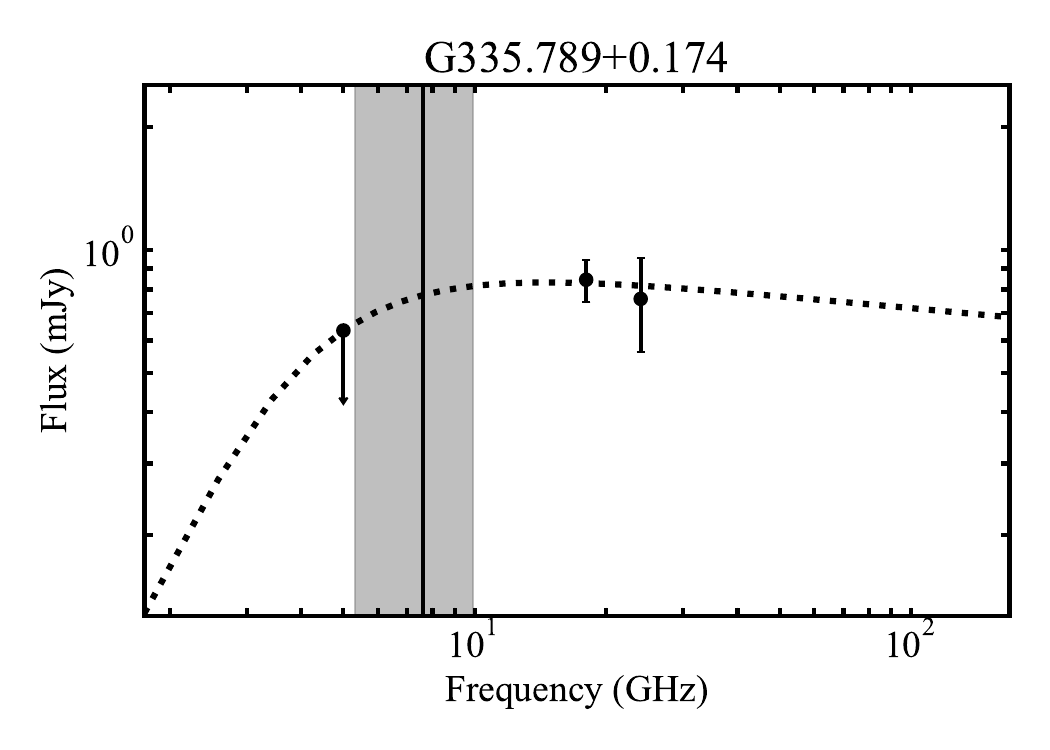}
    \includegraphics[width=0.33\textwidth]{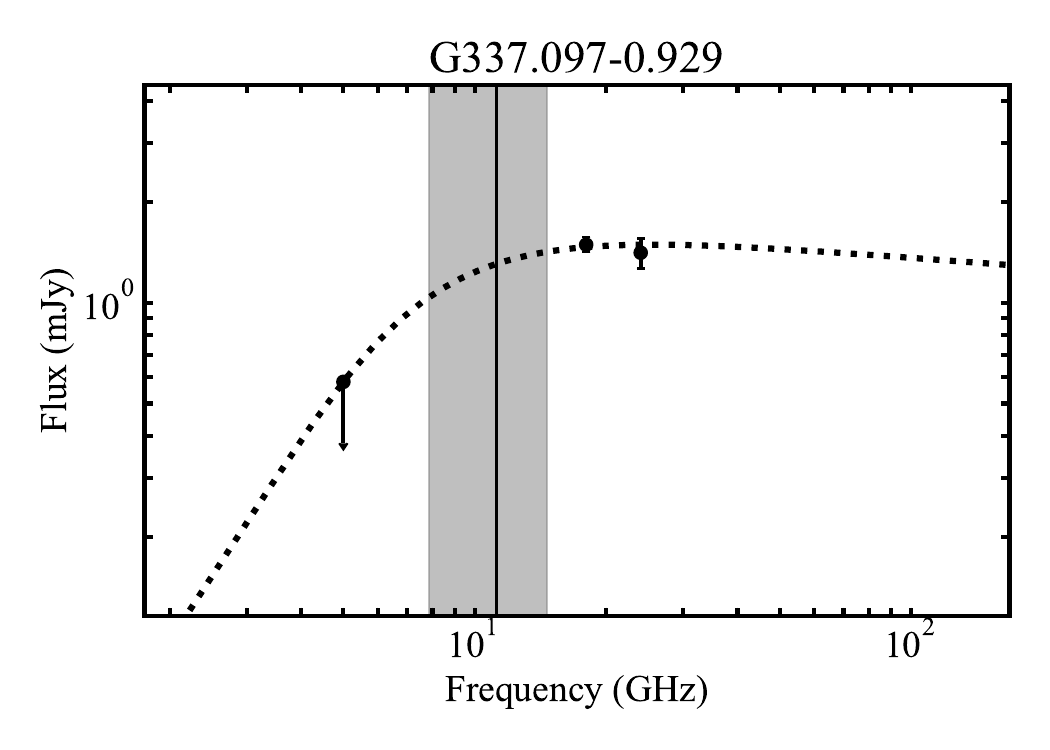}
    \includegraphics[width=0.33\textwidth]{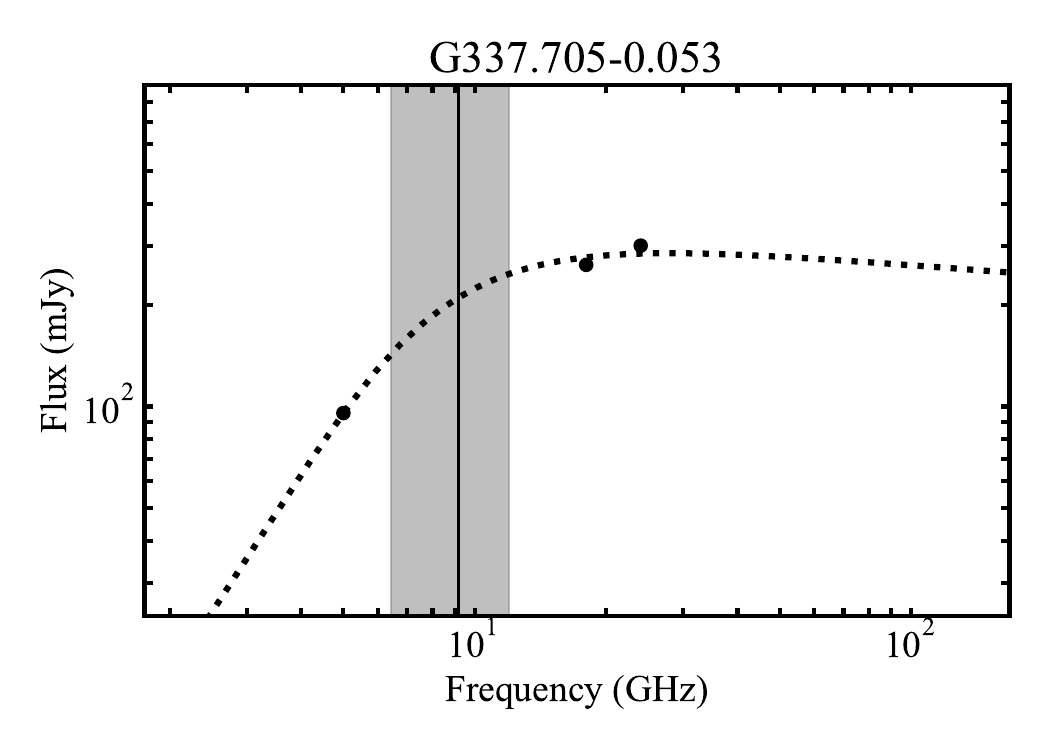}
    \caption{Radio SED fittings for the flux densities between 5 $-$ 24\,GHz. The solid vertical lines shows where the expected turnover frequency of the object lies. The grey shaded area represents the error associated with the turnover frequency}
    \end{figure*}
    \begin{figure*}
    \ContinuedFloat
    \captionsetup{list=off,format=cont}
    \centering
    \includegraphics[width=0.33\textwidth]{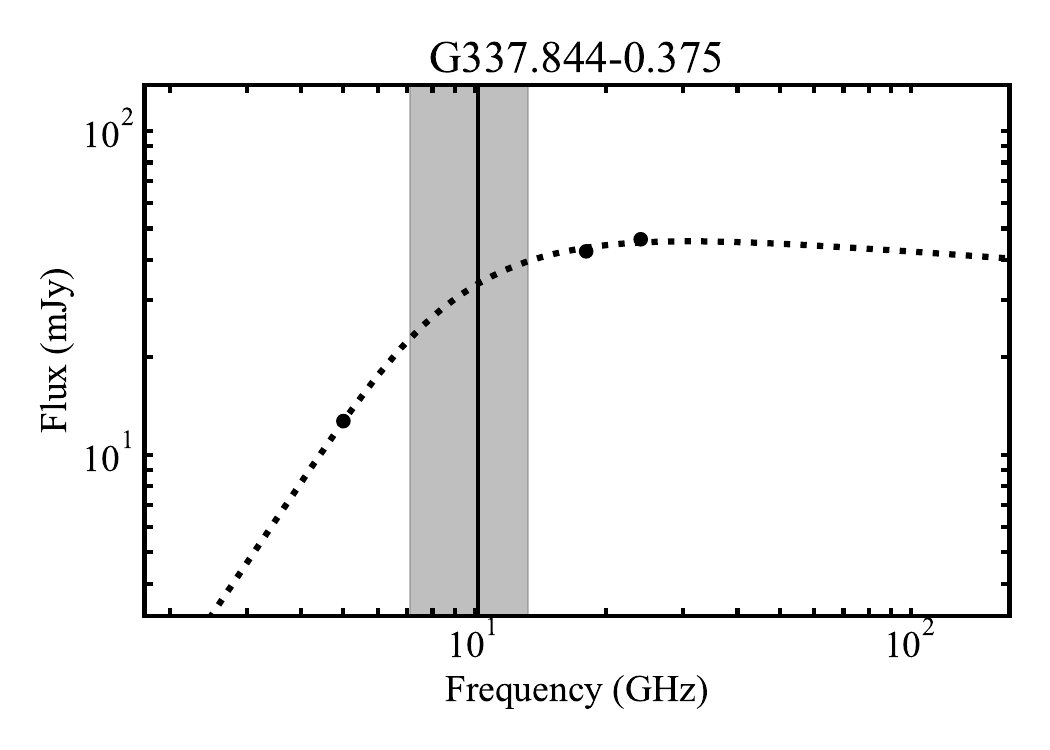}
    \includegraphics[width=0.33\textwidth]{SED/G339.053-0.315.pdf}
    \includegraphics[width=0.33\textwidth]{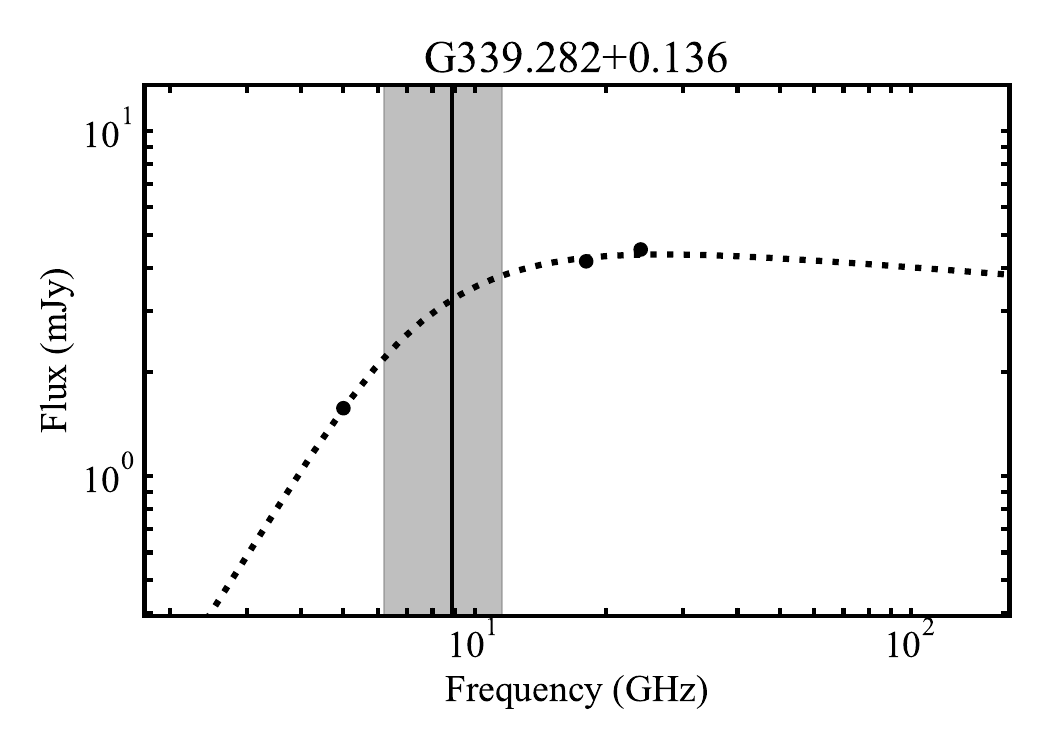}
    \includegraphics[width=0.33\textwidth]{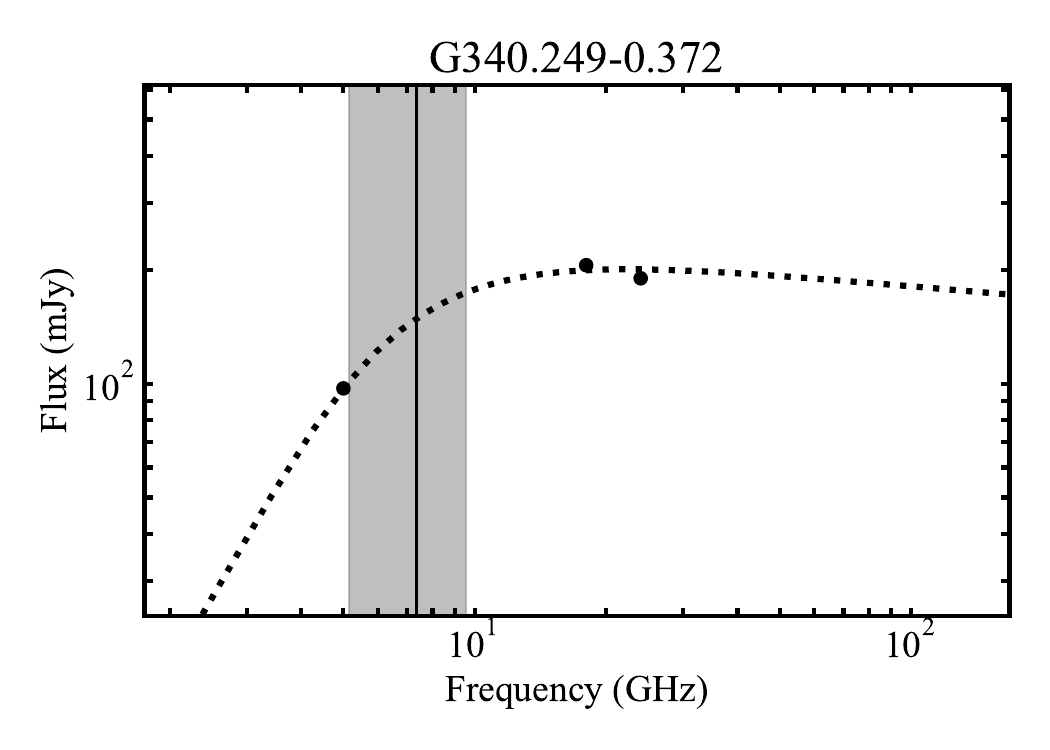}
    \includegraphics[width=0.33\textwidth]{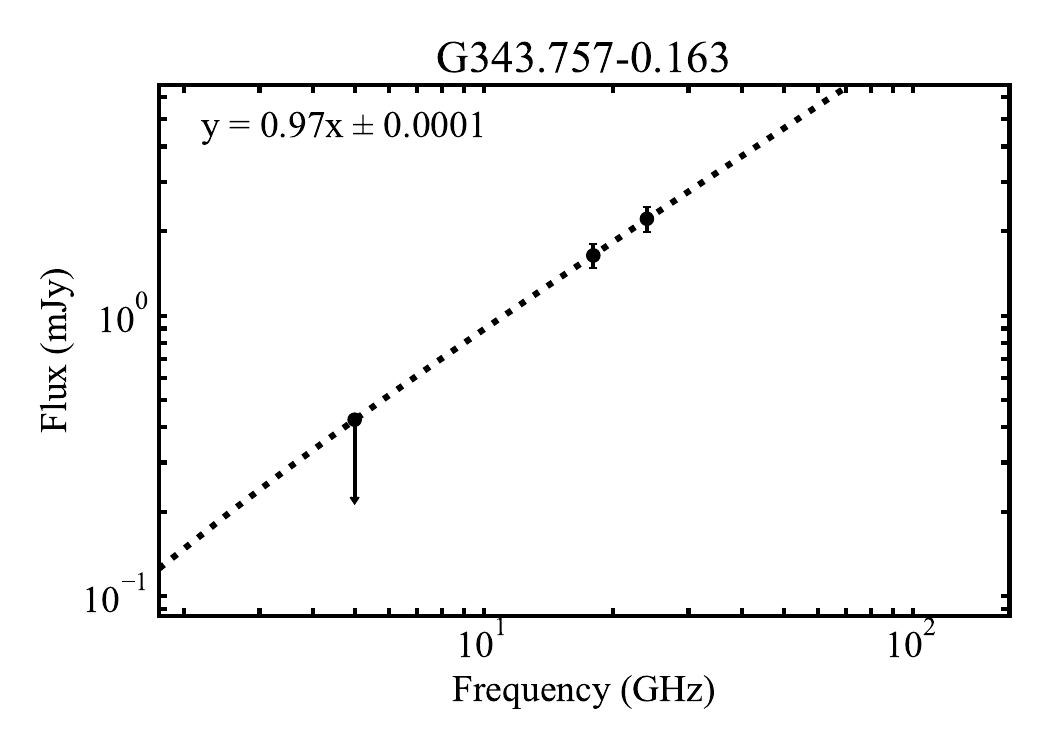}
    \includegraphics[width=0.33\textwidth]{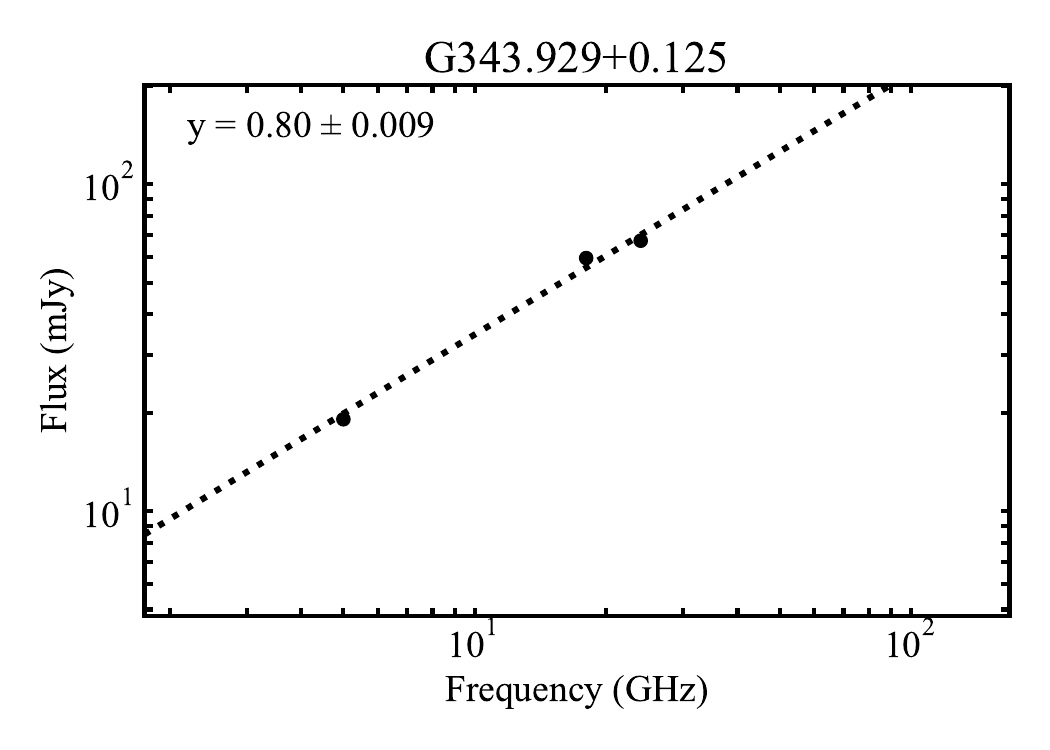}
    \includegraphics[width=0.33\textwidth]{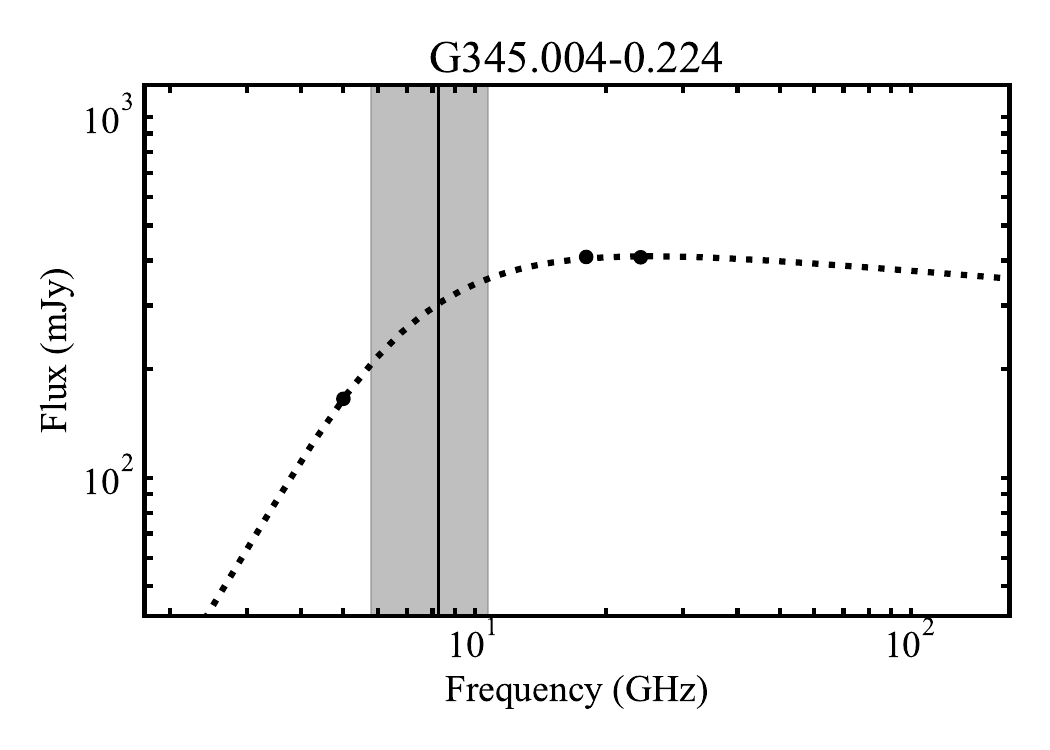}
    \includegraphics[width=0.33\textwidth]{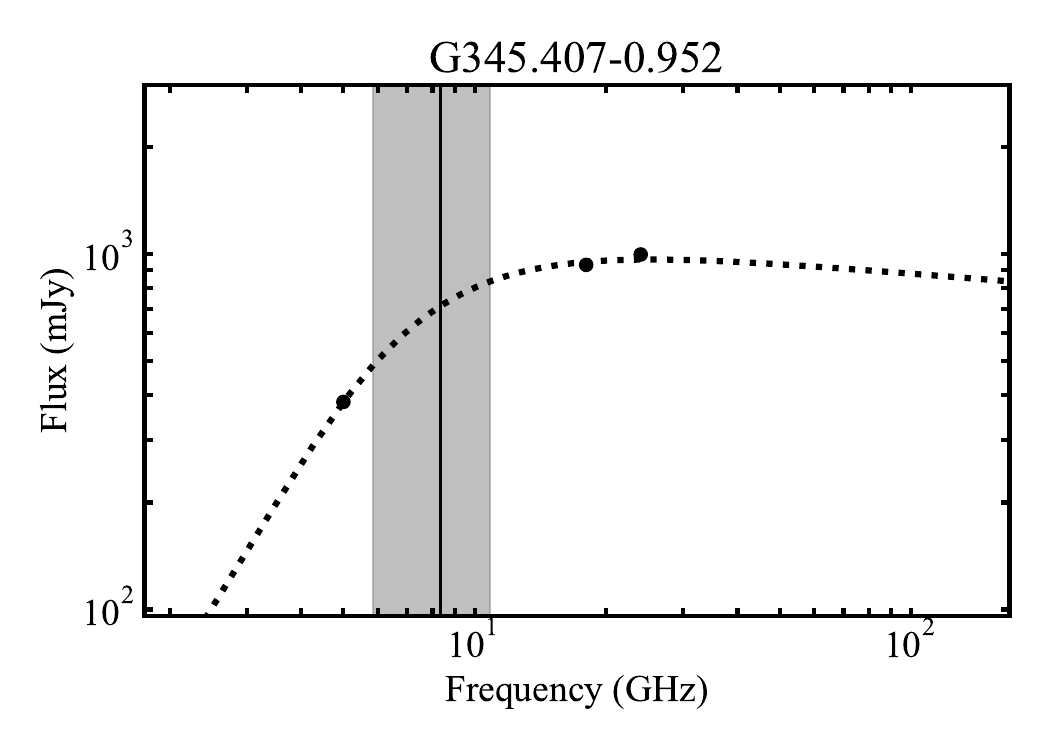}
     \includegraphics[width=0.33\textwidth]{SED/G346.480+0.132.pdf}
    \includegraphics[width=0.33\textwidth]{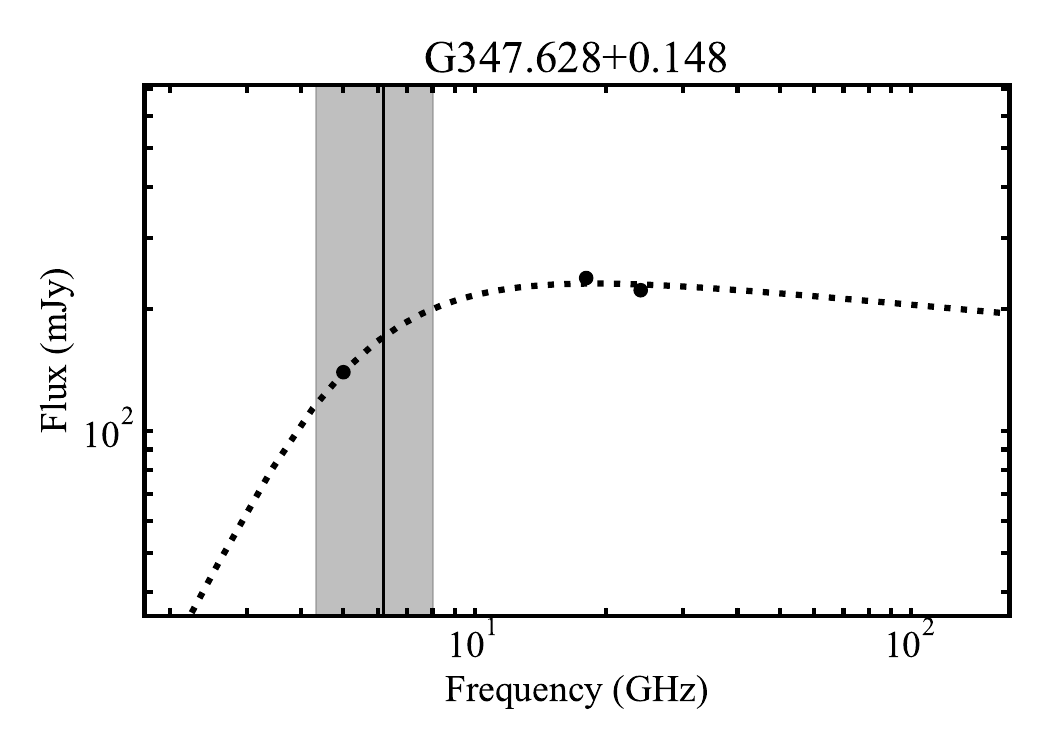}
    \includegraphics[width=0.33\textwidth]{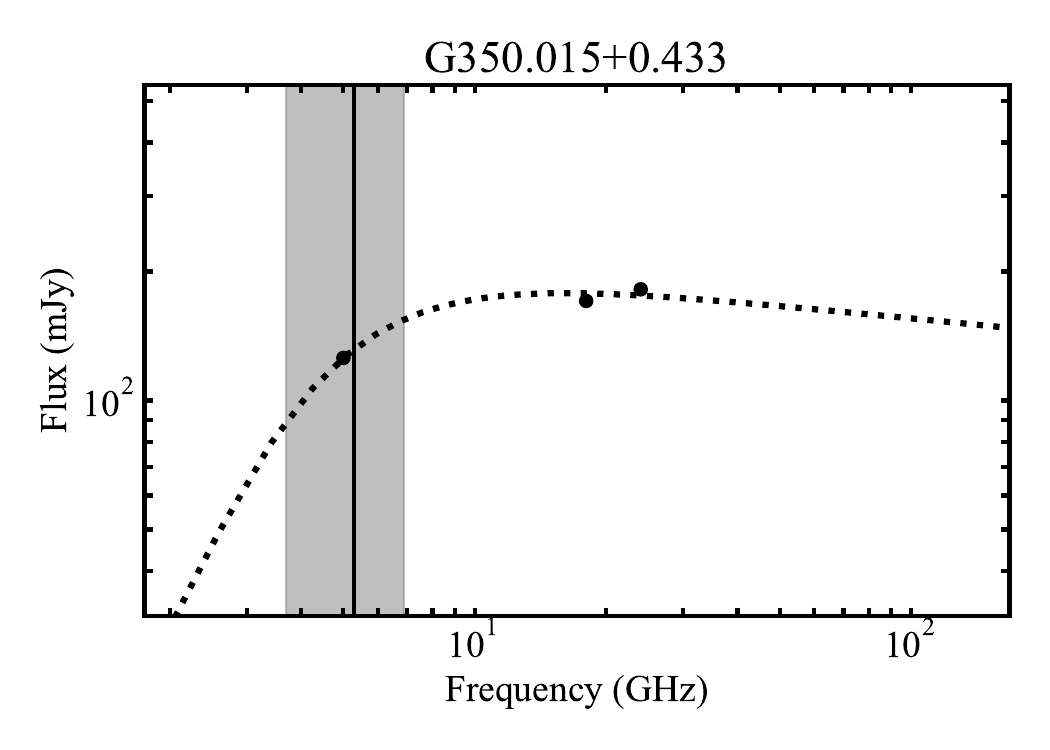}
    \includegraphics[width=0.33\textwidth]{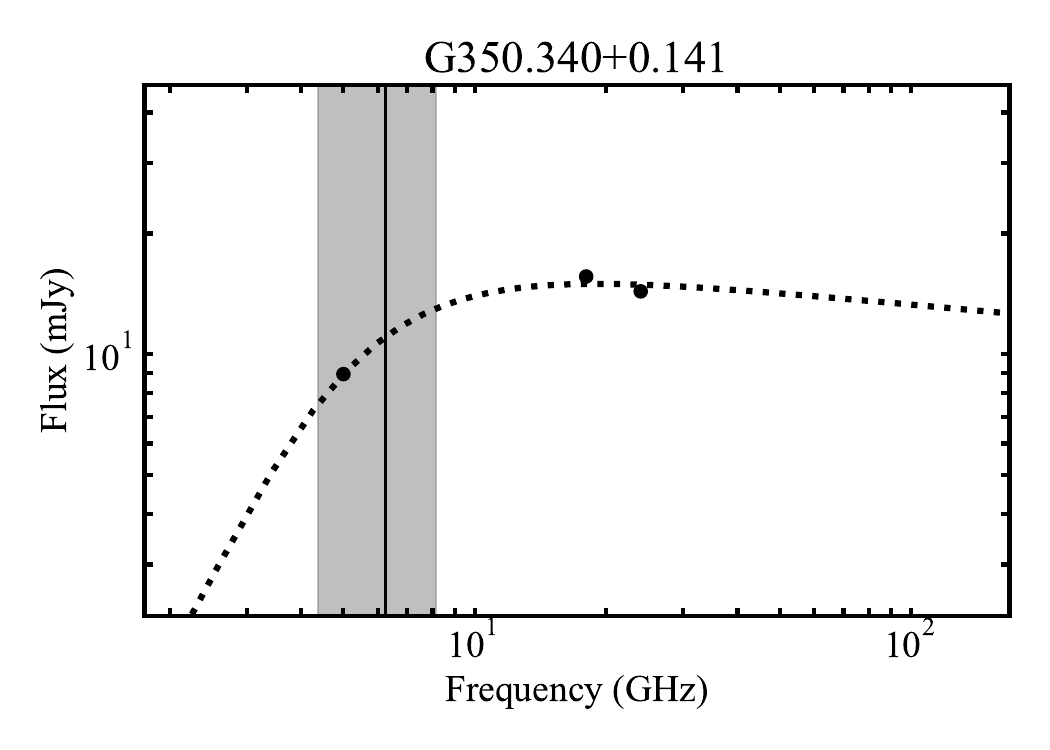}
    \includegraphics[width=0.33\textwidth]{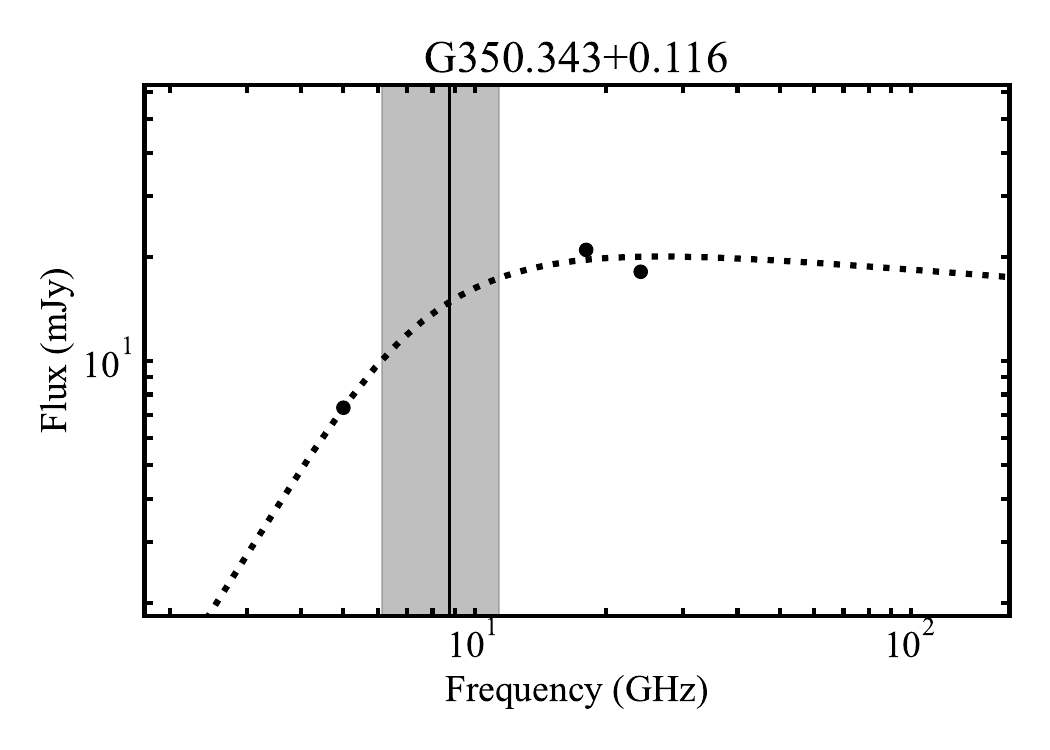}
    \includegraphics[width=0.33\textwidth]{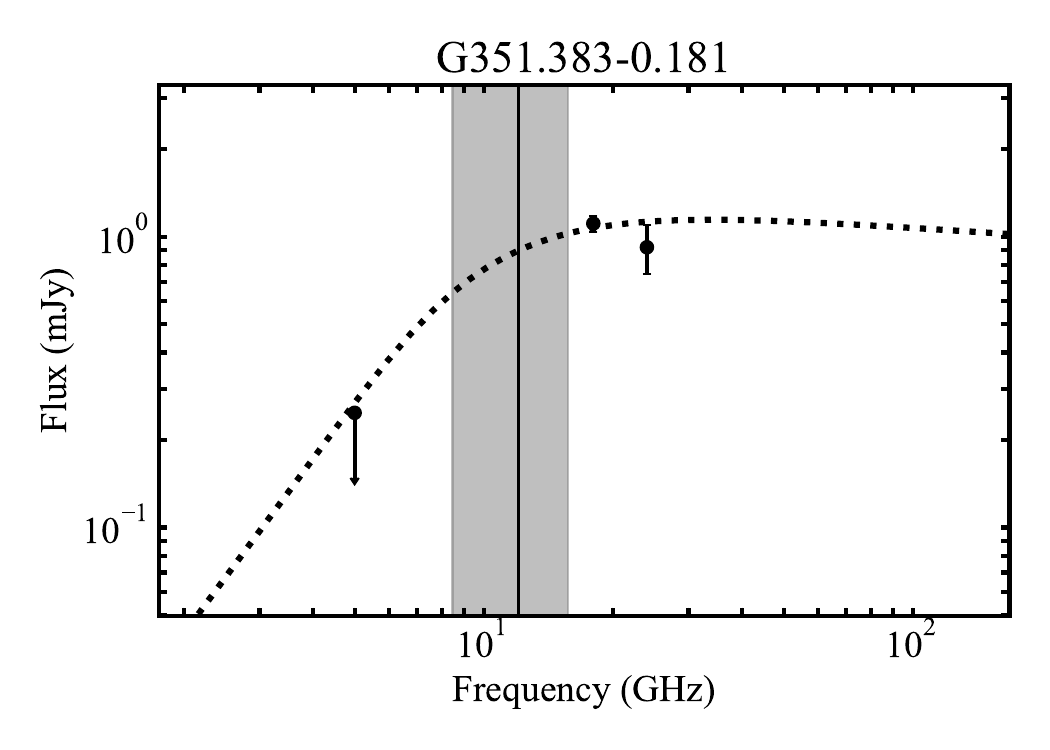}
    \includegraphics[width=0.33\textwidth]{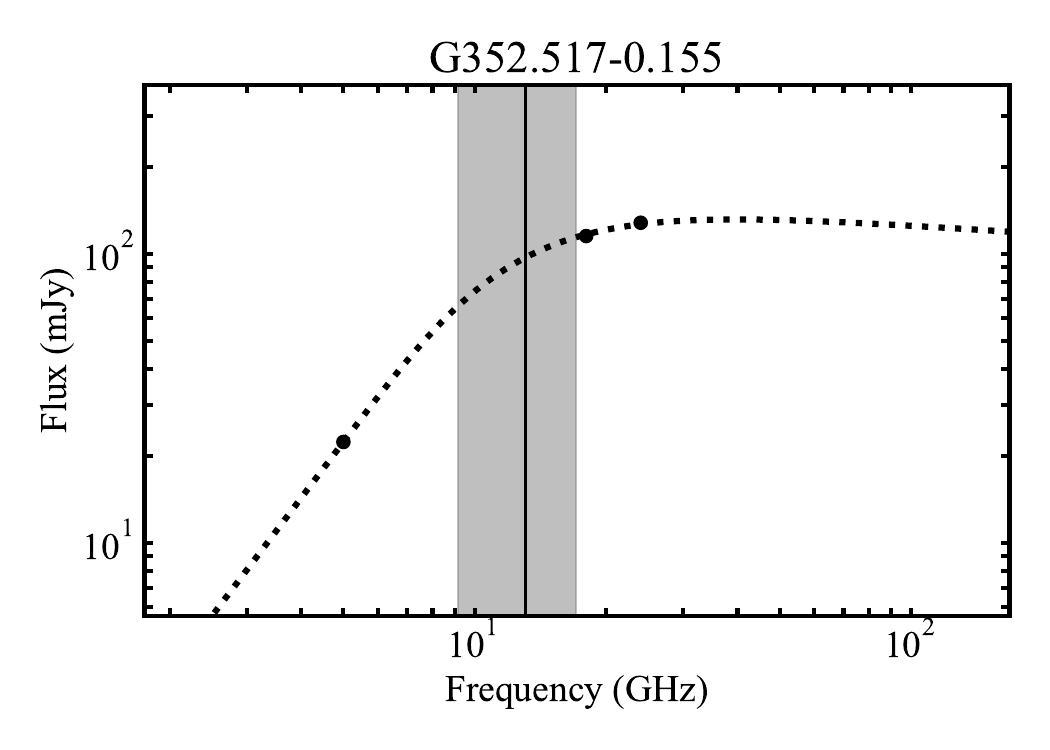}
    \caption{Cont.}
     \label{fig:SED_all}
     \end{figure*}
    \begin{figure*}
    \ContinuedFloat
    \captionsetup{list=off,format=cont}
    \centering
    \includegraphics[width=0.33\textwidth]{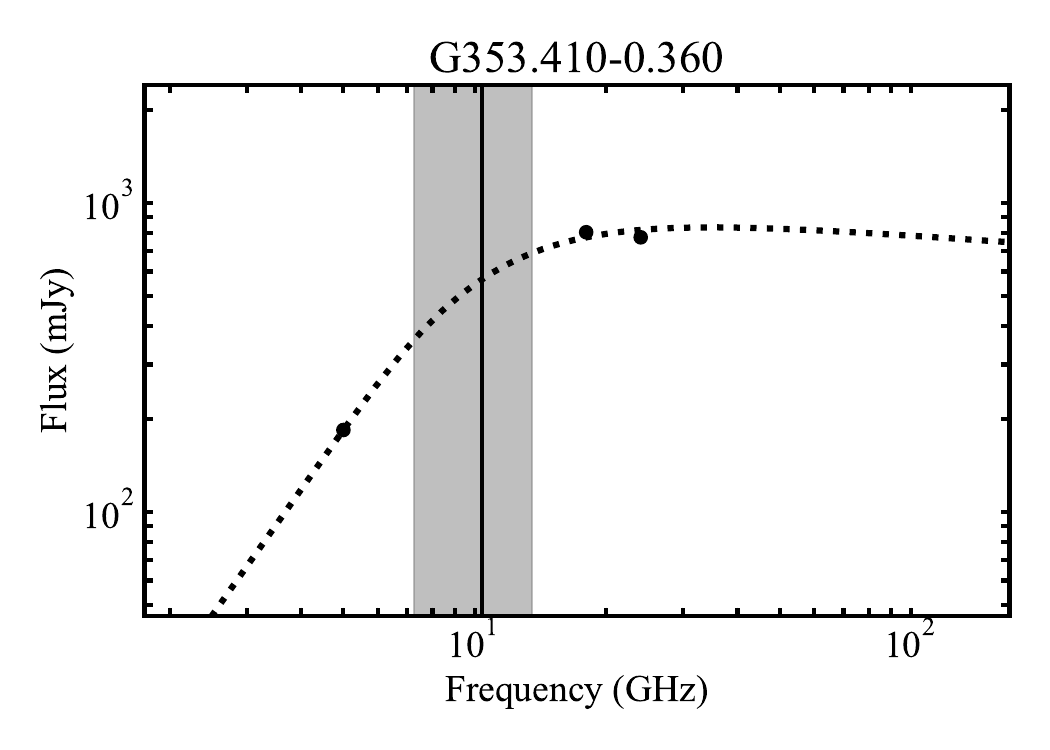}
    \includegraphics[width=0.33\textwidth]{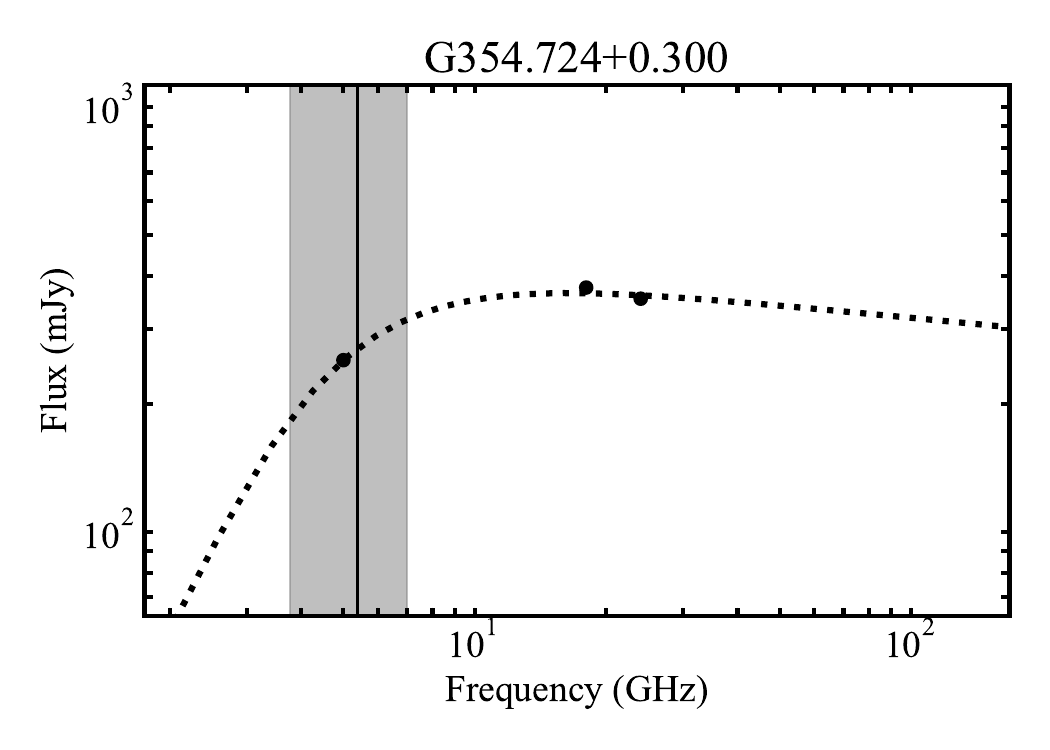}
    \caption{Cont.}
     \label{fig:SED_all}
  \end{figure*}

\section{Optically thick MIR}

\begin{figure*}
  \centering
  \includegraphics[width=0.49\textwidth]{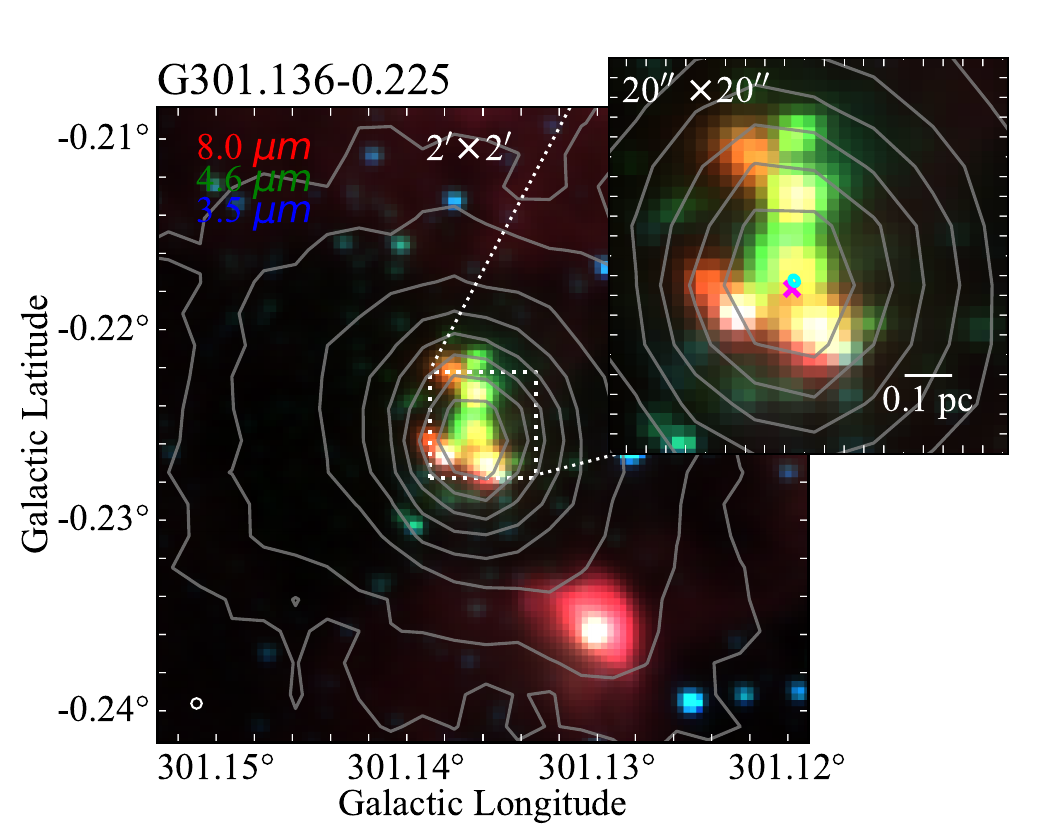}
  \includegraphics[width=0.49\textwidth]{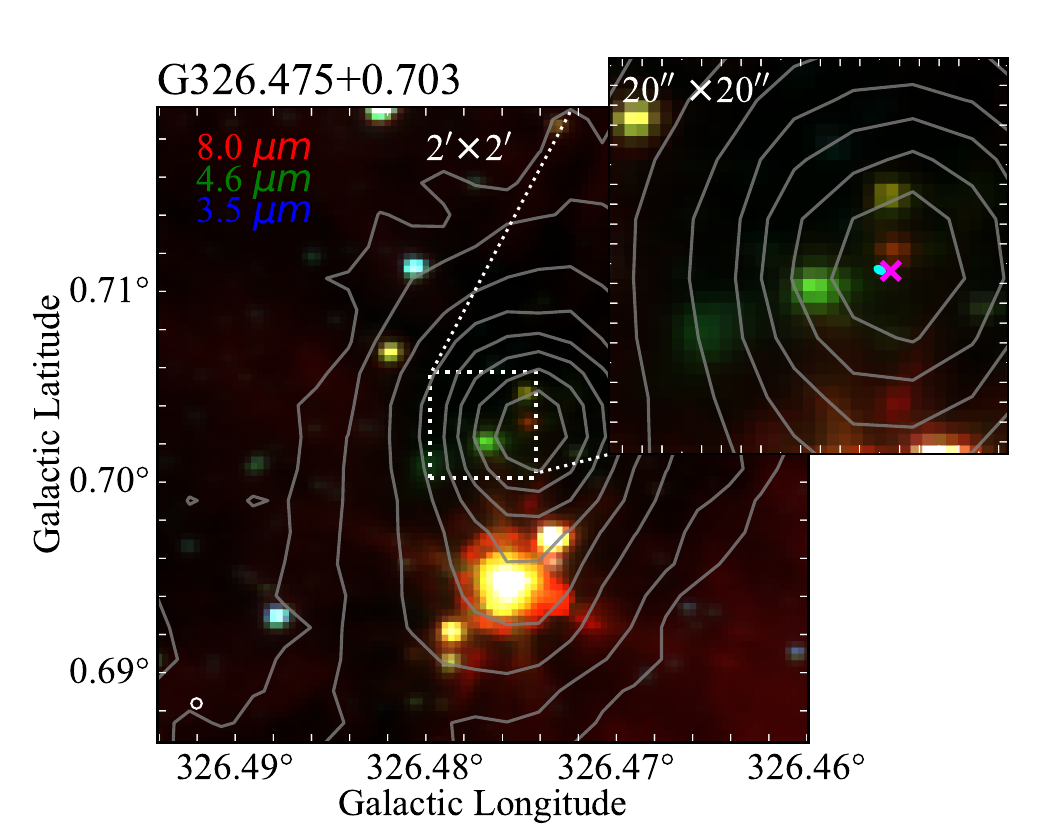}
  \includegraphics[width=0.49\textwidth]{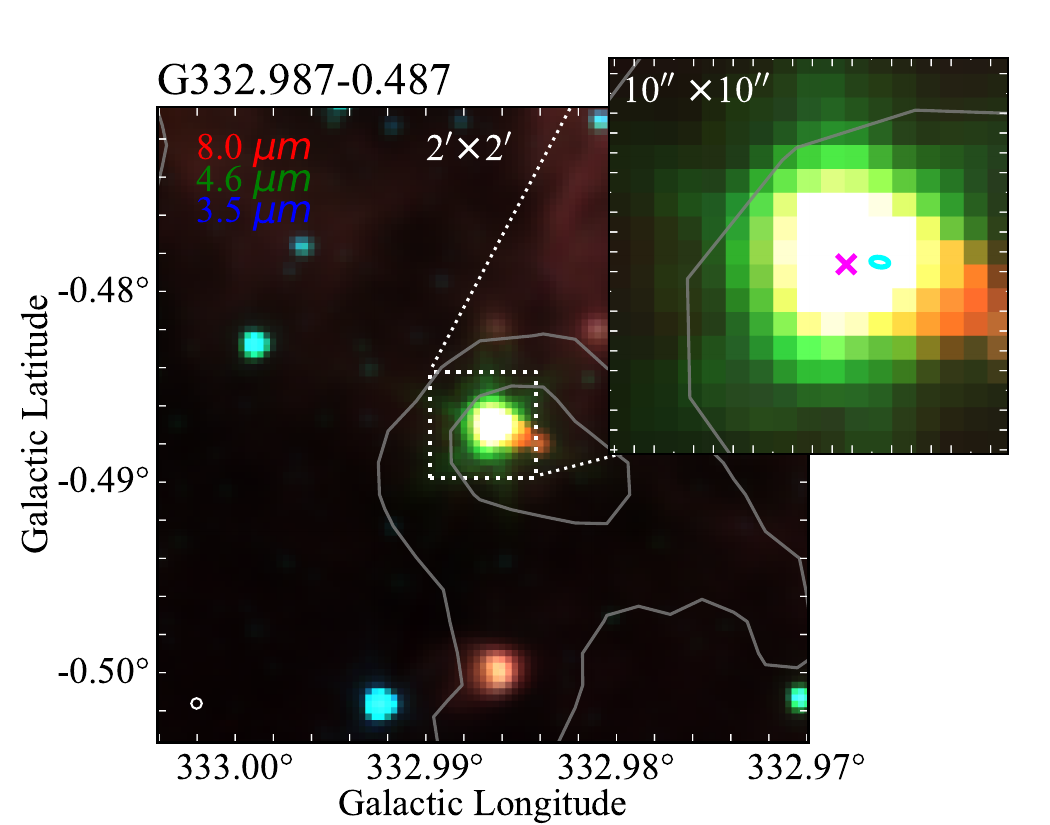}
  \includegraphics[width=0.49\textwidth]{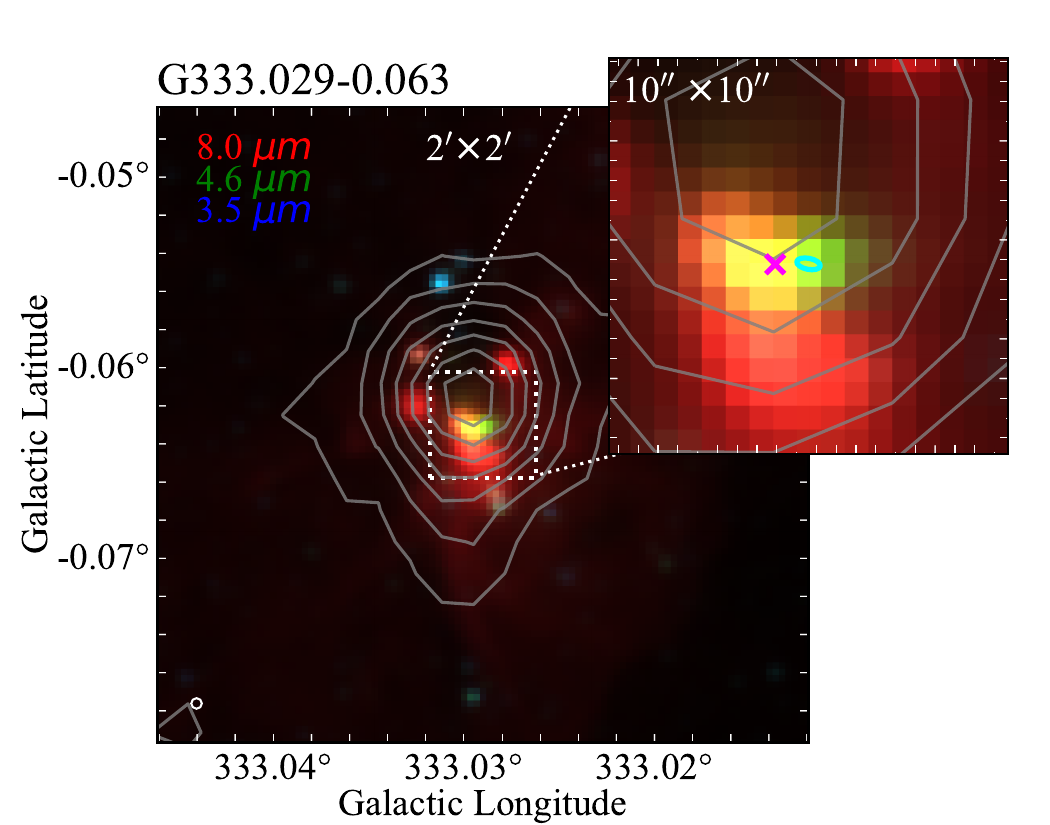}
  \includegraphics[width=0.49\textwidth]{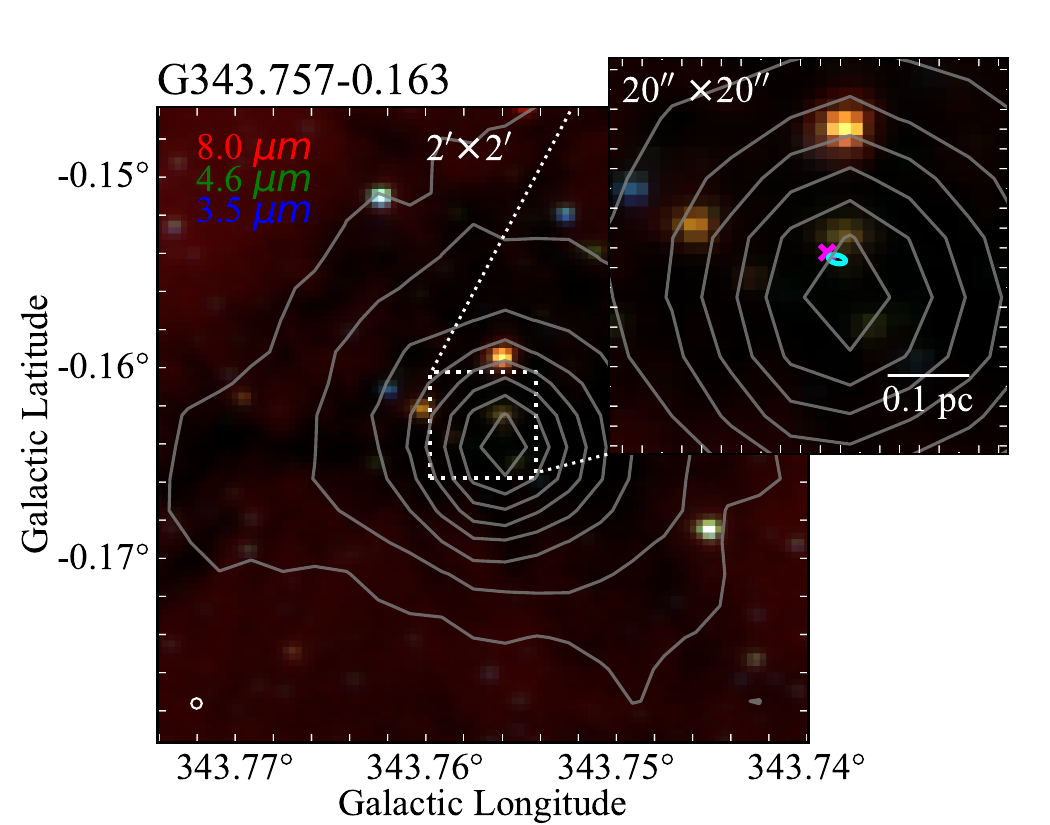}
  \includegraphics[width=0.49\textwidth]{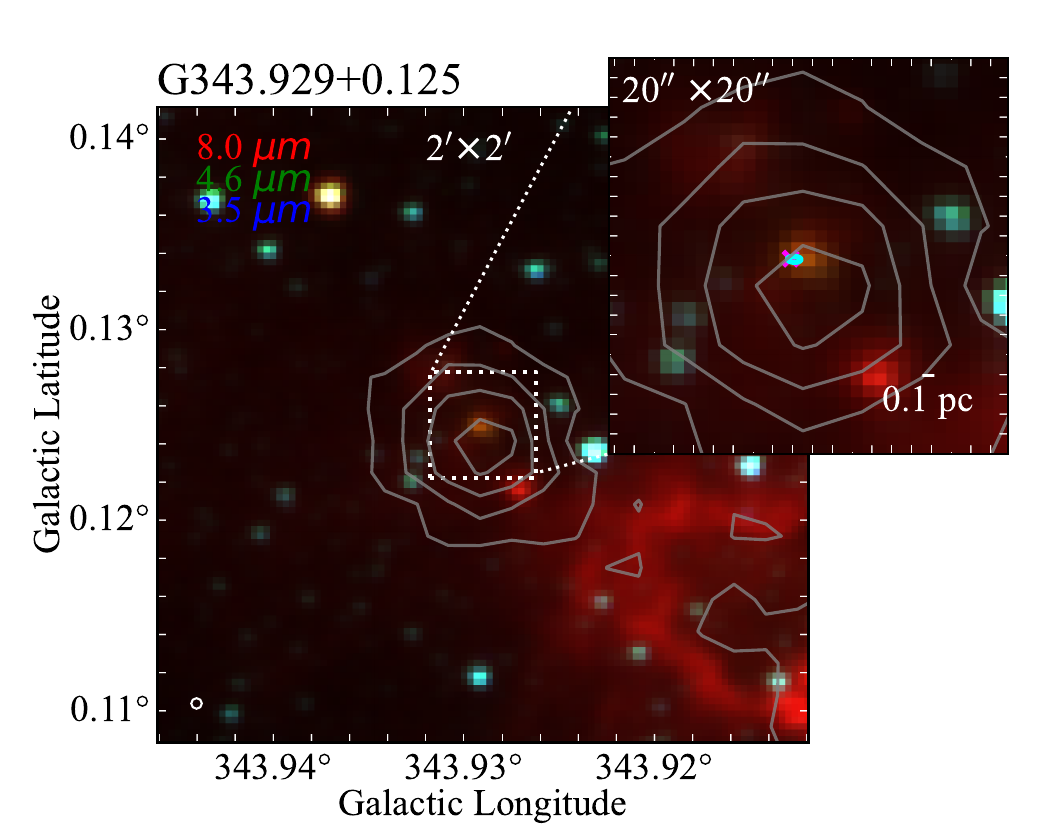}
  \caption{Three-colour composite images of the 11 optically thick compact \hii\ regions. These composite images have been created using Spitzer GLIMPSE 8-\mum\ (red), 4.5-\mum\ (green) and 3.6-\mum\ (blue) bands. The 870-\mum\ dust emission from ATLASGAL is traced by the grey contours \citep{schuller2009}. In the upper-right zoomed panel, the cyan ellipse shows the high-frequency (24-GHz) radio source and the magenta cross shows the position of the MMB methanol maser.}
   \label{fig:rbg_appendix}
   \end{figure*}
    \begin{figure*}
    \ContinuedFloat
    \captionsetup{list=off,format=cont}
    \centering
  \includegraphics[width=0.49\textwidth]{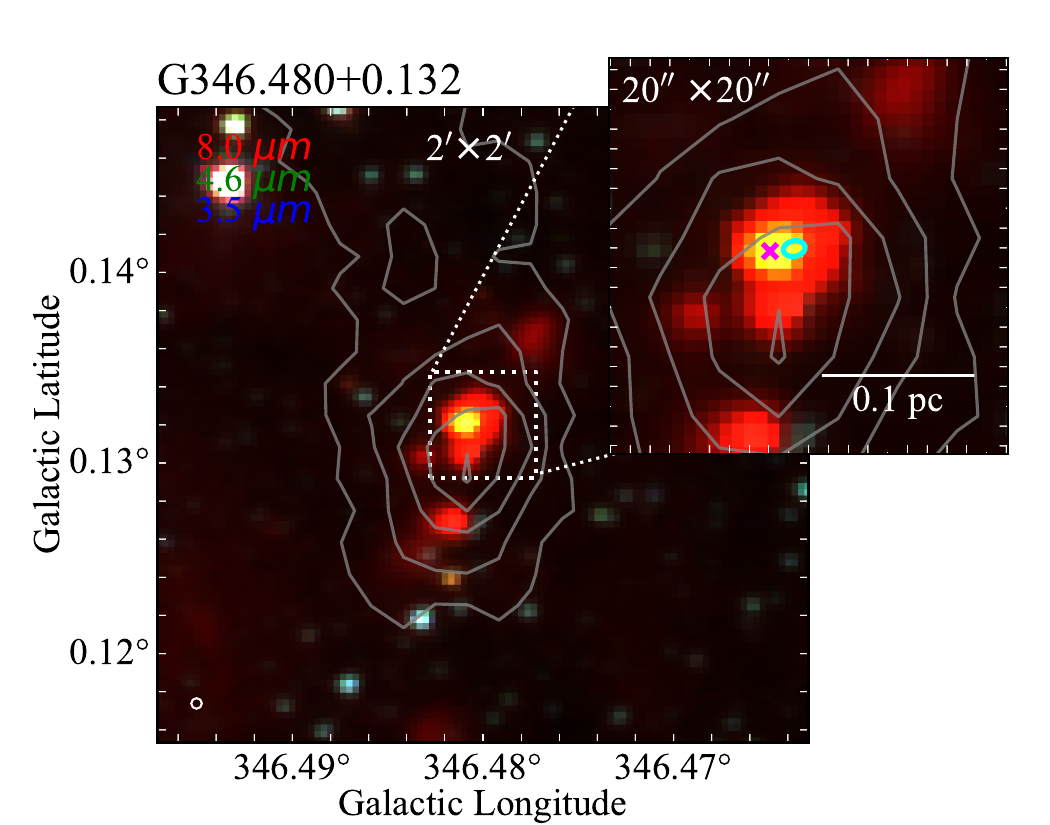}
  \includegraphics[width=0.49\textwidth]{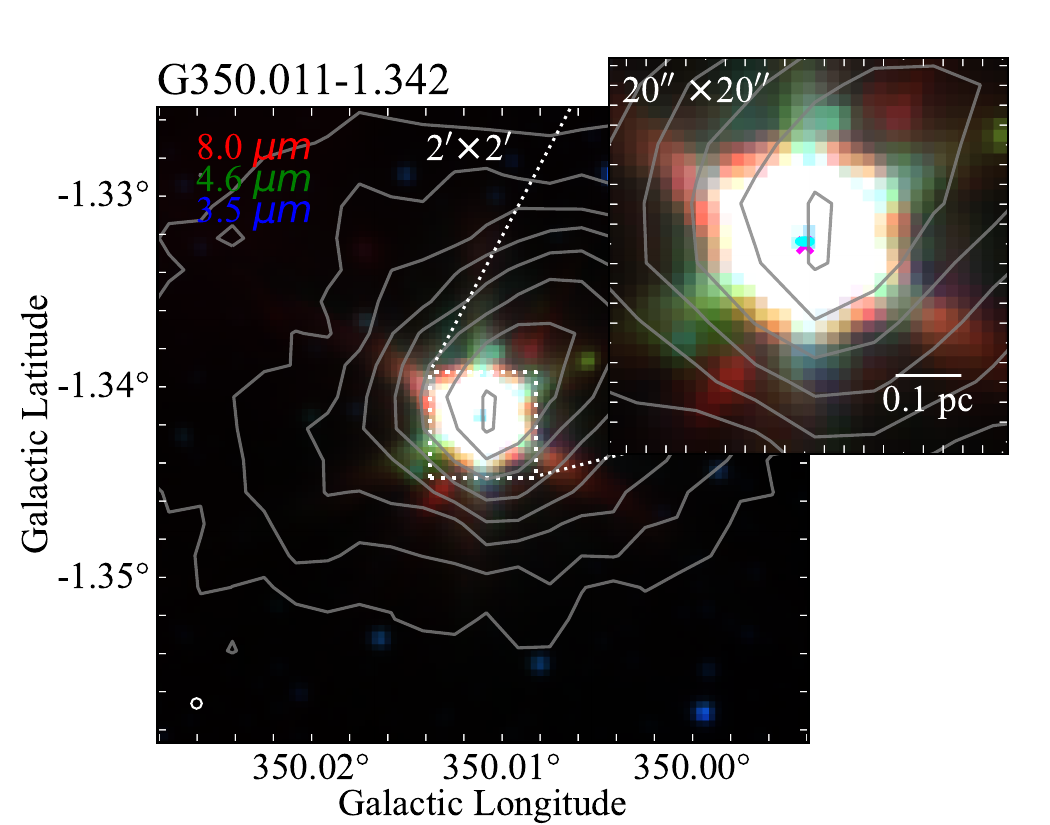}
  \includegraphics[width=0.49\textwidth]{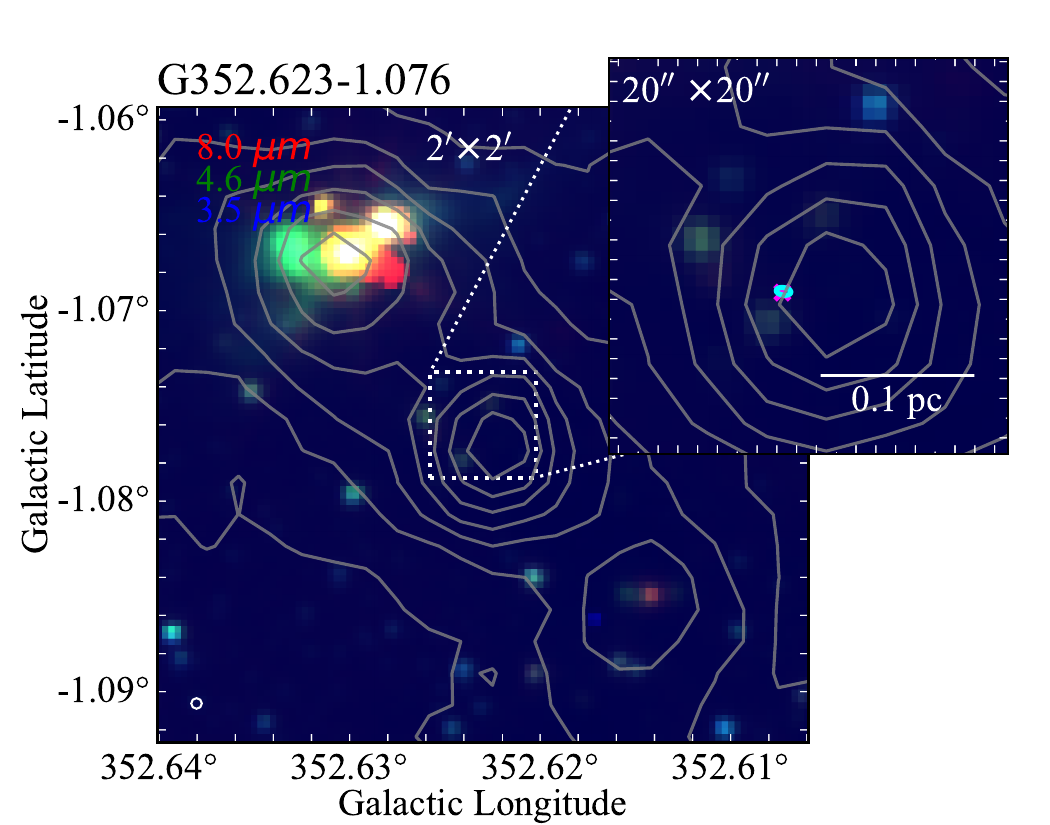}
  \includegraphics[width=0.49\textwidth]{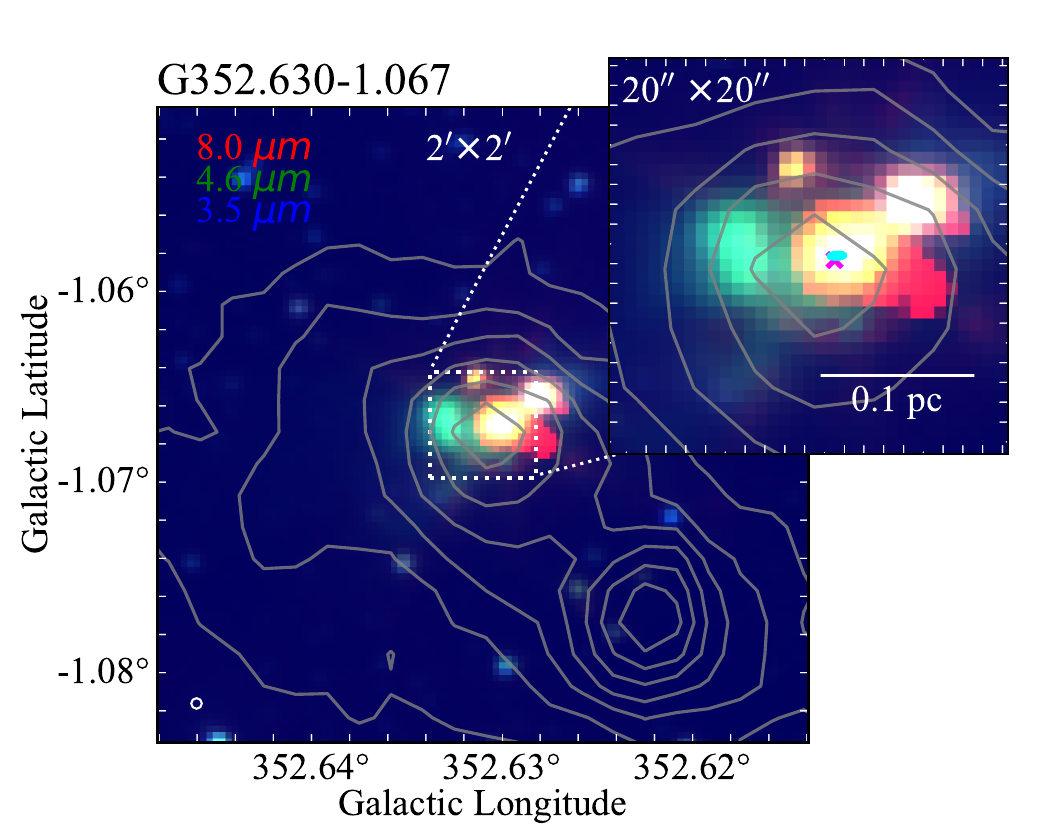}
  \includegraphics[width=0.49\textwidth]{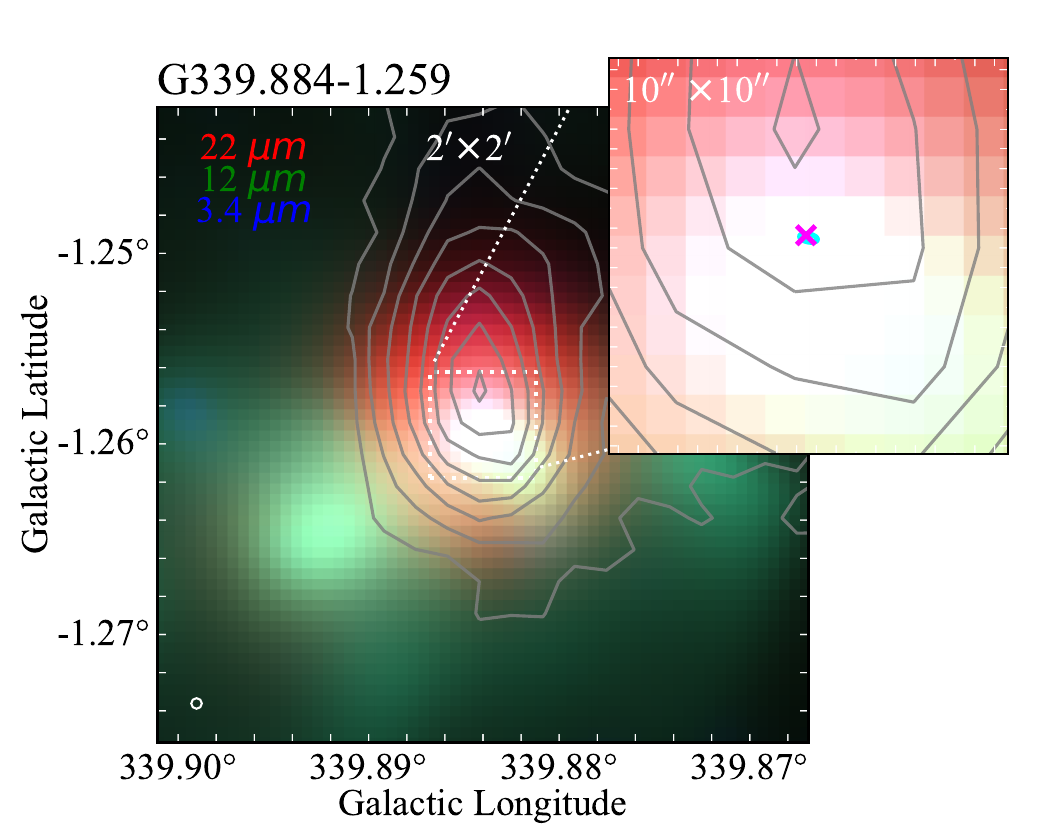}
  \caption{Cont.}
\end{figure*}


\bsp	
\label{lastpage}
\end{document}